\documentclass[%
twocolumn,
superscriptaddress,
longbibliography,
  amsmath,amssymb,
  aps,
 prl,
floatfix,
longbibliography
]{revtex4-2}

\usepackage[colorlinks]{hyperref}
\usepackage{amsmath}
\usepackage[T1]{fontenc}
\usepackage[utf8]{inputenc}
\usepackage{siunitx}
\usepackage{adjustbox}
\usepackage{lmodern}
\usepackage{graphicx}% Include figure files
\usepackage{dcolumn}% Align table columns on decimal point
\usepackage{bm}% bold math
\usepackage[svgnames]{xcolor}
\usepackage[normalem]{ulem}
\newif\ifshowcomments\showcommentstrue

\makeatletter
\newcommand{\titlext}[1]{%
  \onecolumngrid
  \begin{center}
    \large \bfseries #1
  \end{center}
  \vspace{1em}
  \twocolumngrid
}

\makeatletter
\def\maketitle{
\@author@finish
\title@column\titleblock@produce
\suppressfloats[t]}
\makeatother

\newcommand{\bairo}{Ba$_2$IrO$_4$}
\newcommand{\sriro}{Sr$_2$IrO$_4$}

\newcommand{\cdag}{\hat{c}^{\dagger}}
\renewcommand{\c}{\hat{c}^{\phantom{\dagger}}}

\newcommand{\jeff}{$j_{\mathrm{eff}}=1/2$}
\newcommand{\JEFF}{$j_{\mathrm{eff}}=3/2$}

\newcommand{\ri}{\mathbf{r}}
\renewcommand{\k}{\mathbf{k}}

\begin{document}

\preprint{APS}

\title{\texorpdfstring{ Spin-polaron fingerprints in the optical conductivity of iridates}{Spin-polaron fingerprints in the optical conductivity of iridates}}% Force line breaks with \\

\author{Francesco Cassol}
%\email{francesco.cassol@polytechnique.edu}
 \affiliation{Institut de Minéralogie, de Physique des Matériaux et de Cosmochimie, Sorbonne Université,\\CNRS, MNHN, UMR 7590, 4 Place Jussieu, F-75005 Paris, France}
 \affiliation{Centre de Physique Théorique, CNRS, Ecole Polytechnique, Institut Polytechnique de Paris, F-91128 Palaiseau, France}
\author{Léo Gaspard}%
 %\email{leo.gaspard@sorbonne-universite.fr}
  \affiliation{Laboratoire de Chimie et Physique Quantiques, Université de Toulouse, CNRS UMR 5626, \\  118 Route de Narbonne, 31062 Toulouse Cedex 09, France}
 \affiliation{Institut de Minéralogie, de Physique des Matériaux et de Cosmochimie, Sorbonne Université,\\CNRS, MNHN, UMR 7590, 4 Place Jussieu, F-75005 Paris, France}
 \author{Cyril Martins}
%\email{cyril.martins@irsamc.ups-tlse.fr}
  \affiliation{Laboratoire de Chimie et Physique Quantiques, Université de Toulouse, CNRS UMR 5626, \\  118 Route de Narbonne, 31062 Toulouse Cedex 09, France}
\author{Michele Casula}
%\email{michele.casula@sorbonne-universite.fr}
 \affiliation{Institut de Minéralogie, de Physique des Matériaux et de Cosmochimie, Sorbonne Université,\\CNRS, MNHN, UMR 7590, 4 Place Jussieu, F-75005 Paris, France}
\author{Benjamin Lenz}
%\email{benjamin.lenz@sorbonne-universite.fr}
 \affiliation{Institut de Minéralogie, de Physique des Matériaux et de Cosmochimie, Sorbonne Université,\\CNRS, MNHN, UMR 7590, 4 Place Jussieu, F-75005 Paris, France}

\date{\today}% It is always \today, today,

\begin{abstract}

As a consequence of their spin-orbit entangled ground state,  many $5d^{5}$ iridate materials display a peculiar double peak structure in optical transport quantities, such as absorption and conductivity. % 
Their common interpretation is based on the presence of Hubbard subbands in the half-filled \jeff\ manifold.
Herein, we challenge this picture, proposing a scenario based on the presence of spin-polaron (SP) quasiparticles, and assigning a dominant SP character to the first peak.
We illustrate it by taking the materials \bairo\ and \sriro\ as paradigmatic examples, which we investigate within the dynamical mean-field theory and the self-consistent Born approximation. 
Both theories reproduce nontrivial features revealed by angle-resolved photoemission spectroscopy and optical transport measurements, supporting our interpretation. 
In the case of \sriro\, we show how the SP scenario survives in the low-doped regime.
Similar optical transport fingerprints are expected to be found in the wider class of $5d^5$ iridates and more generally in strongly correlated antiferromagnetic regimes, such as those found in cuprates.
\end{abstract}

\maketitle

Among bulk-sensitive experimental techniques, the measurement of optical transport quantities holds a privileged position in the investigation of quantum materials. 
Due to its capability of probing the excited states of a system, optical measurements provide an indirect yet powerful way of accessing information about the unoccupied states in the electronic spectrum \cite{Dressel_Grüner_2002}. 
The optical conductivity thereby contains fingerprints of superconductivity \cite{Molegraaf_2002}, charge-density waves \cite{Katsufuji_1995} or pseudogap phases \cite{Puchkov_1996}, 
and constitutes an essential tool in the study of transition metal oxides
such as high-temperature cuprate superconductors, iron pnictides, or nickelates \cite{Gervais_2002,Millis_2004,Basov_2011,Charnukha_2014}.
For these correlated materials, rationalizing the conductivity represents a nontrivial problem which often requires the use of advanced numerical tools like the GW method \cite{Hedin_1965} or the dynamical mean-field theory (DMFT) \cite{Georges_1996}.

In the case of the perovskite iridates \cite{Rau2016,Bertinshaw2019}, the characteristic of the optical conductivity $\sigma(\omega)$ is a rather universal two-peak structure in the in-plane component.
For its paradigmatic member \sriro\cite{Crawford_1994,Moon2006,Kim_2008,Kim_2009,kim_2012,Kim_2014,Brouet_2015,Dai_2014,Kim_2016,Perkins2014}, it is preserved over a large temperature range and across a magnetic phase transition \cite{moon_2009, Sohn_2014,seo_2017}, under strain \cite{Kim_2016}, extends to the low-doping regime \cite{seo_2017}, and remains robust upon chemical substitutions with Ba \cite{Souri_2017} and Ca \cite{Souri_2016}. 
The presence of both peaks is also detected in the second member of the Ruddlesden-Popper series, Sr$_3$Ir$_2$O$_7$ \cite{Moon2008,Kim_2022}, but only the high-energy peak survives for the series' end member SrIrO$_3$ \cite{Moon2008}. 
However, despite extensive theoretical and experimental investigations of optical transport properties, a fully satisfactory quantitative agreement between theoretical simulations and experiments remained elusive so far.

\begin{figure}[bh]
    \centering
    \includegraphics[width=1.0\linewidth]{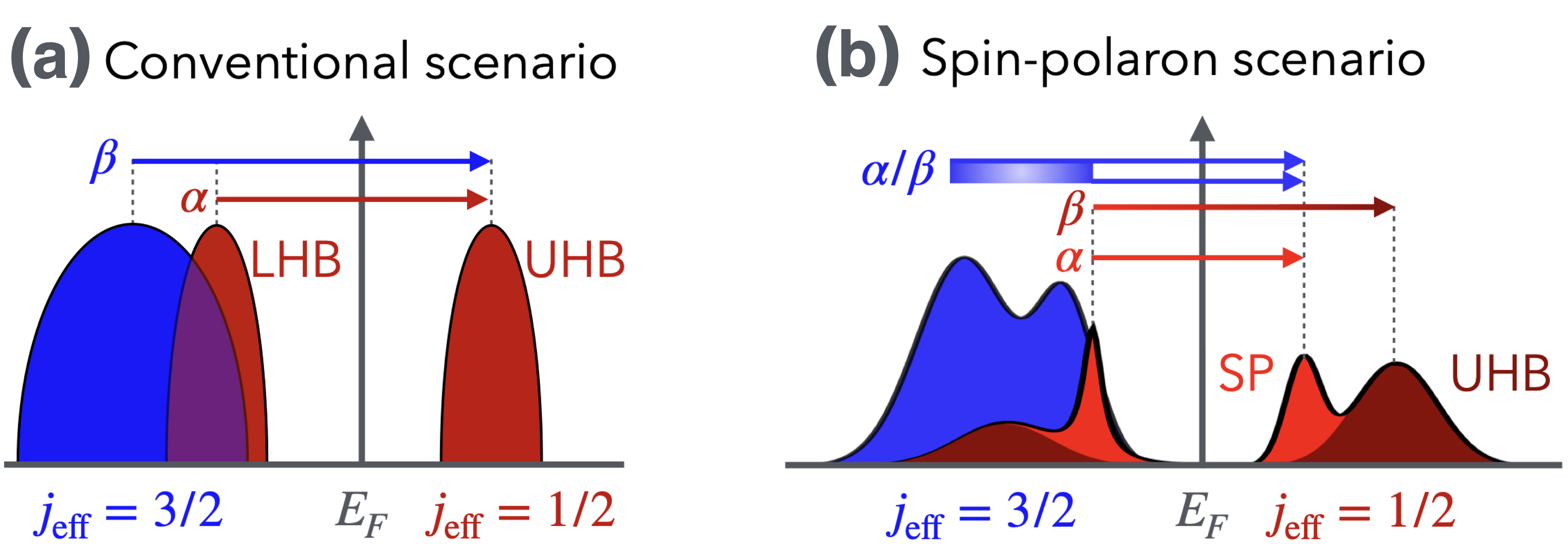}
    \caption{Schematic scenarios for low-energy optical transport quantities involving the iridates' spin-orbit entangled \jeff, \JEFF\ states. In the conventional scenario, the $\alpha$ transition only involves the upper/lower Hubbard bands (UHB/LHB) of the \jeff\ states. In the spin-polaron (SP) scenario, SP quasiparticles are essential in shaping the $\alpha$ and $\beta$ transitions.}
    \label{fig:scenarios}
\end{figure}

Interpretations of the iridates' double peak structure range from charge polarons \cite{Sohn_2014} over spin-orbital excitons \cite{Souri_2017} to the common interpretation in terms of two distinct inter-band transitions of spin-orbit entangled $j_{\mathrm{eff}}=1/2$ and $j_{\mathrm{eff}}=3/2$ states \cite{Moon2008}, as shown in Fig. \ref{fig:scenarios}(a).

In this Letter, we challenge these interpretations by proposing an alternative scenario based on the presence of spin-polaron (SP) quasiparticles within the $j_{\mathrm{eff}}$ states, see Fig.~\ref{fig:scenarios}(b).
By computing optical conductivity for two emblematic iridates, \bairo\ \cite{Okabe_2011,Okabe_2012,Isobe_2012,Moser_2014,Hou_2016,Katukuri_2014} and \sriro, % within linear response using DMFT, 
we assign a mixed orbital character to both peaks and identify the two-peak structure as a fingerprint of the presence of spin-polaron bands.
The SP corresponds to an excess charge dressed by magnetic fluctuations, and its hole counterpart has been proven to explain the low-energy features in the spectrum of cuprates below the Fermi energy \cite{Wang_2015,Bacq_2025}.
The SP scenario thereby underlines the importance of quantum fluctuations and reinforces the analogies between the two families of compounds \cite{kim_2012,DelaTorre_2015, Bertinshaw2019}.

%
%
%
% =============================================

%

We use density functional theory (DFT)+DMFT \cite{Kotliar_2006} to treat the two correlated insulators \bairo\ and \sriro. 
Although some aspects of the physics of the iridates can be described within a single-band \jeff\ model, a quantitative investigation of optical transport properties requires the explicit inclusion of the \JEFF\ bands, to account for interorbital transitions. 
In the case of \bairo, we also verified that the empty $e_g$ bands, crossing the Fermi level $E_F$ within DFT \cite{Cassol_2024}, do not affect the low energy part of $\sigma(\omega)$, as discussed in the Supplemental Material (SM) \cite{SM}.
Therefore, we first construct two Ir-$t_{2g}$-like models with spin-orbit coupling \cite{Pavarini_2021,Cassol_2024}, based on a Wannierization of the DFT bandstructure. 
The effective local Coulomb interaction is written in a generalized Hubbard-Kanamori form \cite{Georges_2013} in the $t_{2g}$ basis.
To achieve agreement with experiments, we adjust the values of the interaction terms computed via constrained random phase approximation \cite{cRPA,cRPA2} to an average intra-orbital interaction of  $U=2.33$ ($1.97$) eV and a Hund's coupling $J=0.22$ ($0.26$) eV for \bairo (\sriro), which are comparable to values in the literature \cite{Martins_2011,Arita_2012,Lenz_2019,Pavarini_2023,Cassol_2024,Choi_2024}. 
The DMFT impurity problem is solved using the continuous-time quantum Monte Carlo solver in the hybridization expansion \cite{CTQMC} implemented in the TRIQS toolkit \cite{TRIQSCTHYB2016,TRIQS2015}.
We perform DMFT calculations in the antiferromagnetic (AFM) phase, incorporating magnetism in a spatial mean-field fashion while allowing for local quantum fluctuations, which give rise to the formation of spin-polarons \cite{Sangiovanni_2006}.
Within DMFT, vertex corrections to the optical conductivity $\sigma(\omega)$ are suppressed \cite{Khurana_1990}, leading to a first-order Kubo formula in the generalized Peierls approach \cite{Tomczak_2009}: 
\begin{equation}\label{eq:optical_cond_awk}
\begin{split}
    \Re \sigma_{ij}(\omega) = \frac{2\pi e^{2} \hbar}{V} & \sum_{\k\sigma}  \int d\omega' \frac{f(\omega') - f(\omega' + \omega)}{\omega} \times \\  & \times \mathfrak{tr} \{ v_{i \k\sigma} A_{\k\sigma} (\omega') v_{j \k\sigma} A_{\k\sigma} (\omega' + \omega) \} .
\end{split}
\end{equation}
Here, V is the unit cell volume, $A_{\k \sigma} (\omega)$ is the orbital- and momentum-resolved  (DMFT) spectral function, $v_{i \k\sigma}$ is the Fermi velocity in the $i$ direction, and $\mathfrak{tr}\{..\}$ denotes the trace in orbital space. Finally, $f(\omega)$ represents the Fermi distribution function for a given temperature.
The polaronic nature of the low-energy states of $A_{\k \sigma} (\omega)$ is verified by effective spin-model calculations within the self-consistent Born approximation (SCBA) \cite{Spin_polaron_t_J,Spin_polaron_t-t1_J}. 
Since the SP feature in the empty part of the spectrum  displays pure \jeff\ character, we perform SCBA in the single-band approximation, taking the corresponding magnon dispersions from previous resonant inelastic x-ray scattering (RIXS) studies on the two compounds \cite{Kim_2014,Clancy_2023}.  
Details of the model construction, its parametrization, DFT+DMFT calculations and the SCBA equations can be found in the SM \cite{SM}.

\begin{figure}
    \centering
    \includegraphics[width=1.0\linewidth]{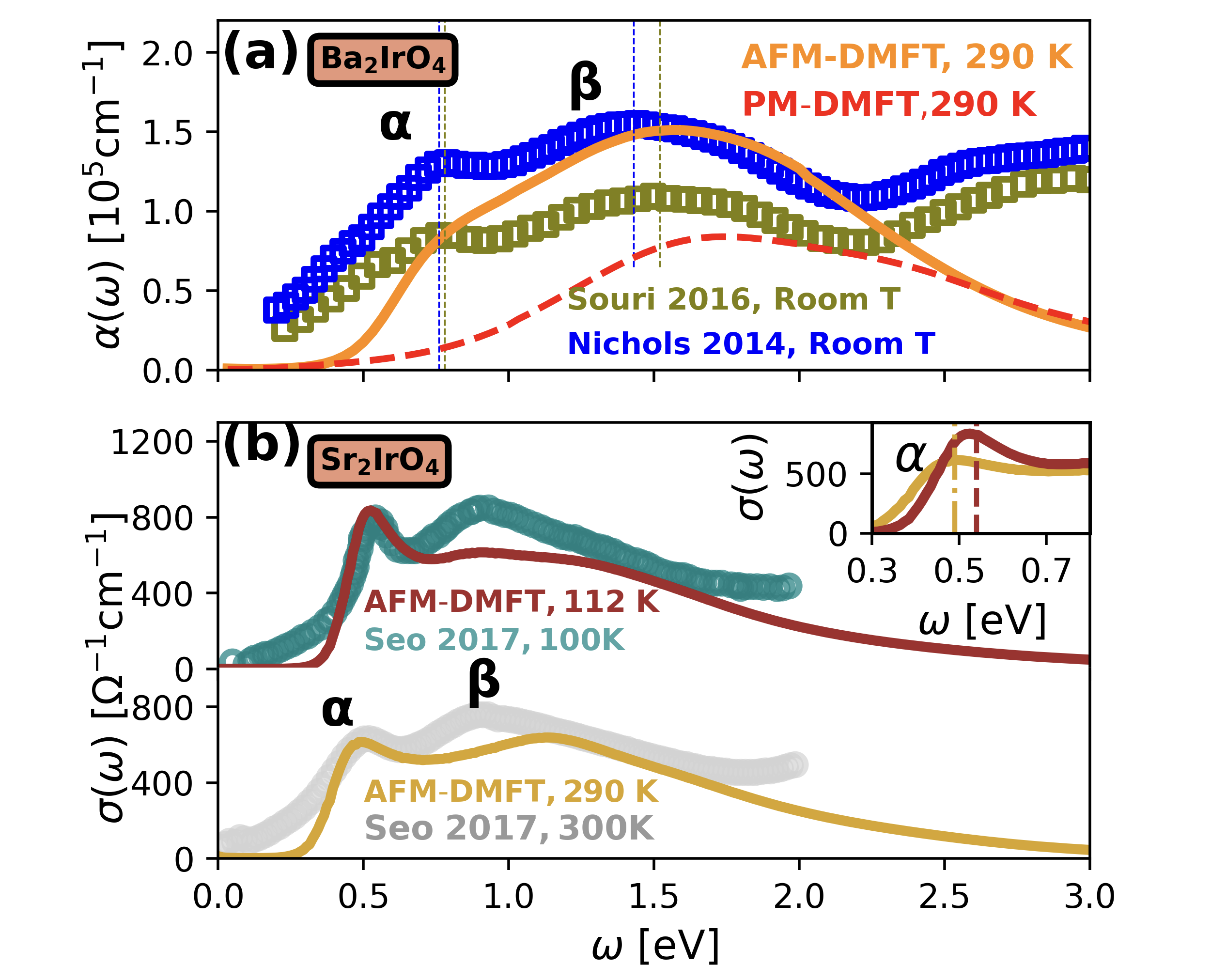}
    \caption{ In-plane optical transport quantities for \bairo\ and \sriro. %: comparison with experiments. 
    (a) Optical absorption of \bairo\ compared with experiment. Data are taken from Refs.~\onlinecite{Nichols_2014,Souri_2016}. For the AFM (PM) calculation, $U$ is set to 2.33 (2.63) eV. (b) Optical conductivity of \sriro\ for $U=1.97$ eV, compared to experiment as a function of temperature. Experimental data extracted from Ref.~\onlinecite{seo_2017}.}
    \label{fig:quantitative_comparison}
\end{figure}

The key results of this work are shown in Fig. \ref{fig:quantitative_comparison}. 
Panel \ref{fig:quantitative_comparison}(a) compares the experimental absorption coefficient $\alpha(\omega)$ (see Eq.\ref{eq:absorption} of SM\cite{SM})  of \bairo\ extracted from Refs.~\onlinecite{Nichols_2014} and~\onlinecite{Souri_2016} as well as the one calculated by single-site AFM-DMFT, showing good agreement.% with experiments.
We reproduce the magnitude of the experimental quantities, capturing the positions of both the $\alpha$ and $\beta$ peaks. 
Including vertex corrections to $\sigma(\omega)$ could further enhance the spectral feature of the $\alpha$ peak, as recently shown in the context of dual-GW \cite{Dasari_2026}.
On the other hand, paramagnetic (PM) DMFT calculations, shown with a dashed red line, only recover a feature comparable to $\beta$, emphasizing the importance of the SP band, absent without a magnetic background, in determining the proper $\sigma(\omega)$. 
We note that in this last case, the Coulomb interaction has to be enhanced to $U=2.63$ eV to trigger the Mott transition.
\begin{figure*}[t]
    \centering
    \includegraphics[width=1\linewidth]{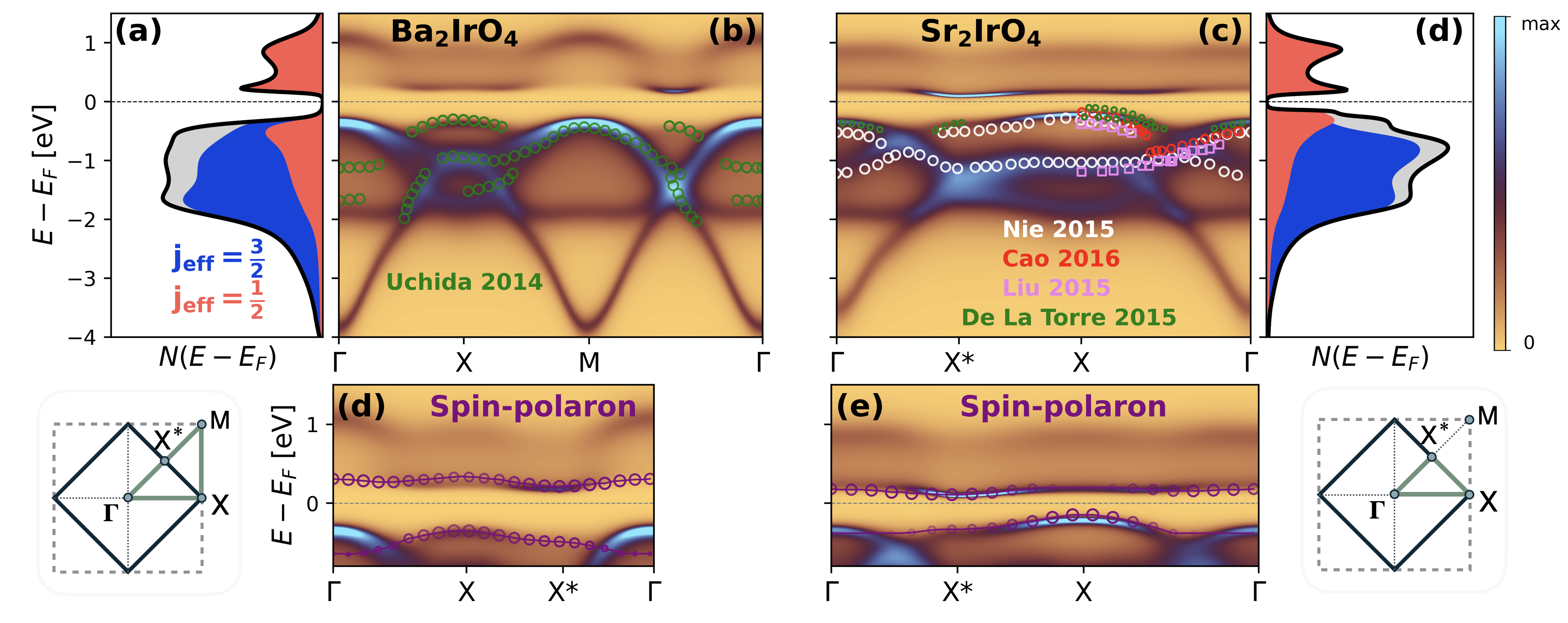}
    \caption{ Electronic structure of \bairo\ and \sriro\  within DFT+AFM-DMFT. Momentum resolved (a) and local (b) spectral function of \bairo. Momentum resolved (c) and local (d) spectral function of \sriro. Experimental dataset are exported from \cite{Uchida_2014} and from \cite{Nie_2015,DelaTorre_2015,Liu_2015,Cao_2016} for \bairo\ and \sriro\ respectively. Comparison between SCBA and AFM-DMFT for \bairo\ (d) and \sriro\ (e). }
    \label{fig:arpes}
\end{figure*}
Switching to \sriro, we can compare the optical conductivity to experiments for different temperatures, see panel \ref{fig:quantitative_comparison}(b). 
The experimental dispersions at room temperature and at $100$ K taken from Ref.~\onlinecite{seo_2017}  are qualitatively reproduced. 
The double peak structure is immediately evident. 
The calculated optical conductivity evolves with temperature: the $\alpha$ peak becomes more intense and presents a blue shift for lower $T$ (see inset), consistent with experiments \cite{moon_2009,Sohn_2014,seo_2017}.
From the optical conductivity, we can also determine the relative permittivity, which fits well with experiments (see SM).
To gain further insight into the origin of the $\alpha$ and $\beta$ peaks, we now analyze the electronic structure itself, as reported in Fig.~\ref{fig:arpes} for both compounds. % in both $\k$-resolved and local versions. 
In the hole part of the spectrum, the local spectral function shows a clear two-peak structure within the upper \jeff\ branch, in agreement with 
the cartoon picture of
Fig.~\ref{fig:scenarios}. 
This can also be appreciated in the $\k$-resolved counterpart, 
which
displays a dispersing high-energy band at $E-E_F\sim 0.8$ (0.7) eV for \bairo (\sriro), 
%which is 
accompanied by a nearly flat low-energy quasiparticle band at $\sim 0.3$ (0.1) eV.
Qualitatively, this structure of the hole spectrum is in good agreement with previous AFM-DMFT \cite{Li_2013} and GW+AFM-DMFT calculations \cite{Choi_2024}. 
In the filled part of the spectrum, we compare our calculations with available angle-resolved photoemission spectroscopy (ARPES). 
For the sake of comparison with experiments, which report the data along the $\Gamma-X-M-\Gamma$ $\k$-path for \bairo\  \cite{Uchida_2014} and along $\Gamma-X^{*}-X-\Gamma$ in the case of \sriro\ \cite{Nie_2015,Liu_2015,Cao_2016,DelaTorre_2015} (see insets). 
Concerning \bairo, in our calculations the point $M$ is equivalent to $\Gamma$,  since we consider the system in a perfect AFM long-range order, while in the experiments $\Gamma$ and $M$ clearly differ.
Except for this disagreement, the other ARPES features are faithfully reproduced.
In \sriro, structural distortions are responsible for a reduction of the Brillouin Zone (BZ). 
Again, we obtain substantial matching over the full $\k$-path. 
Concerning
the energy gap size, within AFM-DMFT we found a value of $\sim150$ meV, which
agrees with 
bulk probes, such as optical absorption \cite{Souri_2016} and the Arrhenius plot of resistivity \cite{Nichols_2014}, but it is smaller than that reported in surface sensitive techniques \cite{Dai_2014,Yan_2015,Brouet_2015}. 
Paramagnetic DMFT calculations fail in reproducing salient features of the electronic structure of \bairo\ and \sriro~\cite{Martins_2011,Cassol_2024}.
However, neither the experimental electronic structure \cite{Uchida_2014,Moser_2014,DelaTorre_2015} nor the optical conductivity measurements \cite{moon_2009,Sohn_2014,seo_2017} show relevant variations across the magnetic transition at $T_N\approx 240$ K \cite{Kim_2008,Okabe_2011}.
This underlines the importance of AFM fluctuations in shaping the spectral function in the PM phase \cite{Martins_2018,Lenz_2019}.
Here, mimicking these fluctuations in the paramagnetic phase by calculations in the AFM phase captures essential features of the spectral function.

Clarifying the microscopic nature of low-energy features in the spectral function is crucial for 
understanding 
the optical conductivity. 
As we had anticipated, they can be traced back to the presence of spin-polarons.
The calculated extra hole/electron self-energy within SCBA is shown in Fig. \ref{fig:arpes}(d,e), where we compare the main polaronic band with the spectral function within AFM-DMFT (see SM \cite{SM} for the full SCBA spectrum). 
For both compounds, SCBA reproduces well the AFM-DMFT dispersion of the \jeff\ band below the Fermi level, and matches the dispersion of the lower-energy band above, revealing their polaronic nature. 
The deviations between SCBA and DMFT can be associated with the 
contribution of
higher-order exchange 
terms
in the spin Hamiltonian, since the compounds are not deep in the Mott regime \cite{Pavarini_2021}, and with the zero-temperature nature of SCBA. 
The presence of spin-polaron excitations has been recently suggested for \sriro \cite{parschke_2017,parschke_2018,parschke_2022}. 
However, the existence of the polaronic band could be checked only in the filled part of the spectrum, where ARPES probes are available. 
On the other hand, the connection between such quasiparticles and the optical spectrum was already discussed for simple model calculations \cite{Taranto_2012}. 
Here, we combine these concepts for both \bairo\ and \sriro, providing a unified theoretical picture. 
%Indeed, since the low-energy quasiparticle peaks in AFM-DMFT can be assigned a spin-polaronic neature and the technique leads to spectra in good agreement with experiments,  the importance of spin-polarons for the optical spectrum. 

%

\begin{figure}
    \centering
    \includegraphics[width=0.99\linewidth]{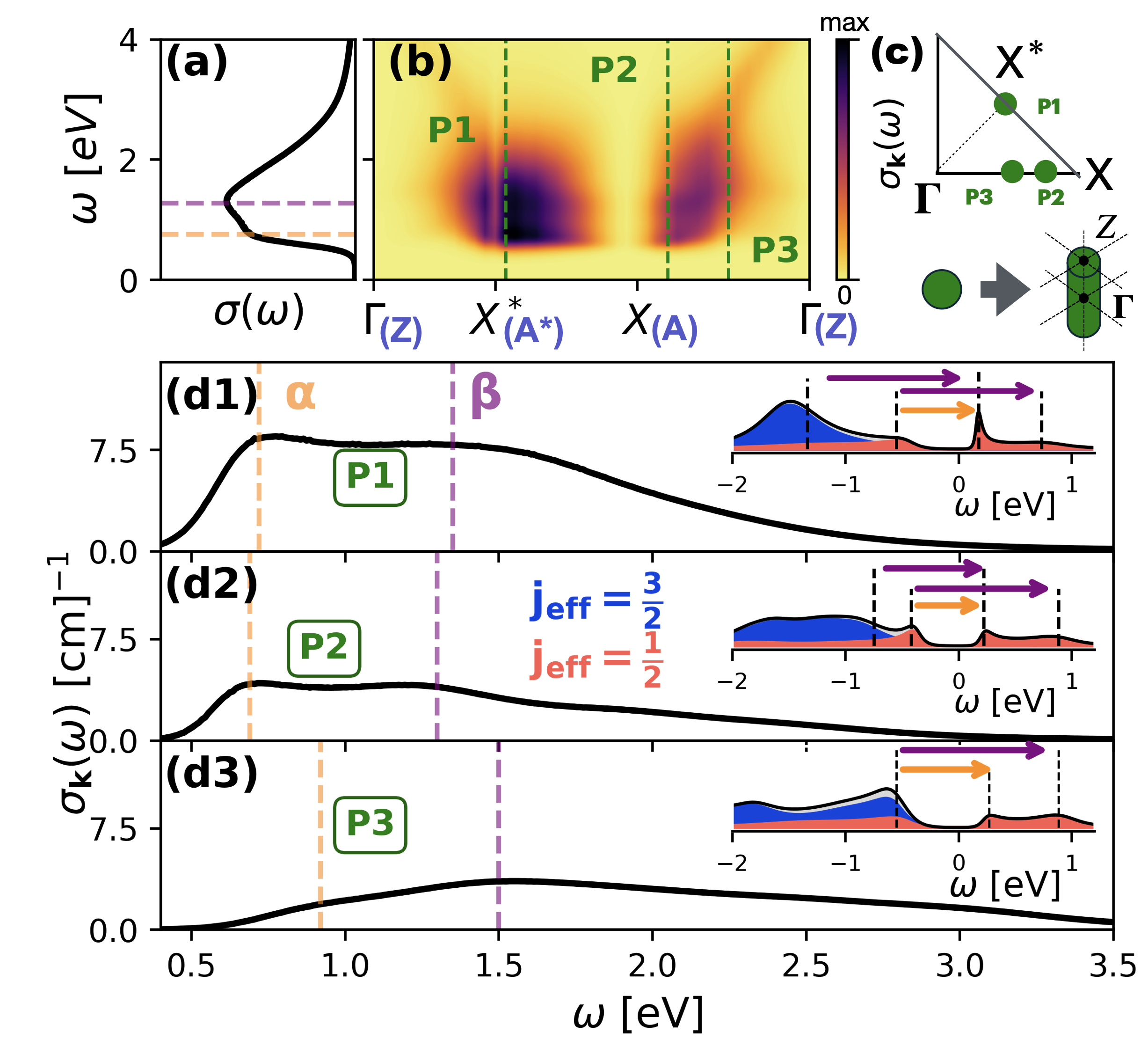}
    \caption{ Relation between $A_{\k}(\omega)$ and $\sigma(\omega)$. Total (a) and 
    momentum selective (b) optical conductivity for \bairo. Sketch of the k-path, selected k-points and the integration cylinder in $k_z$-direction. (d1-d3) $\sigma_{\k}(\omega)$ for selected $\mathbf{k}$ points. In all panels, the inset reports the corresponding spectral function $A_{\mathbf{k}}(\omega)$. Orange (violet) arrows indicate energy excitations 
    consistent with $\alpha$ ($\beta$) corresponding to the selected $\k$-point.  
    }
    \label{fig:k-selectivity}
\end{figure}

%
It is now legitimate to question how the double-band structure of the upper part of the 
spectrum
%eigenspectrum 
affects the shape of the conductivity, which also presents two peaks. 
To assess the importance of having the two clear spectral features (SP and HB) in the unoccupied part of the electronic structure and to investigate the orbital character of the $\alpha$ and $\beta$ peaks, we compute the optical conductivity by restricting the $\k$-integration to specific regions of the BZ, see Fig.\ref{fig:k-selectivity}. 
Since the electronic dispersion is nearly flat outside the $k_xk_y$ plane \cite{Moser_2014,Cassol_2024}, we focus on points along the $\Gamma - X^* - X - \Gamma$ path and integrate along the $k_z$ direction as indicated in Fig.~\ref{fig:k-selectivity}(c) \cite{BZpath}.

Results are shown in Fig.\ref{fig:k-selectivity}(b) for \bairo, but also extend to \sriro. 
For comparison, panel \ref{fig:k-selectivity}(a) displays the total optical conductivity. 
Looking at the dispersion along the $\k$-path, one can notice that the conductivity vanishes at the high-symmetry points $\Gamma$ and $X$. 
This is due to the impact of the Fermi velocities, which vanish in regions where the band structure displays stationary points, see SM \cite{SM}. 
 Since Eq. \ref{eq:optical_cond_awk} involves vertical excitations,   we can compare the shape of $\sigma_{\k}(\omega)$ with the peak positions of $A_{\k}(\omega)$. 
 We select three points marked with vertical green lines in Fig.\ref{fig:k-selectivity}(b), and show the corresponding optical conductivity $\sigma_{\k}(\omega)$ in Fig. \ref{fig:k-selectivity}(d1-d3). 
The corresponding relative density of states (DOS) is shown in the insets, where orange and violet arrows are compatible with optical transitions contributing to the $\alpha$ and $\beta$ peaks, respectively.
\newline
All conductivities feature a double-peak structure.
Point P1 is close to $X^*$, and it is highly representative of the entire BZ. 
As expected within the conventional interpretation, the $\alpha$-peak consists of intra-orbital \jeff\ $\rightarrow$ \jeff\ transitions. In our picture, however, these transitions occur between filled and empty SP quasiparticles, and not between LHB and UHB.
Moreover, due to the double peak structure above the gap, a second type of intra-orbital excitation — from the filled SP to the UHB peak of the \jeff\ band (intra-orbital) — is found to be perfectly consistent with $\beta$. 
On the other hand, excitations from the \JEFF\ to the \jeff\ states (inter-orbital) span a wide energy range due to the broad structure of the \JEFF\ band, which is much lower in energy at this $\k$ point.  
At point P2 \cite{BZpath},
a similar scenario is observed, where the spectral weight of the SP and UHB is comparable in the unoccupied region. 
Moreover, the \JEFF\ manifold is close to the \jeff\ band in the occupied region, leading to non-zero inter-orbital contributions to both the $\alpha$ and $\beta$ peaks. 
Finally, at point P3 \cite{BZpath}, 
the \JEFF\ and \jeff\ peaks in $A_{\mathbf{k}}(\sigma)$ are basically aligned.
This set of calculations validates the SP picture we propose and proves that both peaks feature a mixed orbital character, in contrast to the conventional interpretation, which is only based on \jeff\ Hubbard bands. According to our findings, the $\alpha$ peak mainly involves excitations of SP character.

We note that an AFM-DMFT calculation of the optical conductivity was already presented for \sriro\ in Ref. \onlinecite{Zhang_2013}. 
The low-temperature calculation of Fig. \ref{fig:quantitative_comparison} is in qualitative agreement with the one reported in that work. We notice that the DFT+DMFT calculation reported in Ref.~\onlinecite{Zhang_2013} presents a strongly renormalized band at 0.2 eV, which is consistent with our SP interpretation.
However, such a quasiparticle excitation cannot be captured within an effective single-particle picture such as DFT+U. 

\begin{figure}[tb]
    \centering
    \includegraphics[width=0.99 \linewidth]{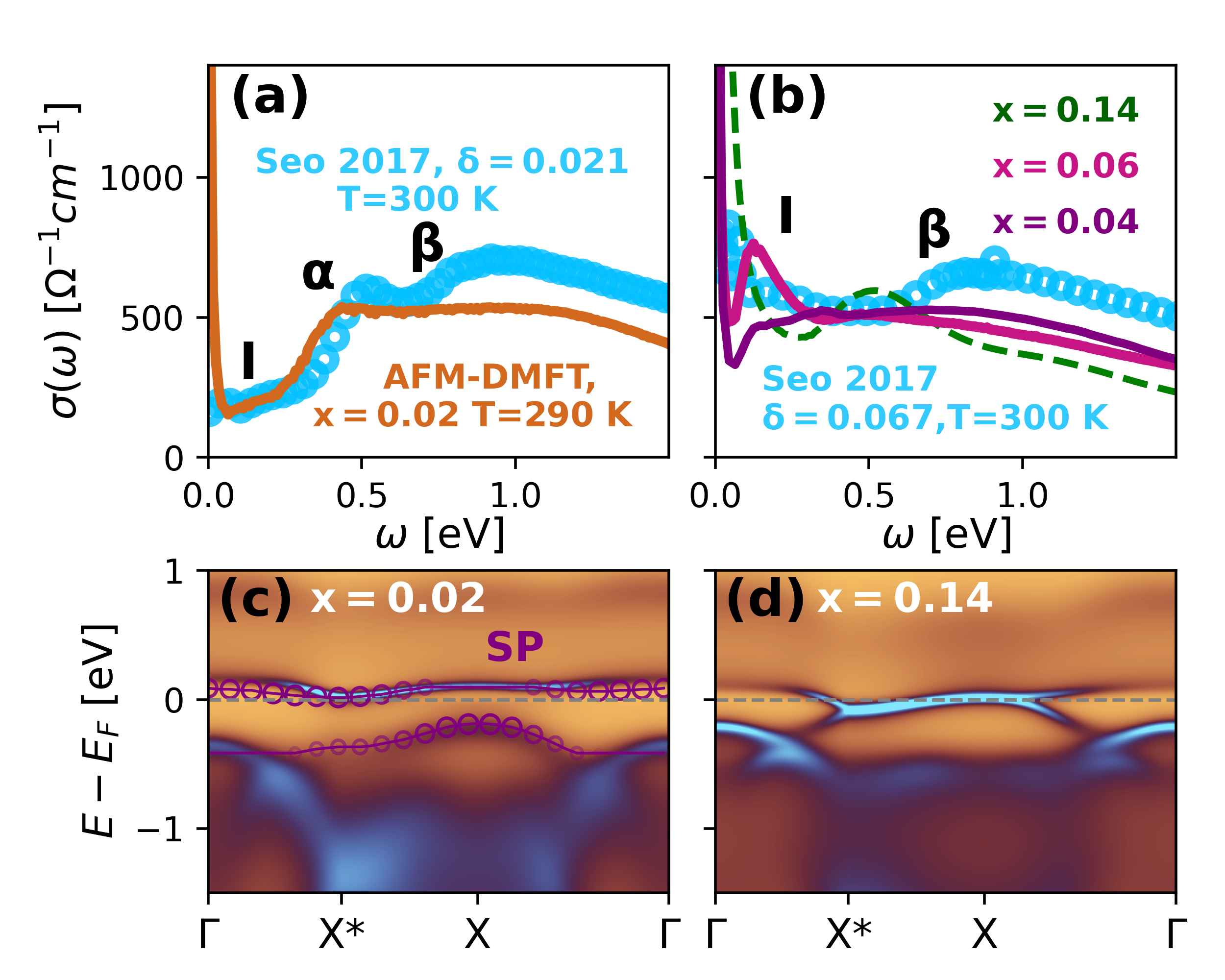}
    \caption{In-plane optical conductivity of electron doped \sriro.  Comparison between AFM-DMFT calculations and experiments \cite{seo_2017} for low (a) and moderate (b) doping levels. In the experimental curves, $\delta$ represents the amount of La substitution. Momentum resolved spectral function at low (c) and intermediate (d) doping levels. To favour a clear representation, a small broadening $\eta=0.01$ has been set in $A_{\k}(\omega)$. For electron doping $x=0.02$ the system is still in the AFM phase displaying spin-polaron features, while for $x=0.14$ it is a paramagnetic correlated metal.}
    \label{fig:doping}
\end{figure}

We conclude this study by discussing the evolution of $\sigma(\omega)$ of \sriro\ upon doping, see Fig.~\ref{fig:doping}. 
In this compound, chemical doping can be obtained either by replacing Ir with Rh/Ru (hole-doped) \cite{Brouet_2015,Louat_2019,Brouet_2021} or by substituting Sr with La (electron-doped) \cite{DelaTorre_2015,Brouet_2015,seo_2017}.
We focus here on the latter as performed in Ref. \onlinecite{seo_2017}, which presents a detailed investigation of the (Sr$_{1-\delta}$La$_\delta$)$_2$IrO$_4$ conductivity as a function of temperature at low and intermediate La-doping. 
Here, we simulate the doped regime, adjusting the chemical potential within the DMFT loop to match the electronic density.
Note that if every La provided an extra electron, this would amount to an electron doping of $x=2\delta$. 
For a La doping of $\delta=0.02$, we find good agreement with experiments for $x=0.02$, suggesting that the effective free carriers are less than $2\delta$. 
Concerning the shape of $\sigma(\omega)$, while both $\alpha$ and $\beta$ survive upon small doping, a new ingap excitation $I$ appears. 
The corresponding AFM-DMFT and SCBA spectral functions are shown in Fig.~\ref{fig:doping}(c), where the upper spin-polaron is promoted to the Fermi level, while the lower one is shifted up, in agreement with ARPES \cite{DelaTorre_2015}. 
Given its low energy, it is clear that the ingap excitation $I$ is purely polaronic.  
The comparison between the calculated conductivity and experiments at higher doping levels does not lead to a quantitative agreement, while still being qualitatively comparable (see, for instance, Fig.\ref{fig:doping}(b)). 
The main experimental features at $\delta=0.067$ show up for a doping range $0.04 \leq x \leq 0.06$ at slightly lower energy. In this range, we observe the polaronic peak $I$, more coherent in our calculations, followed by a peak corresponding to $\beta$.
By further increasing $x$ in our calculations, the AFM phase closes and $I$ is absorbed in the Drude peak, as it can also be observed in the spectral function, see Fig.\ref{fig:doping}(d). 
The corresponding $\sigma(\omega)$ is now composed of three features: a Drude peak, the infrared excitation from and towards the quasiparticle peak at the Fermi level, and a shoulder corresponding to excitation between the incoherent Hubbard branches. 
This is typical for doped correlated insulators, which rapidly become renormalized Fermi liquids with increasing doping \cite{Jarrel1995,Rozenberg95,Toschi2005,Toschi2008,Nicoletti2010}.

In this work, we have computed the optical conductivity of \bairo\ and \sriro, focusing on the in-plane component.
%,achieving good agreement with experiments previously reported. 
%
For \bairo, we correctly reproduced the absorption coefficient. 
Concerning \sriro, we captured the temperature dependence of the optical conductivity.
%
%Through a set of orbital- and $\k$-selective calculations,
Our analysis indicates that the $\alpha$ peak is significantly influenced by spin-polaron effects, while both $\alpha$ and $\beta$ features exhibit a mixed orbital character, particularly for the high-energy peak $\beta$.
%
%Within this framework,
The characteristic double peak shape of $\sigma(\omega)$ originates from a double peak structure in the unoccupied states 
%that can be interpreted as 
stemming from the presence of spin-polarons above the Fermi level, consistent with experimentally measured transport quantities. 
Remarkably, the proposed polaronic picture remains valid in the low-doped regime. 
Since the same scenario involves both \bairo\ and \sriro, we believe that our findings are relevant to understanding the entire class of $5d^5$ perovskite
iridate materials.

Our study suggests that similar spin-polaron fingerprints could be observed in optical transport properties of other strongly correlated AFM materials.
In particular, high-$T_c$ superconducting cuprates are promising candidates, given their similarity with $5d^5$ iridates \cite{DelaTorre_2015,Kim2014F,Peng_2022} and the presence of spin-polaronic quasiparticles in their low-energy spectrum \cite{Grober_2000,Manousakis_2007,Wang_2015,Bacq_2025}.

\vspace{20pt}

\textit{Acknowledgments} - We thank Benjamin Bacq-Labreuil for fruitful discussions on SCBA. We are very grateful to Evgeny Stepanov and Silke Biermann for enlightening discussions concerning the compound investigated in this work. The authors thank Christophe Brun and V\'eronique Brouet for discussions on the experimental aspects of \sriro.
This project was provided with HPC computing and storage resources by GENCI at IDRIS and TGCC thanks to the grants A0170912043 and A0190912043 on the supercomputers Jean Zay (CSL partition) and Joliot Curie (ROME partition).

\textit{Data availability} - The data that support the findings of this article are openly available \cite{cassol_2025_Zenodo}. 


%\bibliography{references}% Produces the bibliography via BibTeX.

\clearpage

%%%%%%%%%%%%%%%%%%%%%%%%%%%%%%%%%%%%%%%%%%%%%%%%%%%%%%%%%%%%%%%%%%%

%%%%%%%%%%%%%%%%%%%%%%%%%%%%%%%%%%%%%%%%%%%%%%%%%%%%%%%%%%%%%%%%%%%

%%%%%%%%%%%%%%%%%%%%%%%%%%%%%%%%%%%%%%%%%%%%%%%%%%%%

\clearpage

%%%%%%%%%%%%%%%%%%%%%%%%SUPPLEMENTAL
\titlext{Supplemental Material to "Spin-polaron fingerprints in the optical conductity of iridates". }

%\maketitle{Supplemental Material}

\section{Computational details}

Density functional theory (DFT) calculations using plane wave basis sets and pseudopotentials were performed with the Quantum Espresso software package~\cite{QuantumEspresso1,QuantumEspresso2} using a GGA functional in the PBE flavor~\cite{PBE1996}. 
We considered norm-conserving pseudopotentials \cite{ONCVPSP2013} from the Pseudodojo library \cite{PseudoDojo2018} in the full (scalar) relativistic flavor for calculations with (without) spin-orbit coupling (SOC). 
For details on the DFT calculations and the construction of the tight-binding model for \bairo,  we refer the reader to Ref. \onlinecite{Cassol_2024}. 

For \sriro, we took as a starting point the measured crystal structure reported in Ref.~\onlinecite{Crawford_1994} and relaxed the atomic positions to a force of $<10^{-3}$ Ry/Bohr. 
We observed an enhancement of the structural distortion upon relaxation. The corresponding distortion angle becomes 13$^{\circ}$, compared to the 11.5$^{\circ}$ measured experimentally \cite{Crawford_1994}. 
The wavefunction energy cutoff was set to 95 Ry. 
We then used a regular $\k$ grid of $8 \times 8 \times 8$ for self-consistent calculations and a $11 \times11\times 11$ grid for subsequent non-self-consistent calculations.

The tight-binding model was obtained via wannierization of the DFT+SOC band structure using maximally localized Wannier functions \cite{MLWF}.
To this end, we employed the Wannier90 package \cite{wannier90} and used an outer (frozen) energy window of [8.8 eV, 12.5] ([10.4 eV, 12.1 eV]), respectively. 
The obtained Wannier function extensions feature large orbital anisotropy. 
Their spread is $\Omega_{d_{xz}}=\Omega_{d_{yz}}=4.17 \ \AA$ and $\Omega_{d_{xy}}=6.66 \ \AA$, respectively.  
This difference can be attributed to the lack of $e_g$ states in the Ir-$t_{2g}$ model. 
In particular, the $d_{xy}$ orbital extends to regions that would be occupied by the $d_{x^2-y^2}$ orbital in an Ir-d model, which explains the large spread of the former.

Dynamical mean-field theory (DMFT) calculations were performed with the TRIQS library \cite{TRIQS2015}. 
The DMFT self-consistency loop was coded taking advantage of the DFTTools module \cite{TRIQSDFTTOOLS2016}, and the impurity problem was solved within the corresponding CT-HYB solver \cite{TRIQSCTHYB2016}, with $28 \times 10^6$ Monte Carlo measurements per iteration. 
To obtain a reliable analytic continuation, we increased the number of cycles to $86 \times 10^6$ for the final iteration. 
DMFT calculations for \bairo\ in the antiferromagnetic (AFM) phase were performed by partitioning the original system into two sublattices $A$ and $B$, hybridizing through the bath. 
The corresponding impurity model was solved for one of the two sublattices. 
The self-energy of the second sublattice, $\Sigma_B(i\omega_n)$, was obtained by applying time reversal symmetry to $\Sigma_A(i\omega_n)$ in the $j_{\mathrm{eff}}$ basis. 
For the values of the interaction we considered, the system spontaneously evolved to the AFM phase once the PM constraint was released, without the need to include any initial symmetry-breaking field. 
In the case of \sriro, which displays structural distortion, the time reversal symmetry operation was combined with a rotation to align with the local IrO$_6$ axis.

The analytical continuation was performed with a quantum causal matrix generalization of the maximum entropy method implemented in the MQEM code \cite{MQEM2018}, with a smearing factor of $5$ meV. 

The optical conductivity was computed via the first order Kubo formula reported in Ref. \onlinecite{Tomczak_2009}, implemented in a homemade code parallelized over the $\k$ points. We employed $\k$-grids of $80 \times 80  \times 40$ points for \bairo\ and of $60 \times 60 \times 60$ in the case of \sriro.

Concerning the application of the self-consistent Born approximation (SCBA) to study spin-polaron (SP) features, we also wrote our own code based on the Refs. \onlinecite{Spin_polaron_t_J,Bacq_2025}  and considered a minimal $t-t'-t''$ single-band \jeff\ model.
We used a $\k$ point grid of $20 \times 20 \times 20$ in the self-consistency loop, and a broadening of $\eta=5 \times 10^{-3} t$, where $t$ is the nearest-neighbor hopping term of the minimal \jeff\ model. 
Numerical values and additional information on the calculations using SCBA are given in the dedicated section of this SM.

\section{Self-consistent Born Approximation}

A spin-polaron is a quasiparticle characterized as an extra particle/hole moving in an antiferromagnetic background, dressed by the magnetic excitations that its passage leaves behind. 
In SCBA, the self-energy of the spin-polaron can be evaluated as \cite{Spin_polaron_t_J,Spin_polaron_t-t1_J}:

\begin{equation}\label{eq:polaronic_self_energy}
    \Sigma^{h/e}(\k, \omega)= \frac{1}{N_{\mathbf{q}}}\sum_{\mathbf{q}}|M_{\mathbf{q},\k}|^{2}G^{h/e}(\k - \mathbf{q},\omega-\omega_{\mathbf{q}})
\end{equation}

where $M_{\mathbf{q},\k}$ is the electron-magnon coupling and the electron/hole Green's function is given by:

\begin{equation}\label{eq:polaronic_gf}
\begin{split}
    G^{h/e} & (\k, \omega)=    \frac{z^2t^{2}}{\omega - \epsilon_{\k} - \Sigma^{h/e}(\k  ,\omega) + i\eta} \ ,
\end{split}
\end{equation}

The fact that the polaronic Green's function depends again on the self-energy itself establishes a condition enabling a self-consistent solution of (\ref{eq:polaronic_self_energy}). 
In Eq. (\ref{eq:polaronic_gf}), $z$ and $t$ represents the coordination number and the nearest neighbor hopping term respectively. The magnon dispersion $\omega_{\mathbf{q}}$ is described in linear spin wave theory and depends on the superexchange parameters $J$, $J'$, and $J''$ characterizing the spin Hamiltonian, %and which are taken from experiments.

\begin{equation}
    \omega_{\mathbf{q}}=(1+n_e)\sqrt{A_{\mathbf{q}}^{2} + B_{\mathbf{q}}^{2}}
\label{wq}
\end{equation}

with:

\begin{equation}
\begin{split}
 A_\mathbf{q} = 2(J  - J^{'} - J^{''} & + J^{'} \cos q_x \cos q_y) + \\  & +J^{''} (\cos 2q_x +  \cos 2q_y),   \end{split}
\label{Aq}
\end{equation}

\begin{equation}
B_q = J' (\cos q_x + \cos q_y)
\label{Bq}
\end{equation}

and where $n_e$ is the corresponding electron doping. 
The parameterization of the spin Hamiltonian, see Table~\ref{tab:hopping}, was taken from previous RIXS experiments \cite{Clancy_2023,kim_2012,Kim_2014}.
Notice that Ref. \onlinecite{Clancy_2023} studied two different types of \bairo\ thin films, grown on PScO$_3$ (PSO) and  GdScO$_3$ (GSO) substrates. Here, we consider the superexchange terms relative to the PSO sample, which, among the two, displays the lattice structure closest to that of the corresponding bulk sample.
Also note that this phenomenological spin model can be justified by analyzing the superexchange interactions and the emerging generalized Heisenberg model \cite{Pavarini_2021}.
In particular, Ref. \onlinecite{Pavarini_2021} showed that for \sriro\ Dzyaloshinskii-Morija terms are reduced to an axial component such that they can be removed via a suitable rotation in pseudo-spin space.
Also, the anisotropy of the Heisenberg exchange terms were found to be small, justifying the simplified model of Eq. (\ref{wq})-(\ref{Bq}).

Moreover, the hopping of extra hole/electrons that do not perturb the antiferromagnetic background are taken into account via $\epsilon_{\k}$. 

\begin{equation}
\begin{split}
  \epsilon^{h/e}_\k =  \pm 4t' \cos k_x \cos k_y + & 2t'' (\cos 2k_x + \cos 2k_y) \\ 
\pm \frac{J_{3s}}{2} & (\cos 2k_x + \cos 2k_y + 4 \cos k_x \cos k_y)
\end{split}
\end{equation}

\begin{figure}[t]
    \centering
    \includegraphics[width=\linewidth]{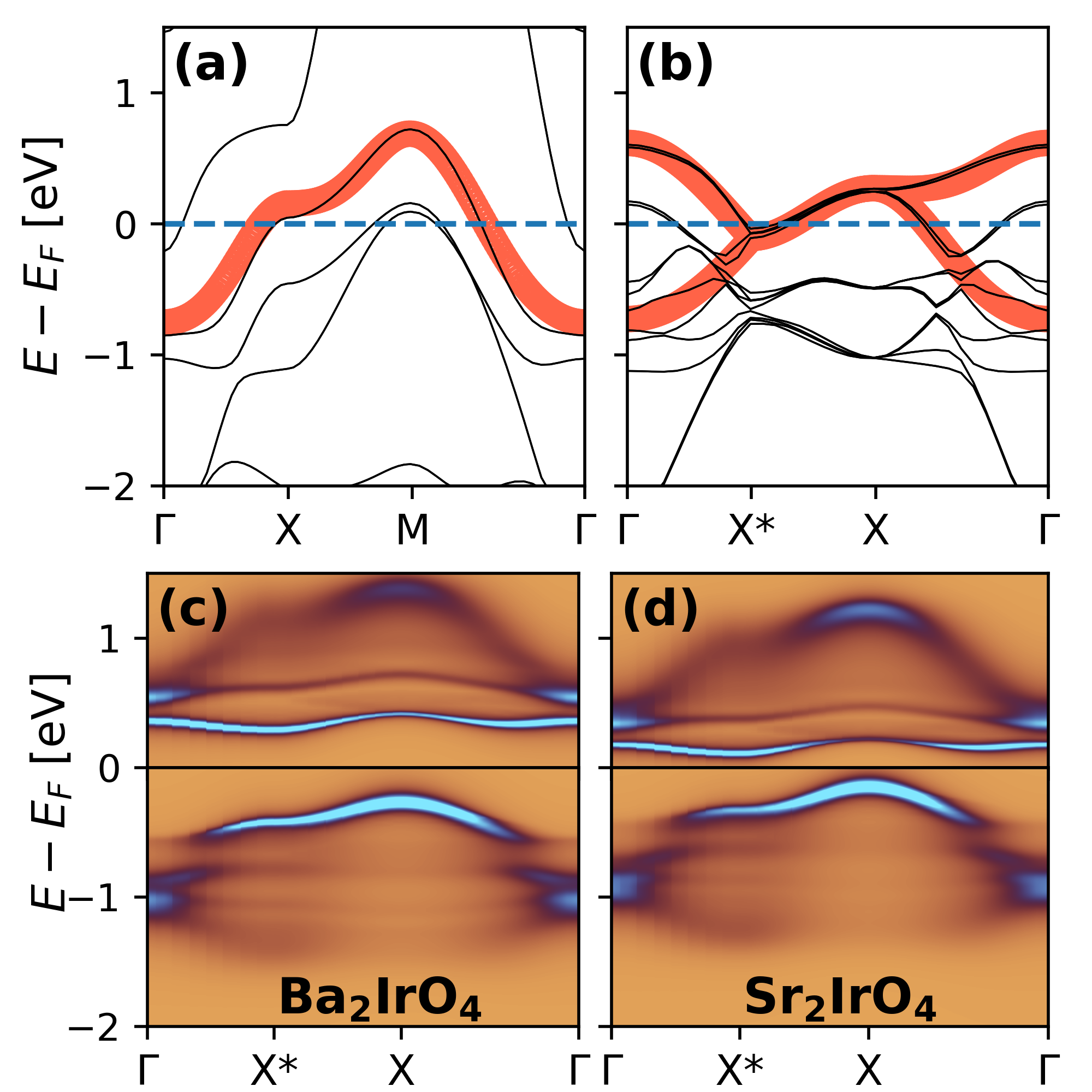}
    \caption{Models and results of SCBA calculations. Minimal $j_{\mathrm{eff}}=1/2$ model for \bairo (a) and \sriro (b). Full SCBA spectrum for \bairo\ (c) and \sriro (d).}
    \label{fig:SM_2}
\end{figure}

where we have added a three-site hopping term, set to $J_{3s}=J$ such that it amounts to the term obtained from a $t/U$ expansion of the Hubbard model in the strong coupling limit \cite{Wang_2015}.

Since the polaronic peak in the hole part of the spectral function of both investigated iridates has a pure \jeff\ character, the SCBA calculations in this work were performed considering a single-band model, corresponding to the \jeff\ manifold.
We thereby neglect the triplet $J=1$ state, which was considered in the spin-polaron studies of \sriro\ in Refs.~\onlinecite{parschke_2017,parschke_2018,parschke_2022}.  
In order to solve the SCBA equation, we therefore construct a $t-t'-t''$ single-band \jeff\ model for both compounds. 
The numerical values are reported in Table~\ref{tab:hopping}.
The comparison between the electronic structure obtained and the DFT reference is reported in Fig.\ref{fig:SM_2}(a) and (b) for \bairo\ and \sriro, respectively.

\begin{table}[t]
    \centering
    \begin{tabular}{|c||ccc||ccc|}
    \hline 
    \hline
       Hopping and Superexchange (meV)  &  $t$  & $t'$ & $t''$ & $J$ & $J'$ & $J''$ \\
         \hline
         \hline
       \bairo  & 180 & 23 & -27 & 85 & -15 & 10 \\
       \hline
       \sriro & 167 & 41 & -27 & 60 & -20 & 15 \\
       \hline
    \end{tabular}
    \caption{Values of the hopping and spin superexchange terms of the minimal single-band \jeff\ model for \bairo\ and \sriro.}
    \label{tab:hopping}
\end{table}

The resulting full SCBA spectrum is shown for both \bairo\ and \sriro\ in Fig. \ref{fig:SM_2}.
They are qualitatively very similar in both the hole and electron parts of their electronic structure.
The lower energy peak in both parts is also the most intense, and it is the one we consider for comparison with AFM-DMFT.

We note, however, that the bare SCBA hole (electron) Green's function needs to be shifted down (up) to match the AFM DMFT calculations. 
In principle, this shift corresponds to $U_{\mathrm{eff}}/2$ where $U_{\mathrm{eff}}$ is the effective Coulomb interaction in the single-band \jeff\ model. 
In our DMFT calculations, the effective Coulomb interaction in this band corresponds to $1.95$ eV ($1.67$ eV) for \bairo\ (\sriro). 
However, the value that we have to use to locate the SCBA bands in correspondence with the DMFT bands is $\tilde{U}_{\mathrm{eff}}=1.52$ eV for \bairo\ and $\tilde{U}_{\mathrm{eff}}=1.26$ eV for \sriro.
The necessity of adjusting the position of the polaronic band to match the DMFT calculations has already been reported in literature \cite{Bacq_2025}, where it was attributed to the different assumptions made in the two theories \cite{Bacq_2025}: 
DMFT captures dynamical local quantum fluctuations, neglecting spatial correlation, while SCBA neglects charge fluctuations, being based on the perfect AFM background. 
More precisely, one key ingredient within DMFT is hybridization to the electronic bath, which leads to an additional traveling channel for the spin-polaron without disturbing the local antiferromagnetic correlations.
This channel is absent in the SCBA.
Secondly, the Gutzwiller projector within SCBA does not allow any double-occupied sites, whereas those are still allowed (but heavily suppressed at strong coupling) within DMFT.
For both reasons, the shift $\tilde{U}_{\mathrm{eff}}$ is smaller than the corresponding effective interaction used in our single-band DMFT calculations.
In the cases studied here, we also notice that SCBA is performed on a single-band framework, where the interaction contains screening effects from the \JEFF\ states, which are naturally incorporated in the impurity problem within DMFT.

Finally, we would like to emphasize that on the DMFT side the inclusion of non-local fluctuations via cluster or diagrammatic extensions of the technique \cite{Klett2020,Schaefer2021} as well as treating dynamically screened interactions~\cite{Werner2010,Pauli2025} changes the gap within the insulating phase. 
As a consequence, to obtain agreement with experiments, the value of the Coulomb interaction needs to be adjusted depending on the level of approximation in the chosen theory.
%Hence, here, we simply assign to each theory the Coulomb interaction that reproduces experiments. 
Here, we used the values of the Coulomb interaction that reproduce experiments for single-site AFM-DMFT.

%%%%%%%%%%%%%%%%%%%%%%%%%%%%%%%%%%%%%

\begin{figure*}[t!]
    \includegraphics[width=\textwidth]{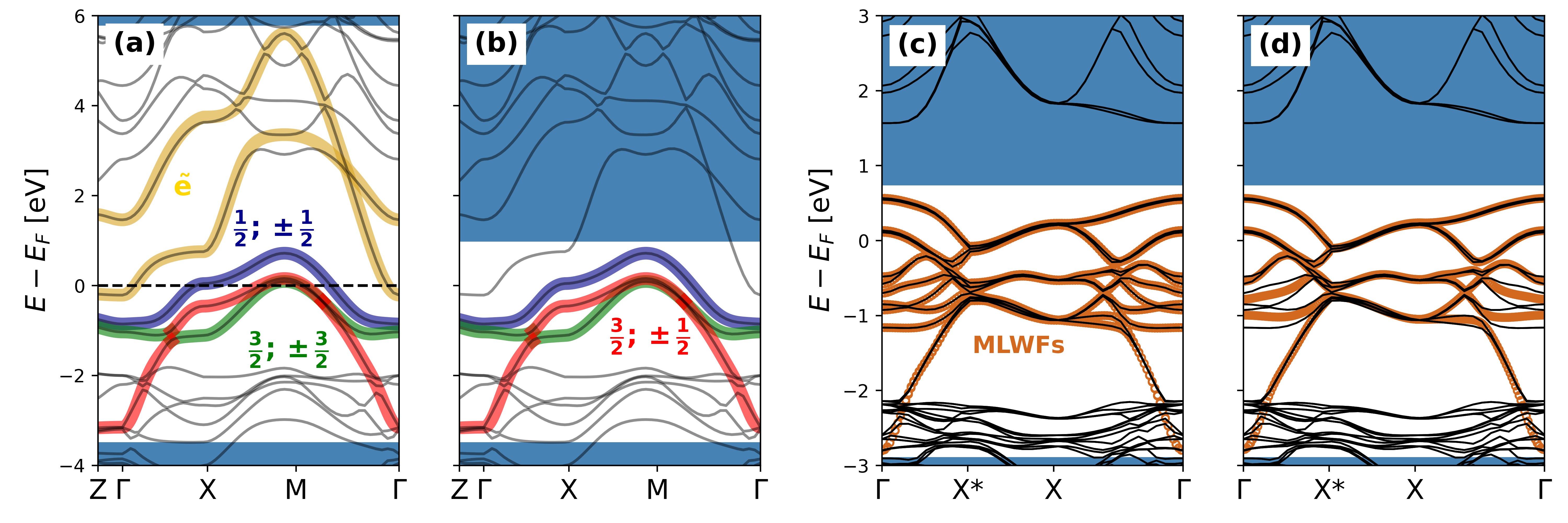}
    \caption{DFT band structure and minimal tight-binding model for \bairo\ and \sriro. (a) DFT band structure and full Ir-$d$ model for \bairo. Different colors highlight the different orbital characters of the minimal set of bands. (b) Three band model Ir-$t_{2g}$ for \bairo. (c) DFT electronic structure and MLWs calculated for \sriro. (d) Layered tight-binding model for \sriro. }
    \label{fig:SM_1}
\end{figure*}

\section{Overview over different Hubbard models}

Here, we report details on the DFT band structure of \bairo\ and \sriro, the tight-binding models used in the DFT+DMFT calculations, and we provide more detailed information on the interacting part of the Hamiltonian. 

Similarly to the computational details, in the case of \bairo, the electronic structure and the models are the same as in  Ref. \onlinecite{Cassol_2024}. 
We mention the key information here for the sake of completeness. 

Fig.\ref{fig:SM_1}(a) reports the DFT band structure of \bairo\ in the PM phase along the high-symmetry points of the corresponding Brillouin zone (BZ). 
In color, we report the dispersion of the Wannier functions of the Ir-$d$ (five-orbital) model, which we use in the next sections to discuss the role played by the $e_g$ bands in the calculation of the optical conductivity. 
The presence of the \JEFF, \jeff\ and $\tilde{e}_g$ manifolds can be clearly distinguished, where the tilde on the latter indicates the presence of SOC (without SOC, those bands are identified as $e_g$). 
Panel \ref{fig:SM_1}(b) reports the same DFT band structure now compared with the Ir-$t_{2g}$ model, which we used in the main text. 
The wannierization was performed on the band structure without SOC, since we calculated the interacting part of the Hamiltonian within the constrained random phase approximation (cRPA) \cite{cRPA,cRPA2} as implemented in the RESPACK code \cite{RESPACK} (see Ref.~\onlinecite{Cassol_2024} for details).
The SOC was then added via a local term in the Hamiltonian, as described in Ref.~\onlinecite{Qiangqiang2023}, with the SOC constant $\lambda=0.31$ eV. 
To perform DFT+AFM-DMFT calculations, we expressed the Ir-$t_{2g}$ Hamiltonian in a new unit cell containing two Ir sites, with unit vectors $\mathbf{v}_1=(a,a,0)$, $\mathbf{v}_2=(-a,a,c)$, $\mathbf{v}_3=(a/2,a/2,c/2)$. 
Here, $a$ and $c$ are the lattice parameters of the \textit{conventional} unit cell of \bairo, and amount to the shorter distance between the Ir sites in the $xy$ plane ($a$) and along the $z$ direction ($c$), respectively. 

Fig. \ref{fig:SM_1}(c) reports the DFT electronic structure of \sriro. 
Due to structural distortions, the $\tilde{e}_g$ bands no longer cross the Fermi level.
With respect to panels \ref{fig:SM_1}(a,b), the bands are backfolded, which results in twice the number of bands when shown along the same $\mathbf{k}$-path. 
Moreover, as a consequence of the rotations of the IrO$_6$ octahedra, the \jeff\ bands with large $d_{xy}$ character bend down around $\Gamma$. 
The fact that the $e_g$ and $d_{xy}$ bands are more affected by the distortion stems from their greater extent in the $xy$ plane, where the distortion takes place. 
Similarly to \bairo, we construct a three-band model, whose dispersion is marked in orange.
In this case, the tight-binding model is obtained by wannierizing the DFT electronic structure with SOC.
As reported in Ref. \onlinecite{Pavarini_2021}, the local one-body Hamiltonian can be expressed in the $\{d_{xz \uparrow},d_{xz \downarrow},d_{yz \uparrow},d_{yz \downarrow},d_{xy \uparrow},d_{xy \downarrow \}}$ basis as:
 \begin{equation}
            H_{loc} = \frac{1}{2} \begin{pmatrix}
            2\varepsilon_{xz} & 0 & -\lambda_z i & 0 & 0 & +\lambda_x i  \\
             0 &  2\varepsilon_{xz} & 0 &  +\lambda_z i  & +\lambda_x i   & 0\\
            +\lambda_z i & 0 & 2\varepsilon_{yz} & 0 & 0 & -\lambda_y  \\
             0 &  -\lambda_z i & 0 &  2\varepsilon_{yz} & +\lambda_y   & 0\\
             0 & -\lambda_x i & 0 & +\lambda_y & 2\varepsilon_{xy} & 0  \\
             -\lambda_x i  &  0 & -\lambda_y & 0& 0   & 2\varepsilon_{xy}\\

        \end{pmatrix} \ ,
\end{equation}\label{eq:local_ham_t2g_sriro}
where the local Ir-$t_{2g}$ energies occupy the diagonal and the values of the SOC coupling constants  $\lambda_z$ and $\lambda_x=\lambda_y$ amount to 0.36 eV and 0.34 eV, respectively, which are slightly larger than in \bairo. The existence of two distinct SOC constants stems here from the absence of the $d_{x^2-y^2}$ orbital in our effective model.
As a result, the Wannier function of $d_{xy}$-type includes some $d_{x^2-y^2}$ component. As mentioned in the Computational Details section, this leads to a more delocalized $d_{xy}$ orbital, and changes the effective SOC constant of this orbital with respect to the $d_{xz/yz}$ orbitals.

In the paramagnetic phase, the primitive cell of \sriro\ contains two layers and four Ir sites, to account for the structural distortion. 
However, in order to account for the canting of the magnetic moments, the AFM conventional cell should contain four layers \cite{Kim_2014,Ye_2013} (8 Ir sites). 
To reduce the complexity of the problem while keeping the crucial ingredients, we consider here a single-layer three-orbital model, namely, we remove the inter-layer hopping terms. 
The impact of this approximation is shown in Fig.\ref{fig:SM_1}(d), where the number of bands in the fit is reduced from 24 to 12. 
Still, the DFT dispersion is perfectly reproduced in the range [$\sim$ -0.5,0.5] eV. 
The only deviation from the DFT reference stems from the fact that, for lower-energy states between -0.5 eV and 1 eV, the actual Wannier fit along the $\Gamma-X^{*}$ and $X-\Gamma$ segments lies in between the bonding and antibonding states generated by the inter-layer hopping. 
In terms of magnetism, our single-layer approximation implies that we assume a magnetic phase of basal type, such as the one of \bairo. 

The interacting part of the Hamiltonian is expressed in the generalized Hubbard-Kanamori form: 

    \begin{align}
            \label{eq:model_int}
         \hat{H}_{\text{int}} =  &\frac{1}{2} \sum_{\sigma} \sum_{ij} U_{ij} \hat{n}_{i\sigma}\hat{n}_{j\bar{\sigma}} + \frac{1}{2} \sum_{\sigma} \sum_{i \neq j} (U_{ij} - J_{ij}) \hat{n}_{i\sigma}\hat{n}_{j\sigma} \nonumber \\&
            - \frac{1}{2} \sum_{\sigma}\sum_{i\neq j} J_{ij} \left[ \cdag_{i\sigma}\c_{i\bar{\sigma}}\cdag_{j\bar{\sigma}}\c_{j\sigma} - \cdag_{i\sigma}\cdag_{i\bar{\sigma}}\c_{j\sigma}\c_{j\bar{\sigma}}\right] \, 
    \end{align}

where the first two terms represent the direct Coulomb interactions, and the last two are spin-flip and pair-hopping terms, respectively.

In the case of \bairo, the Coulomb tensor is again consistent with the one reported in Ref.~\onlinecite{Cassol_2024}, where a small orbital anisotropy enhances the direct interaction on the intra-orbital $d_{xy}$ component, and we have a fully orbital symmetric Hund's coupling. 
To match experiments, we reduced the density-density part of the interaction by 0.2 eV when performing AFM-DMFT. 
The necessity of such a reduction is related to the typical overestimation of the gap obtained when performing mean-field AFM calculations, due to the lack of non-local fluctuations.
On the other hand, for the PM calculations the interaction was enhanced by 0.2 eV (PM absorption Fig.6 of Main Text) and 0.3 eV (Fig.5 of EM),  with respect to the Coulomb tensors computed in Ref. \onlinecite{Cassol_2024} for the Ir-$t_{2g}$  and Ir-$5d$ model, respectively. 

In the case of \sriro\, the Coulomb tensor is not evaluated within cRPA, since the lack of the $e_g$ states artificially increases the spread on $d_{xy}$. 
We therefore simply consider an isotropic tensor setting $U_{ii}=U_{jj}$ and $U'_{ij}=U'_{jk}=U'_{ik}$, for $i,j,k\in\{d_{zx},d_{yz},d_{xy}\}, i\neq j\neq k\neq i$.

For both compounds, we report the average intra(inter)-orbital Coulomb interaction U (U') and the Hund's Coupling $J$ in Tab. \ref{tab:Coulomb_values} used in AFM-DMFT. 
We attribute the reduction of nearly $0.4$ eV of $U$ and $U'$ in \sriro\  compared to \bairo\ to the true 2D nature of the model, which undergoes the metal-insulator transition at weaker interaction strength with respect to a model including inter-layer terms.

\begin{table}[t]
    \centering
    \begin{tabular}{|c||ccc|}
    \hline
    \hline
    Interaction (ev)& \hfill $U$ & $U'$ & $J$ \\
    \hline
    \bairo   &  2.33\ & 1.72 & 0.22 \\

    \sriro   &  1.97 & 1.51 & 0.26\\
    \hline
    \hline
    \end{tabular}
    \caption{Average intra-orbital $U$, inter-orbital $U'$ and exchange interaction $J$ employed for \bairo\ and \sriro.}
    \label{tab:Coulomb_values}
\end{table}

%%%%%%%%%%%%%%%%%%%%%%%%%%%%%%%%%%%%%%%%%%%%%%%%%%%%%%%%%%

\section{Paramagnetism and the role of the $e_g$ states}

Here, we investigate the impact of the $e_g$ bands on the optical conductivity.
In the context of \sriro, structural distortions effectively enhance the splitting between $t_{2g}$ and $e_g$ states, excluding their contribution to the low-energy electronic structure already at the DFT level. 
However, a question remains open in \bairo , where all $d$ states cross the Fermi energy in DFT \cite{Cassol_2024}. 
Therefore, in order to justify our It-$t_{2g}$ model for interpreting the low-energy part of the optical conductivity, we have to identify the contributions of the $e_g$ bands.

\begin{figure}[t!]
    \centering
    \includegraphics[width=\linewidth]{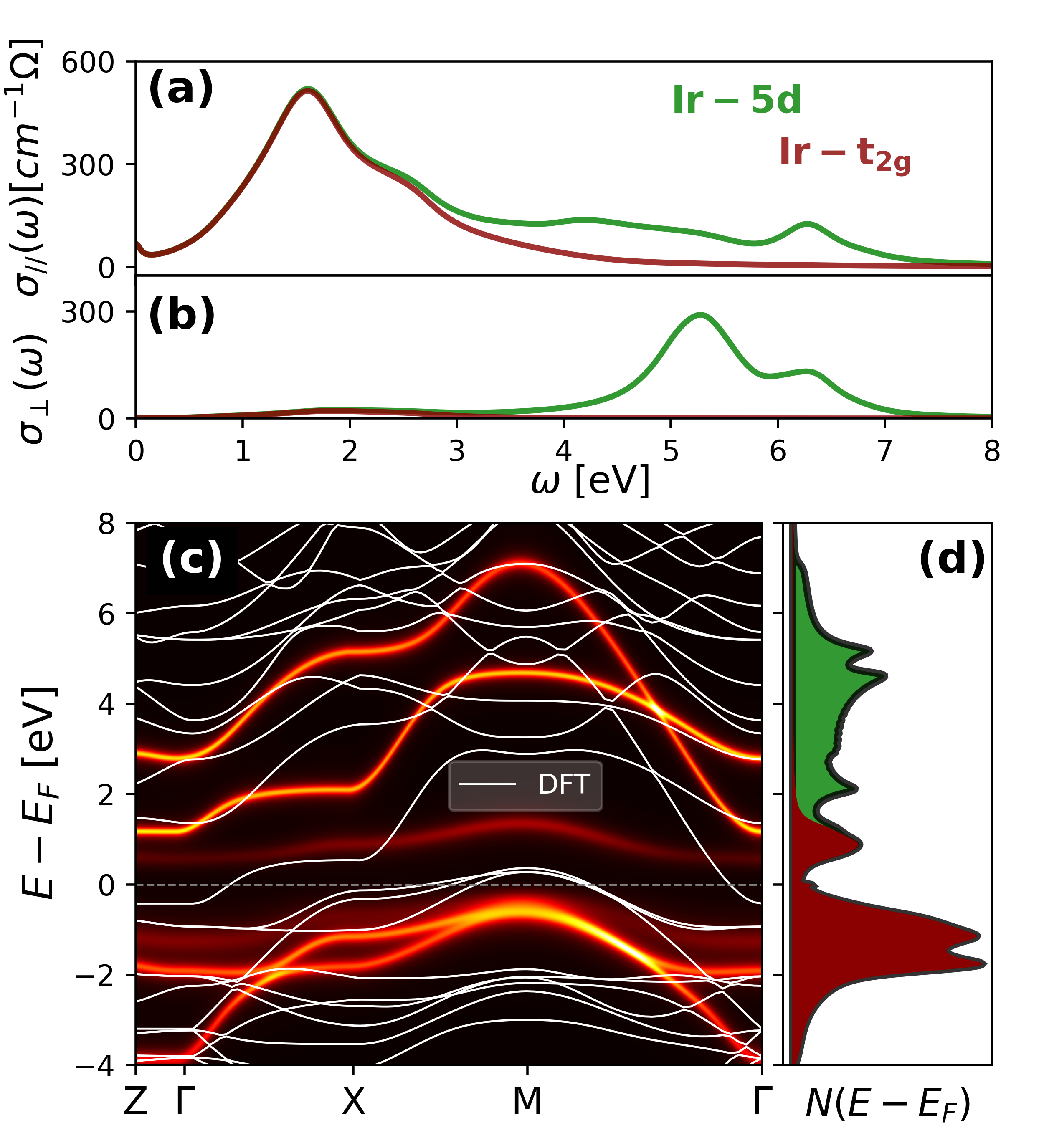}
    \caption{Optical conductivity of \bairo\ in the paramagnetic phase: Ir-$5d$ vs Ir-$t_{2g}$ model. (a) Comparison between the in-plane (a) and out-of-plane (b) conductivity of the full Ir-$d$ and the three-band Ir-$t_{2g}$ models. (c) Momentum resolved spectral function from DFT+DMFT for $\beta=80$ $\text{1/eV}$. (d) Full Ir-$d$ vs Ir-$t_{2g}$ DOS.}
    \label{fig:appendix_2}
\end{figure}

To assess this point, we set $U=2.65$ eV on the $t_{2g}$ states (see previous sections and Ref. \onlinecite{Cassol_2024} for more details) and solve the DMFT equations for the Ir-$5d$ model. Then, we calculate the optical conductivity both in the full manifold and within the $t_{2g}$ subspace, by applying orbital selectivity.  The resulting conductivities and density of states (DOS) are shown in Fig. \ref{fig:appendix_2}, where green and brown curves represent the Ir-$d$ and Ir-$t_{2g}$ quantities, respectively. In the low-frequency regime $\omega<3$ eV, the two curves are essentially identical, proving that the relevant excitations take place within the $j_{\mathrm{eff}}$ states. For $\omega>3$ eV, various excitations can be observed directly involving the different peaks of the $e_g$ states. However, in this frequency range, a direct comparison with experiments is hard to carry out in practice, since transitions involving oxygen $p$ states become important. Since in \bairo\ the position of the $\alpha$ and $\beta$ peaks in optical absorption is 0.78-0.79 eV and 1.43-1.50 eV, respectively, we can safely proceed with a simpler three-band model. 

We note that within this paramagnetic calculation, the low energy part of $\sigma(\omega)$ totally misses the double peak structure both in the full-$d$ and in the $t_{2g}$ case. 
This can be traced back to the absence of spin-polarons, which would require either an AFM DMFT calculation or even an inclusion of non-local fluctuations, which are absent in single-site DMFT, and which are known to restore the proper dispersion of the spectral function $A_{\k}(\omega)$ \cite{Martins_2018,Lenz_2019}.

\section{Fermi velocities}

The optical conductivity depends on the interplay between two main objects: spectral functions and  Fermi velocities, see Eq (\ref{eq:optical_cond_awk}). 
The influence of the former on the conductivity was extensively investigated in the main text. 
Here, we instead focus on the velocities. 

\begin{figure}[t!]
    \centering
    \includegraphics[width=\linewidth]{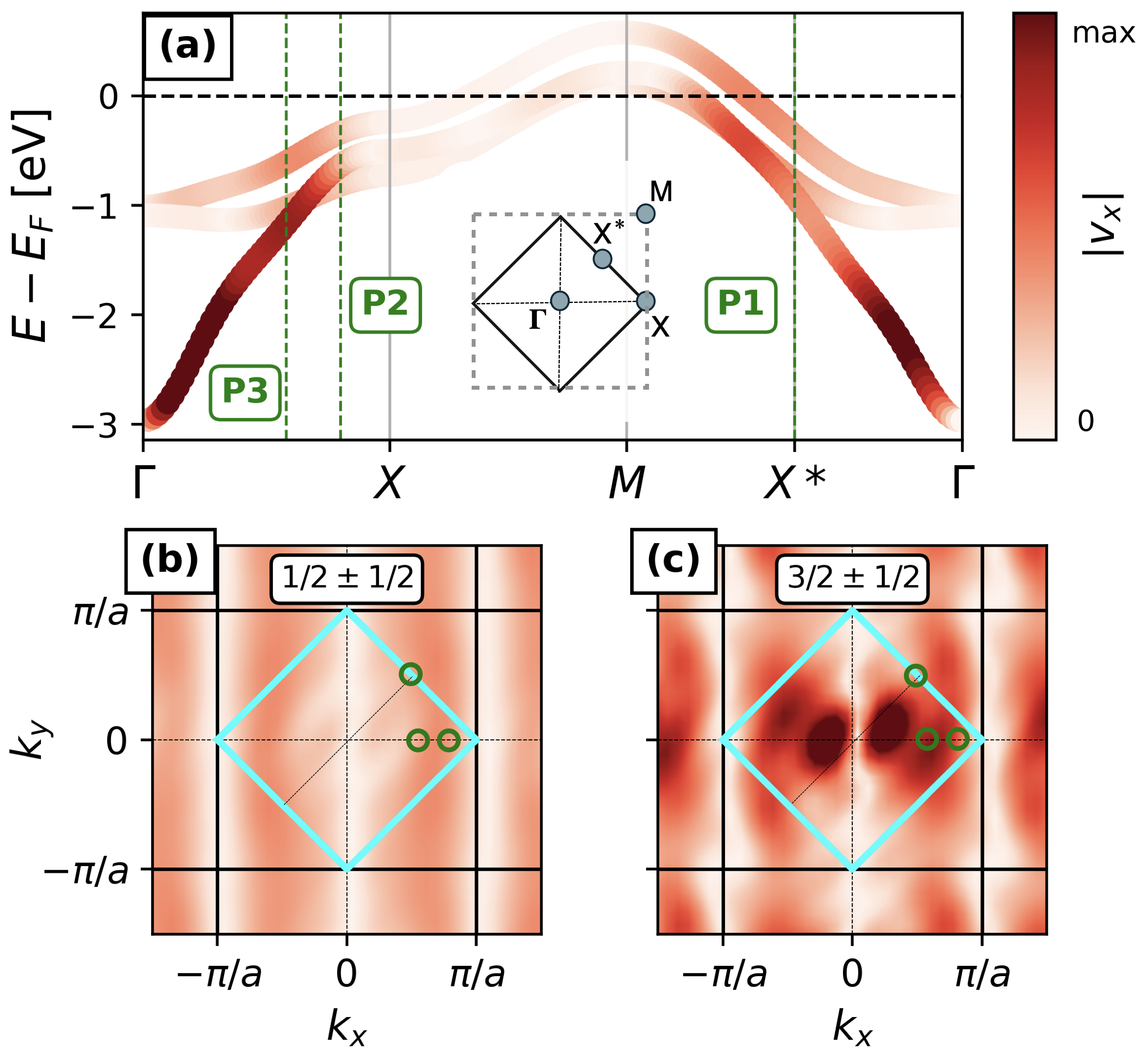}
    \caption{Fermi velocities along the $x$ direction for \bairo. (a) Diagonal elements of the velocity matrix $v_{x}$ plotted as a colormap on diagonal elements of the Hamiltonian $H_{\k}$. The inset displays the BZ in the PM and AFM phase of \bairo, with the corresponding high symmetry points. Fermi velocities in the ($k_x$, $k_y$) plane for the $\frac{1}{2} \pm \frac{1}{2}$ (b) and $\frac{3}{2} \pm \frac{1}{2}$ (c) states. In green we report the points P1, P2 and P3 of the main text used to calculate the momentum resolved conductivity $\sigma_{\k}(\omega)$. }
    \label{fig:appendix_3}
\end{figure}

In case the unit cell contains more than one atom, we employ a generalization of the Fermi velocities beyond the Peierls substitution as developed in Ref.  \onlinecite{Tomczak_2009}:

\begin{equation}\label{eq:Fermi_velocities_gen}
     v_{j\k\sigma}=\left (\partial_{\k_j} H^{AB}_{\k\sigma}-i(\ri_A-\ri_B)_j H^{AB}_{\k\sigma} \right ) \ , 
\end{equation}

where $A$ and $B$ refer to different lattice sites, $\ri_{A}$ and $\ri_{B}$ are the respective lattice positions and both  $v_{j\k\sigma}$ and $H_{\k\sigma}$ are matrices in orbital space, the latter representing the non-interacting reference Hamiltonian in momentum space. 
In (\ref{eq:Fermi_velocities_gen}), the first term amounts to the canonical definition of the velocities, while the second term represents a correction that recovers intra-cell optical transitions. We stress that the inclusion of this second term is crucial to obtain a quantitative agreement with experiments in both \sriro\ and \bairo. 
In Fig. \ref{fig:appendix_3}(a), we display the diagonal elements of the Fermi velocities of \bairo\ as a color map over the corresponding DFT dispersion. 
One can observe that the maximum values of the velocities belong to the $\frac{3}{2} \pm \frac{1}{2}$ state. 
However, since this band is very dispersive, excitations involving this state do not contribute much to the total conductivity. 
Yet, their fingerprint can be recognized in a dispersive feature in $\sigma_{\k}(\omega)$ around the $\Gamma$ point (see Fig.4(b) of the main text).
Moreover, Fig. \ref{fig:appendix_3} confirms that at the high-symmetry points $\Gamma, X$ and $M$, the Fermi velocities vanish since the eigenspectrum exhibits a local stationary point. 
On the other hand, the antinodal point $X^*$ is found in a position where velocities are relatively high for all orbitals, which explains why this region contributes strongly. 
In panel \ref{fig:appendix_2}(b), the velocities of the \jeff\ states are reported over the full BZ. 
The green points correspond to P1, P2 and P3 in the main text, revealing that they are characterized by a large Fermi velocity.

%%%%%%%%%%%%%%%%%%%%%%%%%%%%%%%%%%%%%%%%%%%%%%%%%%%%%%%%%%%%%%%%%%%

\section{Relative permittivity and absorption}

\begin{figure}[t!]
    \centering
    \includegraphics[width=\linewidth]{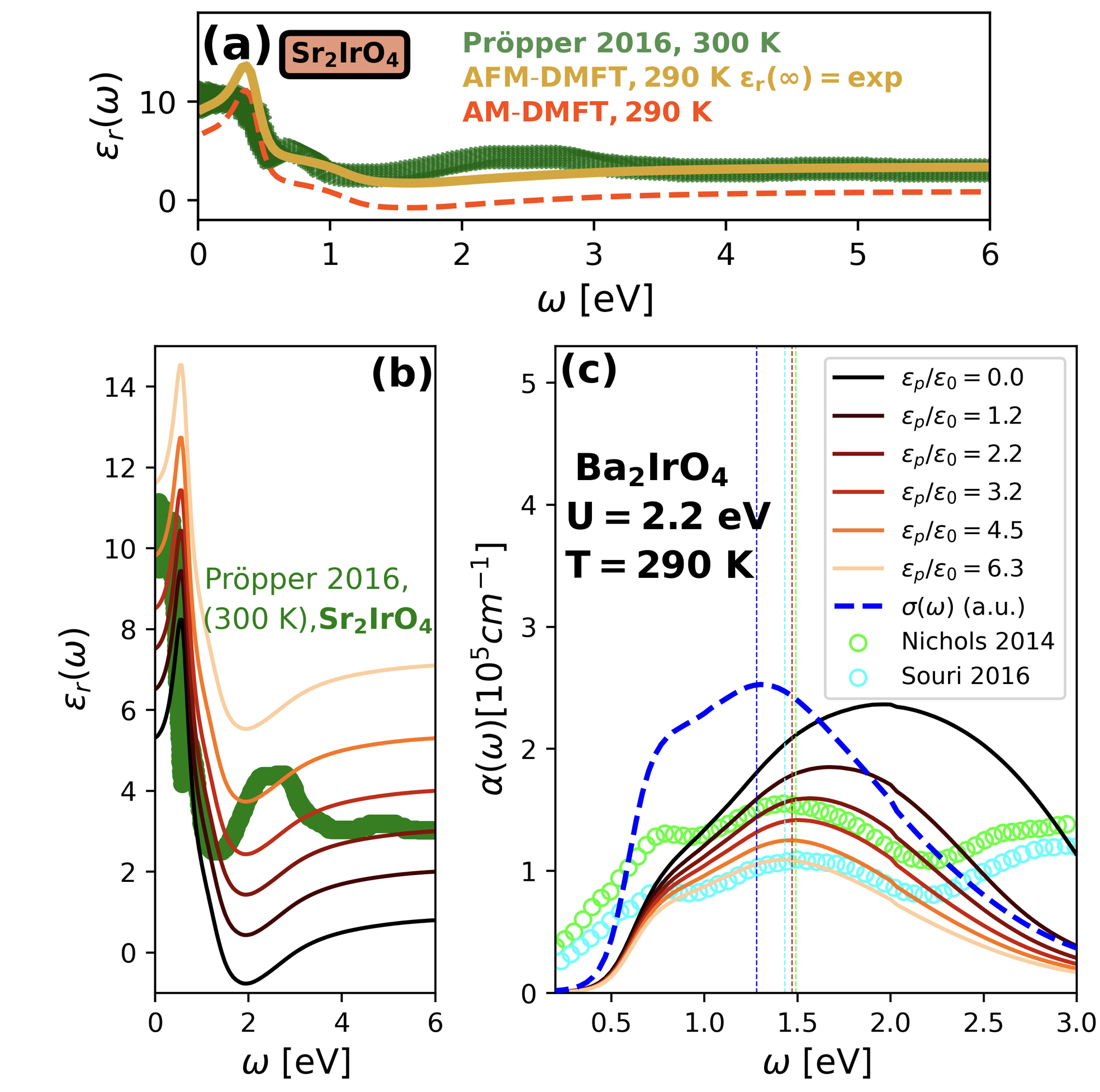}
    \caption{Relative permittivity and optical absorption. (a) Relative permittivity of \sriro\ vs experiments. (b) Dependence of the dielectric function of \bairo\ on $\varepsilon_p$ and comparison with the experimental one relative to \sriro\cite{propper_2016}. (c) The resulting evolution of the optical absorption. Experimental datasets are the same as Fig. \ref{fig:quantitative_comparison}(a).}
    \label{fig:appendix_4}
\end{figure}

From $\sigma(\omega)$, one can evaluate the relative permittivity $\varepsilon_r(\omega)$. 
In DMFT, only the electronic contribution to transport quantities is taken into account. 
However, in solid-state systems, there will be additional contributions, for instance from phonons, impurities, or defects. 
To compensate for their absence in our calculation, we shift the bare DMFT permittivity to align the high-frequency limit with the experimental data: $\varepsilon_r(\omega)=\varepsilon_r^{\mathrm{DMFT}}(\omega)+\varepsilon_p$ such that $\varepsilon_r(\omega\rightarrow\infty) = \varepsilon_r^{\mathrm{exp}}(\omega\rightarrow\infty)$, see Fig.~\ref{fig:appendix_4}(a). 
In doing so, we achieve good quantitative agreement in the low-frequency region, accurately reproducing the zero-frequency magnitude and the positions of the two shoulders corresponding to the peaks $\alpha$ and $\beta$ in conductivity.

However, in case of \bairo, the only experimental data available in the literature \cite{Nichols_2014,Souri_2016} is the optical absorption $\alpha (\omega)$, which can be calculated from $\varepsilon_r(\omega)$ by:

\begin{equation}
    \alpha(\omega)= \sqrt{2}\frac{\omega}{c} \sqrt{|\varepsilon_r(\omega)|- \Re \varepsilon_{r}{(\omega)}}.
    \label{eq:absorption}
\end{equation}
 Given the similarity between the two compounds, we assume that the high-frequency tail of the permittivity of \bairo\ is the same as that of \sriro, setting a constant static shift $\varepsilon_{p}=2.5\varepsilon_{0}$ (see Fig.\ref{fig:appendix_4}(b)). 
The dependence of the optical absorption on $\varepsilon_p$ is reported in panel (c). 
The corresponding optical conductivity of \bairo\ is reported in arbitrary units for comparison. 
First, we notice a reduction in the overall magnitude for increasing $\varepsilon_p$. 
Concerning the shape,  while the lower energy peak remains at the same position upon variation of $\varepsilon_p$, and its position coincides with the $\alpha$ peak in conductivity, the higher energy peak undergoes a blue shift when increasing  $\varepsilon_p$. 
Although there is no experimental evidence that the high-frequency tail of the relative permittivity of \bairo\ is equal to that of \sriro\, we notice that for all shifts $\varepsilon_p$ that make the dielectric function positive defined, both the magnitude and the peak position remain in good agreement with the experiments. 
Hence, the degree of arbitrariness that we have when assigning a given value to $\varepsilon_p$ does not reduce the quality of the analysis.

%%%%%%%%%%%%%%%%%%%%%%% ENDSUPPLEMENTAL

%


\begin{thebibliography}{103}%
\makeatletter
\providecommand \@ifxundefined [1]{%
 \@ifx{#1\undefined}
}%
\providecommand \@ifnum [1]{%
 \ifnum #1\expandafter \@firstoftwo
 \else \expandafter \@secondoftwo
 \fi
}%
\providecommand \@ifx [1]{%
 \ifx #1\expandafter \@firstoftwo
 \else \expandafter \@secondoftwo
 \fi
}%
\providecommand \natexlab [1]{#1}%
\providecommand \enquote  [1]{``#1''}%
\providecommand \bibnamefont  [1]{#1}%
\providecommand \bibfnamefont [1]{#1}%
\providecommand \citenamefont [1]{#1}%
\providecommand \href@noop [0]{\@secondoftwo}%
\providecommand \href [0]{\begingroup \@sanitize@url \@href}%
\providecommand \@href[1]{\@@startlink{#1}\@@href}%
\providecommand \@@href[1]{\endgroup#1\@@endlink}%
\providecommand \@sanitize@url [0]{\catcode `\\12\catcode `\$12\catcode `\&12\catcode `\#12\catcode `\^12\catcode `\_12\catcode `\%12\relax}%
\providecommand \@@startlink[1]{}%
\providecommand \@@endlink[0]{}%
\providecommand \url  [0]{\begingroup\@sanitize@url \@url }%
\providecommand \@url [1]{\endgroup\@href {#1}{\urlprefix }}%
\providecommand \urlprefix  [0]{URL }%
\providecommand \Eprint [0]{\href }%
\providecommand \doibase [0]{https://doi.org/}%
\providecommand \selectlanguage [0]{\@gobble}%
\providecommand \bibinfo  [0]{\@secondoftwo}%
\providecommand \bibfield  [0]{\@secondoftwo}%
\providecommand \translation [1]{[#1]}%
\providecommand \BibitemOpen [0]{}%
\providecommand \bibitemStop [0]{}%
\providecommand \bibitemNoStop [0]{.\EOS\space}%
\providecommand \EOS [0]{\spacefactor3000\relax}%
\providecommand \BibitemShut  [1]{\csname bibitem#1\endcsname}%
\let\auto@bib@innerbib\@empty
%</preamble>
\bibitem [{\citenamefont {Dressel}\ and\ \citenamefont {Grüner}(2002)}]{Dressel_Grüner_2002}%
  \BibitemOpen
  \bibfield  {author} {\bibinfo {author} {\bibfnamefont {M.}~\bibnamefont {Dressel}}\ and\ \bibinfo {author} {\bibfnamefont {G.}~\bibnamefont {Grüner}},\ }\href@noop {} {\emph {\bibinfo {title} {{Electrodynamics of Solids: Optical Properties of Electrons in Matter}}}}\ (\bibinfo  {publisher} {Cambridge University Press},\ \bibinfo {year} {2002})\BibitemShut {NoStop}%
\bibitem [{\citenamefont {Molegraaf}\ \emph {et~al.}(2002)\citenamefont {Molegraaf}, \citenamefont {Presura}, \citenamefont {van~der Marel}, \citenamefont {Kes},\ and\ \citenamefont {Li}}]{Molegraaf_2002}%
  \BibitemOpen
  \bibfield  {author} {\bibinfo {author} {\bibfnamefont {H.~J.~A.}\ \bibnamefont {Molegraaf}}, \bibinfo {author} {\bibfnamefont {C.}~\bibnamefont {Presura}}, \bibinfo {author} {\bibfnamefont {D.}~\bibnamefont {van~der Marel}}, \bibinfo {author} {\bibfnamefont {P.~H.}\ \bibnamefont {Kes}},\ and\ \bibinfo {author} {\bibfnamefont {M.}~\bibnamefont {Li}},\ }\bibfield  {title} {\bibinfo {title} {{Superconductivity-Induced Transfer of In-Plane Spectral Weight in $\text{Bi}_2\text{Sr}_2\text{CaCu}_2\text{O}_{8+\delta}$}},\ }\href {https://www.science.org/doi/abs/10.1126/science.1069947} {\bibfield  {journal} {\bibinfo  {journal} {Science}\ }\textbf {\bibinfo {volume} {295}},\ \bibinfo {pages} {2239} (\bibinfo {year} {2002})}\BibitemShut {NoStop}%
\bibitem [{\citenamefont {Katsufuji}\ \emph {et~al.}(1995)\citenamefont {Katsufuji}, \citenamefont {Okimoto}, \citenamefont {Arima}, \citenamefont {Tokura},\ and\ \citenamefont {Torrance}}]{Katsufuji_1995}%
  \BibitemOpen
  \bibfield  {author} {\bibinfo {author} {\bibfnamefont {T.}~\bibnamefont {Katsufuji}}, \bibinfo {author} {\bibfnamefont {Y.}~\bibnamefont {Okimoto}}, \bibinfo {author} {\bibfnamefont {T.}~\bibnamefont {Arima}}, \bibinfo {author} {\bibfnamefont {Y.}~\bibnamefont {Tokura}},\ and\ \bibinfo {author} {\bibfnamefont {J.~B.}\ \bibnamefont {Torrance}},\ }\bibfield  {title} {\bibinfo {title} {{Optical spectroscopy of the metal-insulator transition in ${\mathrm{NdNiO}}_{3}$}},\ }\href {https://doi.org/10.1103/PhysRevB.51.4830} {\bibfield  {journal} {\bibinfo  {journal} {Phys. Rev. B}\ }\textbf {\bibinfo {volume} {51}},\ \bibinfo {pages} {4830} (\bibinfo {year} {1995})}\BibitemShut {NoStop}%
\bibitem [{\citenamefont {Puchkov}\ \emph {et~al.}(1996)\citenamefont {Puchkov}, \citenamefont {Basov},\ and\ \citenamefont {Timusk}}]{Puchkov_1996}%
  \BibitemOpen
  \bibfield  {author} {\bibinfo {author} {\bibfnamefont {A.~V.}\ \bibnamefont {Puchkov}}, \bibinfo {author} {\bibfnamefont {D.~N.}\ \bibnamefont {Basov}},\ and\ \bibinfo {author} {\bibfnamefont {T.}~\bibnamefont {Timusk}},\ }\bibfield  {title} {\bibinfo {title} {{The pseudogap state in high- superconductors: an infrared study}},\ }\href {https://doi.org/10.1088/0953-8984/8/48/023} {\bibfield  {journal} {\bibinfo  {journal} {Journal of Physics: Condensed Matter}\ }\textbf {\bibinfo {volume} {8}},\ \bibinfo {pages} {10049} (\bibinfo {year} {1996})}\BibitemShut {NoStop}%
\bibitem [{\citenamefont {Gervais}(2002)}]{Gervais_2002}%
  \BibitemOpen
  \bibfield  {author} {\bibinfo {author} {\bibfnamefont {F.}~\bibnamefont {Gervais}},\ }\bibfield  {title} {\bibinfo {title} {{Optical conductivity of oxides}},\ }\href {https://doi.org/https://doi.org/10.1016/S0927-796X(02)00073-6} {\bibfield  {journal} {\bibinfo  {journal} {Materials Science and Engineering: R: Reports}\ }\textbf {\bibinfo {volume} {39}},\ \bibinfo {pages} {29} (\bibinfo {year} {2002})}\BibitemShut {NoStop}%
\bibitem [{\citenamefont {Millis}(2004)}]{Millis_2004}%
  \BibitemOpen
  \bibfield  {author} {\bibinfo {author} {\bibfnamefont {A.}~\bibnamefont {Millis}},\ }\bibfield  {title} {\bibinfo {title} {{Optical conductivity and correlated electron physics}},\ }\href@noop {} {\bibfield  {journal} {\bibinfo  {journal} {Strong interactions in low dimensions}\ ,\ \bibinfo {pages} {195}} (\bibinfo {year} {2004})}\BibitemShut {NoStop}%
\bibitem [{\citenamefont {Basov}\ \emph {et~al.}(2011)\citenamefont {Basov}, \citenamefont {Averitt}, \citenamefont {van~der Marel}, \citenamefont {Dressel},\ and\ \citenamefont {Haule}}]{Basov_2011}%
  \BibitemOpen
  \bibfield  {author} {\bibinfo {author} {\bibfnamefont {D.~N.}\ \bibnamefont {Basov}}, \bibinfo {author} {\bibfnamefont {R.~D.}\ \bibnamefont {Averitt}}, \bibinfo {author} {\bibfnamefont {D.}~\bibnamefont {van~der Marel}}, \bibinfo {author} {\bibfnamefont {M.}~\bibnamefont {Dressel}},\ and\ \bibinfo {author} {\bibfnamefont {K.}~\bibnamefont {Haule}},\ }\bibfield  {title} {\bibinfo {title} {{Electrodynamics of correlated electron materials}},\ }\href {https://doi.org/10.1103/RevModPhys.83.471} {\bibfield  {journal} {\bibinfo  {journal} {Rev. Mod. Phys.}\ }\textbf {\bibinfo {volume} {83}},\ \bibinfo {pages} {471} (\bibinfo {year} {2011})}\BibitemShut {NoStop}%
\bibitem [{\citenamefont {Charnukha}(2014)}]{Charnukha_2014}%
  \BibitemOpen
  \bibfield  {author} {\bibinfo {author} {\bibfnamefont {A.}~\bibnamefont {Charnukha}},\ }\bibfield  {title} {\bibinfo {title} {{Optical conductivity of iron-based superconductors}},\ }\href {https://doi.org/10.1088/0953-8984/26/25/253203} {\bibfield  {journal} {\bibinfo  {journal} {Journal of Physics: Condensed Matter}\ }\textbf {\bibinfo {volume} {26}},\ \bibinfo {pages} {253203} (\bibinfo {year} {2014})}\BibitemShut {NoStop}%
\bibitem [{\citenamefont {Hedin}(1965)}]{Hedin_1965}%
  \BibitemOpen
  \bibfield  {author} {\bibinfo {author} {\bibfnamefont {L.}~\bibnamefont {Hedin}},\ }\bibfield  {title} {\bibinfo {title} {{New Method for Calculating the One-Particle Green's Function with Application to the Electron-Gas Problem}},\ }\href {https://doi.org/10.1103/PhysRev.139.A796} {\bibfield  {journal} {\bibinfo  {journal} {Phys. Rev.}\ }\textbf {\bibinfo {volume} {139}},\ \bibinfo {pages} {A796} (\bibinfo {year} {1965})}\BibitemShut {NoStop}%
\bibitem [{\citenamefont {Georges}\ \emph {et~al.}(1996)\citenamefont {Georges}, \citenamefont {Kotliar}, \citenamefont {Krauth},\ and\ \citenamefont {Rozenberg}}]{Georges_1996}%
  \BibitemOpen
  \bibfield  {author} {\bibinfo {author} {\bibfnamefont {A.}~\bibnamefont {Georges}}, \bibinfo {author} {\bibfnamefont {G.}~\bibnamefont {Kotliar}}, \bibinfo {author} {\bibfnamefont {W.}~\bibnamefont {Krauth}},\ and\ \bibinfo {author} {\bibfnamefont {M.~J.}\ \bibnamefont {Rozenberg}},\ }\bibfield  {title} {\bibinfo {title} {{Dynamical mean-field theory of strongly correlated fermion systems and the limit of infinite dimensions}},\ }\href {https://doi.org/10.1103/RevModPhys.68.13} {\bibfield  {journal} {\bibinfo  {journal} {Rev. Mod. Phys.}\ }\textbf {\bibinfo {volume} {68}},\ \bibinfo {pages} {13} (\bibinfo {year} {1996})}\BibitemShut {NoStop}%
\bibitem [{\citenamefont {Rau}\ \emph {et~al.}(2016)\citenamefont {Rau}, \citenamefont {Lee},\ and\ \citenamefont {Kee}}]{Rau2016}%
  \BibitemOpen
  \bibfield  {author} {\bibinfo {author} {\bibfnamefont {J.~G.}\ \bibnamefont {Rau}}, \bibinfo {author} {\bibfnamefont {E.~K.-H.}\ \bibnamefont {Lee}},\ and\ \bibinfo {author} {\bibfnamefont {H.-Y.}\ \bibnamefont {Kee}},\ }\bibfield  {title} {\bibinfo {title} {Spin-orbit physics giving rise to novel phases in correlated systems: Iridates and related materials},\ }\href {https://doi.org/https://doi.org/10.1146/annurev-conmatphys-031115-011319} {\bibfield  {journal} {\bibinfo  {journal} {Annual Review of Condensed Matter Physics}\ }\textbf {\bibinfo {volume} {7}},\ \bibinfo {pages} {195} (\bibinfo {year} {2016})}\BibitemShut {NoStop}%
\bibitem [{\citenamefont {Bertinshaw}\ \emph {et~al.}(2019)\citenamefont {Bertinshaw}, \citenamefont {Kim}, \citenamefont {Khaliullin},\ and\ \citenamefont {Kim}}]{Bertinshaw2019}%
  \BibitemOpen
  \bibfield  {author} {\bibinfo {author} {\bibfnamefont {J.}~\bibnamefont {Bertinshaw}}, \bibinfo {author} {\bibfnamefont {Y.}~\bibnamefont {Kim}}, \bibinfo {author} {\bibfnamefont {G.}~\bibnamefont {Khaliullin}},\ and\ \bibinfo {author} {\bibfnamefont {B.}~\bibnamefont {Kim}},\ }\bibfield  {title} {\bibinfo {title} {Square lattice iridates},\ }\href {https://doi.org/https://doi.org/10.1146/annurev-conmatphys-031218-013113} {\bibfield  {journal} {\bibinfo  {journal} {Annual Review of Condensed Matter Physics}\ }\textbf {\bibinfo {volume} {10}},\ \bibinfo {pages} {315} (\bibinfo {year} {2019})}\BibitemShut {NoStop}%
\bibitem [{\citenamefont {Crawford}\ \emph {et~al.}(1994)\citenamefont {Crawford}, \citenamefont {Subramanian}, \citenamefont {Harlow}, \citenamefont {Fernandez-Baca}, \citenamefont {Wang},\ and\ \citenamefont {Johnston}}]{Crawford_1994}%
  \BibitemOpen
  \bibfield  {author} {\bibinfo {author} {\bibfnamefont {M.~K.}\ \bibnamefont {Crawford}}, \bibinfo {author} {\bibfnamefont {M.~A.}\ \bibnamefont {Subramanian}}, \bibinfo {author} {\bibfnamefont {R.~L.}\ \bibnamefont {Harlow}}, \bibinfo {author} {\bibfnamefont {J.~A.}\ \bibnamefont {Fernandez-Baca}}, \bibinfo {author} {\bibfnamefont {Z.~R.}\ \bibnamefont {Wang}},\ and\ \bibinfo {author} {\bibfnamefont {D.~C.}\ \bibnamefont {Johnston}},\ }\bibfield  {title} {\bibinfo {title} {{Structural and magnetic studies of $\mathrm{Sr}_{2}$$\mathrm{IrO}_{4}$}},\ }\href {https://doi.org/10.1103/PhysRevB.49.9198} {\bibfield  {journal} {\bibinfo  {journal} {Phys. Rev. B}\ }\textbf {\bibinfo {volume} {49}},\ \bibinfo {pages} {9198} (\bibinfo {year} {1994})}\BibitemShut {NoStop}%
\bibitem [{\citenamefont {Moon}\ \emph {et~al.}(2006)\citenamefont {Moon}, \citenamefont {Kim}, \citenamefont {Kim}, \citenamefont {Lee}, \citenamefont {Kim}, \citenamefont {Park}, \citenamefont {Kim}, \citenamefont {Oh}, \citenamefont {Nakatsuji}, \citenamefont {Maeno}, \citenamefont {Nagai}, \citenamefont {Ikeda}, \citenamefont {Cao},\ and\ \citenamefont {Noh}}]{Moon2006}%
  \BibitemOpen
  \bibfield  {author} {\bibinfo {author} {\bibfnamefont {S.~J.}\ \bibnamefont {Moon}}, \bibinfo {author} {\bibfnamefont {M.~W.}\ \bibnamefont {Kim}}, \bibinfo {author} {\bibfnamefont {K.~W.}\ \bibnamefont {Kim}}, \bibinfo {author} {\bibfnamefont {Y.~S.}\ \bibnamefont {Lee}}, \bibinfo {author} {\bibfnamefont {J.-Y.}\ \bibnamefont {Kim}}, \bibinfo {author} {\bibfnamefont {J.-H.}\ \bibnamefont {Park}}, \bibinfo {author} {\bibfnamefont {B.~J.}\ \bibnamefont {Kim}}, \bibinfo {author} {\bibfnamefont {S.-J.}\ \bibnamefont {Oh}}, \bibinfo {author} {\bibfnamefont {S.}~\bibnamefont {Nakatsuji}}, \bibinfo {author} {\bibfnamefont {Y.}~\bibnamefont {Maeno}}, \bibinfo {author} {\bibfnamefont {I.}~\bibnamefont {Nagai}}, \bibinfo {author} {\bibfnamefont {S.~I.}\ \bibnamefont {Ikeda}}, \bibinfo {author} {\bibfnamefont {G.}~\bibnamefont {Cao}},\ and\ \bibinfo {author} {\bibfnamefont {T.~W.}\ \bibnamefont {Noh}},\ }\bibfield  {title} {\bibinfo {title} {{Electronic structures of layered perovskite
  ${\mathrm{Sr}}_{2}M{\mathrm{O}}_{4}$ ($M=\mathrm{Ru}$, Rh, and Ir)}},\ }\href {https://doi.org/10.1103/PhysRevB.74.113104} {\bibfield  {journal} {\bibinfo  {journal} {Phys. Rev. B}\ }\textbf {\bibinfo {volume} {74}},\ \bibinfo {pages} {113104} (\bibinfo {year} {2006})}\BibitemShut {NoStop}%
\bibitem [{\citenamefont {Kim}\ \emph {et~al.}(2008)\citenamefont {Kim}, \citenamefont {Jin}, \citenamefont {Moon}, \citenamefont {Kim}, \citenamefont {Park}, \citenamefont {Leem}, \citenamefont {Yu}, \citenamefont {Noh}, \citenamefont {Kim}, \citenamefont {Oh}, \citenamefont {Park}, \citenamefont {Durairaj}, \citenamefont {Cao},\ and\ \citenamefont {Rotenberg}}]{Kim_2008}%
  \BibitemOpen
  \bibfield  {author} {\bibinfo {author} {\bibfnamefont {B.~J.}\ \bibnamefont {Kim}}, \bibinfo {author} {\bibfnamefont {H.}~\bibnamefont {Jin}}, \bibinfo {author} {\bibfnamefont {S.~J.}\ \bibnamefont {Moon}}, \bibinfo {author} {\bibfnamefont {J.-Y.}\ \bibnamefont {Kim}}, \bibinfo {author} {\bibfnamefont {B.-G.}\ \bibnamefont {Park}}, \bibinfo {author} {\bibfnamefont {C.~S.}\ \bibnamefont {Leem}}, \bibinfo {author} {\bibfnamefont {J.}~\bibnamefont {Yu}}, \bibinfo {author} {\bibfnamefont {T.~W.}\ \bibnamefont {Noh}}, \bibinfo {author} {\bibfnamefont {C.}~\bibnamefont {Kim}}, \bibinfo {author} {\bibfnamefont {S.-J.}\ \bibnamefont {Oh}}, \bibinfo {author} {\bibfnamefont {J.-H.}\ \bibnamefont {Park}}, \bibinfo {author} {\bibfnamefont {V.}~\bibnamefont {Durairaj}}, \bibinfo {author} {\bibfnamefont {G.}~\bibnamefont {Cao}},\ and\ \bibinfo {author} {\bibfnamefont {E.}~\bibnamefont {Rotenberg}},\ }\bibfield  {title} {\bibinfo {title} {{Novel ${J}_{\mathrm{eff}}=1/2$ Mott State Induced by Relativistic Spin-Orbit
  Coupling in $\text{Ba}_2 \text{IrO}_4$}},\ }\href {https://link.aps.org/doi/10.1103/PhysRevLett.101.076402} {\bibfield  {journal} {\bibinfo  {journal} {Phys. Rev. Lett.}\ }\textbf {\bibinfo {volume} {101}},\ \bibinfo {pages} {076402} (\bibinfo {year} {2008})}\BibitemShut {NoStop}%
\bibitem [{\citenamefont {Kim}\ \emph {et~al.}(2009)\citenamefont {Kim}, \citenamefont {Ohsumi}, \citenamefont {Komesu}, \citenamefont {Sakai}, \citenamefont {Morita}, \citenamefont {Takagi},\ and\ \citenamefont {Arima}}]{Kim_2009}%
  \BibitemOpen
  \bibfield  {author} {\bibinfo {author} {\bibfnamefont {J.}~\bibnamefont {Kim}}, \bibinfo {author} {\bibfnamefont {H.}~\bibnamefont {Ohsumi}}, \bibinfo {author} {\bibfnamefont {T.}~\bibnamefont {Komesu}}, \bibinfo {author} {\bibfnamefont {S.}~\bibnamefont {Sakai}}, \bibinfo {author} {\bibfnamefont {T.}~\bibnamefont {Morita}}, \bibinfo {author} {\bibfnamefont {H.}~\bibnamefont {Takagi}},\ and\ \bibinfo {author} {\bibfnamefont {T.}~\bibnamefont {Arima}},\ }\bibfield  {title} {\bibinfo {title} {{Phase-sensitive observation of a spin–orbital Mott state in $\text{Sr}_2 \text{IrO}_4$}},\ }\href {https://doi.org/10.1126/science.1167106} {\bibfield  {journal} {\bibinfo  {journal} {Science}\ }\textbf {\bibinfo {volume} {323}},\ \bibinfo {pages} {1239} (\bibinfo {year} {2009})}\BibitemShut {NoStop}%
\bibitem [{\citenamefont {Kim}\ \emph {et~al.}(2012)\citenamefont {Kim}, \citenamefont {Casa}, \citenamefont {Upton}, \citenamefont {Gog}, \citenamefont {Kim}, \citenamefont {Mitchell}, \citenamefont {Van~Veenendaal}, \citenamefont {Daghofer}, \citenamefont {Van Den~Brink}, \citenamefont {Khaliullin},\ and\ \citenamefont {Kim}}]{kim_2012}%
  \BibitemOpen
  \bibfield  {author} {\bibinfo {author} {\bibfnamefont {J.}~\bibnamefont {Kim}}, \bibinfo {author} {\bibfnamefont {D.}~\bibnamefont {Casa}}, \bibinfo {author} {\bibfnamefont {M.~H.}\ \bibnamefont {Upton}}, \bibinfo {author} {\bibfnamefont {T.}~\bibnamefont {Gog}}, \bibinfo {author} {\bibfnamefont {Y.-J.}\ \bibnamefont {Kim}}, \bibinfo {author} {\bibfnamefont {J.~F.}\ \bibnamefont {Mitchell}}, \bibinfo {author} {\bibfnamefont {M.}~\bibnamefont {Van~Veenendaal}}, \bibinfo {author} {\bibfnamefont {M.}~\bibnamefont {Daghofer}}, \bibinfo {author} {\bibfnamefont {J.}~\bibnamefont {Van Den~Brink}}, \bibinfo {author} {\bibfnamefont {G.}~\bibnamefont {Khaliullin}},\ and\ \bibinfo {author} {\bibfnamefont {B.~J.}\ \bibnamefont {Kim}},\ }\bibfield  {title} {\bibinfo {title} {{Magnetic Excitation Spectra of $\text{Sr}_2 \text{IrO}_4$ Probed by Resonant Inelastic X-Ray Scattering: Establishing Links to Cuprate Superconductors}},\ }\href {https://doi.org/10.1103/PhysRevLett.108.177003} {\bibfield  {journal} {\bibinfo
  {journal} {Physical Review Letters}\ }\textbf {\bibinfo {volume} {108}},\ \bibinfo {pages} {177003} (\bibinfo {year} {2012})}\BibitemShut {NoStop}%
\bibitem [{\citenamefont {Kim}\ \emph {et~al.}(2014{\natexlab{a}})\citenamefont {Kim}, \citenamefont {Daghofer}, \citenamefont {Said}, \citenamefont {Gog}, \citenamefont {van~den Brink}, \citenamefont {Khaliullin},\ and\ \citenamefont {Kim}}]{Kim_2014}%
  \BibitemOpen
  \bibfield  {author} {\bibinfo {author} {\bibfnamefont {J.}~\bibnamefont {Kim}}, \bibinfo {author} {\bibfnamefont {M.}~\bibnamefont {Daghofer}}, \bibinfo {author} {\bibfnamefont {A.~H.}\ \bibnamefont {Said}}, \bibinfo {author} {\bibfnamefont {T.}~\bibnamefont {Gog}}, \bibinfo {author} {\bibfnamefont {J.}~\bibnamefont {van~den Brink}}, \bibinfo {author} {\bibfnamefont {G.}~\bibnamefont {Khaliullin}},\ and\ \bibinfo {author} {\bibfnamefont {B.~J.}\ \bibnamefont {Kim}},\ }\bibfield  {title} {\bibinfo {title} {{Excitonic quasiparticles in a spin-orbit Mott insulator}},\ }\href {https://doi.org/10.1038/ncomms5453} {\bibfield  {journal} {\bibinfo  {journal} {Nature Communications}\ }\textbf {\bibinfo {volume} {5}},\ \bibinfo {pages} {4453} (\bibinfo {year} {2014}{\natexlab{a}})}\BibitemShut {NoStop}%
\bibitem [{\citenamefont {Brouet}\ \emph {et~al.}(2015)\citenamefont {Brouet}, \citenamefont {Mansart}, \citenamefont {Perfetti}, \citenamefont {Piovera}, \citenamefont {Vobornik}, \citenamefont {Le~F\`evre}, \citenamefont {Bertran}, \citenamefont {Riggs}, \citenamefont {Shapiro}, \citenamefont {Giraldo-Gallo},\ and\ \citenamefont {Fisher}}]{Brouet_2015}%
  \BibitemOpen
  \bibfield  {author} {\bibinfo {author} {\bibfnamefont {V.}~\bibnamefont {Brouet}}, \bibinfo {author} {\bibfnamefont {J.}~\bibnamefont {Mansart}}, \bibinfo {author} {\bibfnamefont {L.}~\bibnamefont {Perfetti}}, \bibinfo {author} {\bibfnamefont {C.}~\bibnamefont {Piovera}}, \bibinfo {author} {\bibfnamefont {I.}~\bibnamefont {Vobornik}}, \bibinfo {author} {\bibfnamefont {P.}~\bibnamefont {Le~F\`evre}}, \bibinfo {author} {\bibfnamefont {F.~m.~c.}\ \bibnamefont {Bertran}}, \bibinfo {author} {\bibfnamefont {S.~C.}\ \bibnamefont {Riggs}}, \bibinfo {author} {\bibfnamefont {M.~C.}\ \bibnamefont {Shapiro}}, \bibinfo {author} {\bibfnamefont {P.}~\bibnamefont {Giraldo-Gallo}},\ and\ \bibinfo {author} {\bibfnamefont {I.~R.}\ \bibnamefont {Fisher}},\ }\bibfield  {title} {\bibinfo {title} {{Transfer of spectral weight across the gap of ${\mathrm{Sr}}_{2}{\mathrm{IrO}}_{4}$ induced by La doping}},\ }\href {https://doi.org/10.1103/PhysRevB.92.081117} {\bibfield  {journal} {\bibinfo  {journal} {Phys. Rev. B}\ }\textbf
  {\bibinfo {volume} {92}},\ \bibinfo {pages} {081117} (\bibinfo {year} {2015})}\BibitemShut {NoStop}%
\bibitem [{\citenamefont {Dai}\ \emph {et~al.}(2014)\citenamefont {Dai}, \citenamefont {Calleja}, \citenamefont {Cao},\ and\ \citenamefont {McElroy}}]{Dai_2014}%
  \BibitemOpen
  \bibfield  {author} {\bibinfo {author} {\bibfnamefont {J.}~\bibnamefont {Dai}}, \bibinfo {author} {\bibfnamefont {E.}~\bibnamefont {Calleja}}, \bibinfo {author} {\bibfnamefont {G.}~\bibnamefont {Cao}},\ and\ \bibinfo {author} {\bibfnamefont {K.}~\bibnamefont {McElroy}},\ }\bibfield  {title} {\bibinfo {title} {{Local density of states study of a spin-orbit-coupling induced Mott insulator ${\mathrm{Sr}}_{2}\mathrm{Ir}{\mathrm{O}}_{4}$}},\ }\href {https://doi.org/10.1103/PhysRevB.90.041102} {\bibfield  {journal} {\bibinfo  {journal} {Phys. Rev. B}\ }\textbf {\bibinfo {volume} {90}},\ \bibinfo {pages} {041102} (\bibinfo {year} {2014})}\BibitemShut {NoStop}%
\bibitem [{\citenamefont {Kim}\ \emph {et~al.}(2016)\citenamefont {Kim}, \citenamefont {Kim}, \citenamefont {Kim},\ and\ \citenamefont {Min}}]{Kim_2016}%
  \BibitemOpen
  \bibfield  {author} {\bibinfo {author} {\bibfnamefont {B.}~\bibnamefont {Kim}}, \bibinfo {author} {\bibfnamefont {B.~H.}\ \bibnamefont {Kim}}, \bibinfo {author} {\bibfnamefont {K.}~\bibnamefont {Kim}},\ and\ \bibinfo {author} {\bibfnamefont {B.~I.}\ \bibnamefont {Min}},\ }\bibfield  {title} {\bibinfo {title} {{Substrate-tuning of correlated spin-orbit oxides revealed by optical conductivity calculations}},\ }\href {https://doi.org/10.1038/srep27095} {\bibfield  {journal} {\bibinfo  {journal} {Scientific Reports}\ }\textbf {\bibinfo {volume} {6}},\ \bibinfo {pages} {27095} (\bibinfo {year} {2016})}\BibitemShut {NoStop}%
\bibitem [{\citenamefont {Perkins}\ \emph {et~al.}(2014)\citenamefont {Perkins}, \citenamefont {Sizyuk},\ and\ \citenamefont {W\"olfle}}]{Perkins2014}%
  \BibitemOpen
  \bibfield  {author} {\bibinfo {author} {\bibfnamefont {N.~B.}\ \bibnamefont {Perkins}}, \bibinfo {author} {\bibfnamefont {Y.}~\bibnamefont {Sizyuk}},\ and\ \bibinfo {author} {\bibfnamefont {P.}~\bibnamefont {W\"olfle}},\ }\bibfield  {title} {\bibinfo {title} {{Interplay of many-body and single-particle interactions in iridates and rhodates}},\ }\href {https://doi.org/10.1103/PhysRevB.89.035143} {\bibfield  {journal} {\bibinfo  {journal} {Phys. Rev. B}\ }\textbf {\bibinfo {volume} {89}},\ \bibinfo {pages} {035143} (\bibinfo {year} {2014})}\BibitemShut {NoStop}%
\bibitem [{\citenamefont {Moon}\ \emph {et~al.}(2009)\citenamefont {Moon}, \citenamefont {Jin}, \citenamefont {Choi}, \citenamefont {Lee}, \citenamefont {Seo}, \citenamefont {Yu}, \citenamefont {Cao}, \citenamefont {Noh},\ and\ \citenamefont {Lee}}]{moon_2009}%
  \BibitemOpen
  \bibfield  {author} {\bibinfo {author} {\bibfnamefont {S.~J.}\ \bibnamefont {Moon}}, \bibinfo {author} {\bibfnamefont {H.}~\bibnamefont {Jin}}, \bibinfo {author} {\bibfnamefont {W.~S.}\ \bibnamefont {Choi}}, \bibinfo {author} {\bibfnamefont {J.~S.}\ \bibnamefont {Lee}}, \bibinfo {author} {\bibfnamefont {S.~S.~A.}\ \bibnamefont {Seo}}, \bibinfo {author} {\bibfnamefont {J.}~\bibnamefont {Yu}}, \bibinfo {author} {\bibfnamefont {G.}~\bibnamefont {Cao}}, \bibinfo {author} {\bibfnamefont {T.~W.}\ \bibnamefont {Noh}},\ and\ \bibinfo {author} {\bibfnamefont {Y.~S.}\ \bibnamefont {Lee}},\ }\bibfield  {title} {\bibinfo {title} {{Temperature dependence of the electronic structure of the ${J}_{\text{eff}}=1/2$ Mott insulator $\text{Sr}_{2}\text{IrO}_{4}$ studied by optical spectroscopy studied by optical spectroscopy}},\ }\href {https://doi.org/10.1103/PhysRevB.80.195110} {\bibfield  {journal} {\bibinfo  {journal} {Physical Review B}\ }\textbf {\bibinfo {volume} {80}},\ \bibinfo {pages} {195110} (\bibinfo {year}
  {2009})},\ \bibinfo {note} {publisher: American Physical Society}\BibitemShut {NoStop}%
\bibitem [{\citenamefont {Sohn}\ \emph {et~al.}(2014)\citenamefont {Sohn}, \citenamefont {Lee}, \citenamefont {Park}, \citenamefont {Noh}, \citenamefont {Yoo}, \citenamefont {Moon}, \citenamefont {Kim}, \citenamefont {Qi}, \citenamefont {Cao}, \citenamefont {Cho},\ and\ \citenamefont {Noh}}]{Sohn_2014}%
  \BibitemOpen
  \bibfield  {author} {\bibinfo {author} {\bibfnamefont {C.~H.}\ \bibnamefont {Sohn}}, \bibinfo {author} {\bibfnamefont {M.-C.}\ \bibnamefont {Lee}}, \bibinfo {author} {\bibfnamefont {H.~J.}\ \bibnamefont {Park}}, \bibinfo {author} {\bibfnamefont {K.~J.}\ \bibnamefont {Noh}}, \bibinfo {author} {\bibfnamefont {H.~K.}\ \bibnamefont {Yoo}}, \bibinfo {author} {\bibfnamefont {S.~J.}\ \bibnamefont {Moon}}, \bibinfo {author} {\bibfnamefont {K.~W.}\ \bibnamefont {Kim}}, \bibinfo {author} {\bibfnamefont {T.~F.}\ \bibnamefont {Qi}}, \bibinfo {author} {\bibfnamefont {G.}~\bibnamefont {Cao}}, \bibinfo {author} {\bibfnamefont {D.-Y.}\ \bibnamefont {Cho}},\ and\ \bibinfo {author} {\bibfnamefont {T.~W.}\ \bibnamefont {Noh}},\ }\bibfield  {title} {\bibinfo {title} {{Orbital-dependent polaron formation in the relativistic Mott insulator $\mathrm{Sr}_{2}\mathrm{IrO}_{4}$}},\ }\href {https://doi.org/10.1103/PhysRevB.90.041105} {\bibfield  {journal} {\bibinfo  {journal} {Phys. Rev. B}\ }\textbf {\bibinfo {volume} {90}},\
  \bibinfo {pages} {041105} (\bibinfo {year} {2014})}\BibitemShut {NoStop}%
\bibitem [{\citenamefont {Seo}\ \emph {et~al.}(2017)\citenamefont {Seo}, \citenamefont {Ahn}, \citenamefont {Song}, \citenamefont {Chen}, \citenamefont {Wilson},\ and\ \citenamefont {Moon}}]{seo_2017}%
  \BibitemOpen
  \bibfield  {author} {\bibinfo {author} {\bibfnamefont {J.~H.}\ \bibnamefont {Seo}}, \bibinfo {author} {\bibfnamefont {G.~H.}\ \bibnamefont {Ahn}}, \bibinfo {author} {\bibfnamefont {S.~J.}\ \bibnamefont {Song}}, \bibinfo {author} {\bibfnamefont {X.}~\bibnamefont {Chen}}, \bibinfo {author} {\bibfnamefont {S.~D.}\ \bibnamefont {Wilson}},\ and\ \bibinfo {author} {\bibfnamefont {S.~J.}\ \bibnamefont {Moon}},\ }\bibfield  {title} {\bibinfo {title} {{Infrared probe of pseudogap in electron-doped $\text{Sr}_2 \text{IrO}_4$}},\ }\href {https://doi.org/10.1038/s41598-017-10725-z} {\bibfield  {journal} {\bibinfo  {journal} {Scientific Reports}\ }\textbf {\bibinfo {volume} {7}},\ \bibinfo {pages} {10494} (\bibinfo {year} {2017})},\ \bibinfo {note} {publisher: Nature Publishing Group}\BibitemShut {NoStop}%
\bibitem [{\citenamefont {Souri}\ \emph {et~al.}(2017)\citenamefont {Souri}, \citenamefont {Kim}, \citenamefont {Gruenewald}, \citenamefont {Connell}, \citenamefont {Thompson}, \citenamefont {Nichols}, \citenamefont {Terzic}, \citenamefont {Min}, \citenamefont {Cao}, \citenamefont {Brill},\ and\ \citenamefont {Seo}}]{Souri_2017}%
  \BibitemOpen
  \bibfield  {author} {\bibinfo {author} {\bibfnamefont {M.}~\bibnamefont {Souri}}, \bibinfo {author} {\bibfnamefont {B.~H.}\ \bibnamefont {Kim}}, \bibinfo {author} {\bibfnamefont {J.~H.}\ \bibnamefont {Gruenewald}}, \bibinfo {author} {\bibfnamefont {J.~G.}\ \bibnamefont {Connell}}, \bibinfo {author} {\bibfnamefont {J.}~\bibnamefont {Thompson}}, \bibinfo {author} {\bibfnamefont {J.}~\bibnamefont {Nichols}}, \bibinfo {author} {\bibfnamefont {J.}~\bibnamefont {Terzic}}, \bibinfo {author} {\bibfnamefont {B.~I.}\ \bibnamefont {Min}}, \bibinfo {author} {\bibfnamefont {G.}~\bibnamefont {Cao}}, \bibinfo {author} {\bibfnamefont {J.~W.}\ \bibnamefont {Brill}},\ and\ \bibinfo {author} {\bibfnamefont {A.}~\bibnamefont {Seo}},\ }\bibfield  {title} {\bibinfo {title} {{Optical signatures of spin-orbit exciton in bandwidth-controlled $\mathrm{S}{\mathrm{r}}_{2}\mathrm{IrO}_{4}$ epitaxial films via high-concentration Ca and Ba doping}},\ }\href {https://doi.org/10.1103/PhysRevB.95.235125} {\bibfield  {journal} {\bibinfo
  {journal} {Phys. Rev. B}\ }\textbf {\bibinfo {volume} {95}},\ \bibinfo {pages} {235125} (\bibinfo {year} {2017})}\BibitemShut {NoStop}%
\bibitem [{\citenamefont {Souri}\ \emph {et~al.}(2016)\citenamefont {Souri}, \citenamefont {Gruenewald}, \citenamefont {Terzic}, \citenamefont {Brill}, \citenamefont {Cao},\ and\ \citenamefont {Seo}}]{Souri_2016}%
  \BibitemOpen
  \bibfield  {author} {\bibinfo {author} {\bibfnamefont {M.}~\bibnamefont {Souri}}, \bibinfo {author} {\bibfnamefont {J.~H.}\ \bibnamefont {Gruenewald}}, \bibinfo {author} {\bibfnamefont {J.}~\bibnamefont {Terzic}}, \bibinfo {author} {\bibfnamefont {J.~W.}\ \bibnamefont {Brill}}, \bibinfo {author} {\bibfnamefont {G.}~\bibnamefont {Cao}},\ and\ \bibinfo {author} {\bibfnamefont {S.~S.~A.}\ \bibnamefont {Seo}},\ }\bibfield  {title} {\bibinfo {title} {{Investigations of metastable $\text{Ca}_2 \text{IrO}_4$ epitaxial thin-films: systematic comparison with $\text{Sr}_2 \text{IrO}_4$ and $\text{Ba}_2 \text{IrO}_4$}},\ }\href {https://doi.org/10.1038/srep25967} {\bibfield  {journal} {\bibinfo  {journal} {Scientific Reports}\ }\textbf {\bibinfo {volume} {6}},\ \bibinfo {pages} {25967} (\bibinfo {year} {2016})},\ \bibinfo {note} {publisher: Nature Publishing Group}\BibitemShut {NoStop}%
\bibitem [{\citenamefont {Moon}\ \emph {et~al.}(2008)\citenamefont {Moon}, \citenamefont {Jin}, \citenamefont {Kim}, \citenamefont {Choi}, \citenamefont {Lee}, \citenamefont {Yu}, \citenamefont {Cao}, \citenamefont {Sumi}, \citenamefont {Funakubo}, \citenamefont {Bernhard},\ and\ \citenamefont {Noh}}]{Moon2008}%
  \BibitemOpen
  \bibfield  {author} {\bibinfo {author} {\bibfnamefont {S.~J.}\ \bibnamefont {Moon}}, \bibinfo {author} {\bibfnamefont {H.}~\bibnamefont {Jin}}, \bibinfo {author} {\bibfnamefont {K.~W.}\ \bibnamefont {Kim}}, \bibinfo {author} {\bibfnamefont {W.~S.}\ \bibnamefont {Choi}}, \bibinfo {author} {\bibfnamefont {Y.~S.}\ \bibnamefont {Lee}}, \bibinfo {author} {\bibfnamefont {J.}~\bibnamefont {Yu}}, \bibinfo {author} {\bibfnamefont {G.}~\bibnamefont {Cao}}, \bibinfo {author} {\bibfnamefont {A.}~\bibnamefont {Sumi}}, \bibinfo {author} {\bibfnamefont {H.}~\bibnamefont {Funakubo}}, \bibinfo {author} {\bibfnamefont {C.}~\bibnamefont {Bernhard}},\ and\ \bibinfo {author} {\bibfnamefont {T.~W.}\ \bibnamefont {Noh}},\ }\bibfield  {title} {\bibinfo {title} {{Dimensionality-Controlled Insulator-Metal Transition and Correlated Metallic State in $5d$ Transition Metal Oxides ${\mathrm{Sr}}_{n+1}{\mathrm{Ir}}_{n}{\mathrm{O}}_{3n+1}$ ($n=1$, 2, and $\ensuremath{\infty}$)}},\ }\href {https://doi.org/10.1103/PhysRevLett.101.226402}
  {\bibfield  {journal} {\bibinfo  {journal} {Phys. Rev. Lett.}\ }\textbf {\bibinfo {volume} {101}},\ \bibinfo {pages} {226402} (\bibinfo {year} {2008})}\BibitemShut {NoStop}%
\bibitem [{\citenamefont {Kim}\ \emph {et~al.}(2022)\citenamefont {Kim}, \citenamefont {Ahn}, \citenamefont {Schmehr}, \citenamefont {Wilson},\ and\ \citenamefont {Moon}}]{Kim_2022}%
  \BibitemOpen
  \bibfield  {author} {\bibinfo {author} {\bibfnamefont {D.}~\bibnamefont {Kim}}, \bibinfo {author} {\bibfnamefont {G.}~\bibnamefont {Ahn}}, \bibinfo {author} {\bibfnamefont {J.}~\bibnamefont {Schmehr}}, \bibinfo {author} {\bibfnamefont {S.~D.}\ \bibnamefont {Wilson}},\ and\ \bibinfo {author} {\bibfnamefont {S.~J.}\ \bibnamefont {Moon}},\ }\bibfield  {title} {\bibinfo {title} {{Effects of the on-site energy on the electronic response of Sr${}_3$(Ir$_{1-x}$Mn$_{x}$)$_2$O$_7$}},\ }\href {https://doi.org/10.1038/s41598-022-23593-z} {\bibfield  {journal} {\bibinfo  {journal} {Scientific Reports}\ }\textbf {\bibinfo {volume} {12}},\ \bibinfo {pages} {18957} (\bibinfo {year} {2022})}\BibitemShut {NoStop}%
\bibitem [{\citenamefont {Okabe}\ \emph {et~al.}(2011)\citenamefont {Okabe}, \citenamefont {Isobe}, \citenamefont {{Takayama-Muromachi}}, \citenamefont {Koda}, \citenamefont {Takeshita}, \citenamefont {Hiraishi}, \citenamefont {Miyazaki}, \citenamefont {Kadono}, \citenamefont {Miyake},\ and\ \citenamefont {Akimitsu}}]{Okabe_2011}%
  \BibitemOpen
  \bibfield  {author} {\bibinfo {author} {\bibfnamefont {H.}~\bibnamefont {Okabe}}, \bibinfo {author} {\bibfnamefont {M.}~\bibnamefont {Isobe}}, \bibinfo {author} {\bibfnamefont {E.}~\bibnamefont {{Takayama-Muromachi}}}, \bibinfo {author} {\bibfnamefont {A.}~\bibnamefont {Koda}}, \bibinfo {author} {\bibfnamefont {S.}~\bibnamefont {Takeshita}}, \bibinfo {author} {\bibfnamefont {M.}~\bibnamefont {Hiraishi}}, \bibinfo {author} {\bibfnamefont {M.}~\bibnamefont {Miyazaki}}, \bibinfo {author} {\bibfnamefont {R.}~\bibnamefont {Kadono}}, \bibinfo {author} {\bibfnamefont {Y.}~\bibnamefont {Miyake}},\ and\ \bibinfo {author} {\bibfnamefont {J.}~\bibnamefont {Akimitsu}},\ }\bibfield  {title} {\bibinfo {title} {{$\text{Ba}_2 \text{IrO}_4$: {{A}} Spin-Orbit {{Mott}} Insulating Quasi-Two-Dimensional Antiferromagnet}},\ }\href {https://doi.org/10.1103/PhysRevB.83.155118} {\bibfield  {journal} {\bibinfo  {journal} {Phys. Rev. B}\ }\textbf {\bibinfo {volume} {83}},\ \bibinfo {pages} {155118} (\bibinfo {year}
  {2011})}\BibitemShut {NoStop}%
\bibitem [{\citenamefont {Okabe}\ \emph {et~al.}(2012)\citenamefont {Okabe}, \citenamefont {Isobe}, \citenamefont {Takayama-Muromachi}, \citenamefont {Koda}, \citenamefont {Takeshita}, \citenamefont {Hiraishi}, \citenamefont {Miyazaki}, \citenamefont {Kadono}, \citenamefont {Miyake},\ and\ \citenamefont {Akimitsu}}]{Okabe_2012}%
  \BibitemOpen
  \bibfield  {author} {\bibinfo {author} {\bibfnamefont {H.}~\bibnamefont {Okabe}}, \bibinfo {author} {\bibfnamefont {M.}~\bibnamefont {Isobe}}, \bibinfo {author} {\bibfnamefont {E.}~\bibnamefont {Takayama-Muromachi}}, \bibinfo {author} {\bibfnamefont {A.}~\bibnamefont {Koda}}, \bibinfo {author} {\bibfnamefont {S.}~\bibnamefont {Takeshita}}, \bibinfo {author} {\bibfnamefont {M.}~\bibnamefont {Hiraishi}}, \bibinfo {author} {\bibfnamefont {M.}~\bibnamefont {Miyazaki}}, \bibinfo {author} {\bibfnamefont {R.}~\bibnamefont {Kadono}}, \bibinfo {author} {\bibfnamefont {Y.}~\bibnamefont {Miyake}},\ and\ \bibinfo {author} {\bibfnamefont {J.}~\bibnamefont {Akimitsu}},\ }\bibfield  {title} {\bibinfo {title} {{Magnetic ordering in spin-orbit Mott insulator $\text{Ba}_2 \text{IrO}_4$ probed by $\mu$SR}},\ }\href {https://dx.doi.org/10.1088/1742-6596/400/3/032071} {\bibfield  {journal} {\bibinfo  {journal} {Journal of Physics: Conference Series}\ }\textbf {\bibinfo {volume} {400}},\ \bibinfo {pages} {032071} (\bibinfo
  {year} {2012})}\BibitemShut {NoStop}%
\bibitem [{\citenamefont {Isobe}\ \emph {et~al.}(2012)\citenamefont {Isobe}, \citenamefont {Okabe}, \citenamefont {Takayama-Muromachi}, \citenamefont {Koda}, \citenamefont {Takeshita}, \citenamefont {Hiraishi}, \citenamefont {Miyazaki}, \citenamefont {Kadono}, \citenamefont {Miyake},\ and\ \citenamefont {Akimitsu}}]{Isobe_2012}%
  \BibitemOpen
  \bibfield  {author} {\bibinfo {author} {\bibfnamefont {M.}~\bibnamefont {Isobe}}, \bibinfo {author} {\bibfnamefont {H.}~\bibnamefont {Okabe}}, \bibinfo {author} {\bibfnamefont {E.}~\bibnamefont {Takayama-Muromachi}}, \bibinfo {author} {\bibfnamefont {A.}~\bibnamefont {Koda}}, \bibinfo {author} {\bibfnamefont {S.}~\bibnamefont {Takeshita}}, \bibinfo {author} {\bibfnamefont {M.}~\bibnamefont {Hiraishi}}, \bibinfo {author} {\bibfnamefont {M.}~\bibnamefont {Miyazaki}}, \bibinfo {author} {\bibfnamefont {R.}~\bibnamefont {Kadono}}, \bibinfo {author} {\bibfnamefont {Y.}~\bibnamefont {Miyake}},\ and\ \bibinfo {author} {\bibfnamefont {J.}~\bibnamefont {Akimitsu}},\ }\bibfield  {title} {\bibinfo {title} {{Spin-Orbit Mott State in the Novel Quasi-2D Antiferromagnet $\text{Ba}_2 \text{IrO}_4$}},\ }\href {https://doi.org/10.1088/1742-6596/400/3/032028} {\bibfield  {journal} {\bibinfo  {journal} {Journal of Physics: Conference Series}\ }\textbf {\bibinfo {volume} {400}},\ \bibinfo {pages} {032028} (\bibinfo {year}
  {2012})}\BibitemShut {NoStop}%
\bibitem [{\citenamefont {Moser}\ \emph {et~al.}(2014)\citenamefont {Moser}, \citenamefont {Moreschini}, \citenamefont {Ebrahimi}, \citenamefont {Piazza}, \citenamefont {Isobe}, \citenamefont {Okabe}, \citenamefont {Akimitsu}, \citenamefont {Mazurenko}, \citenamefont {Kim}, \citenamefont {Bostwick}, \citenamefont {Rotenberg}, \citenamefont {Chang}, \citenamefont {R{\o}nnow},\ and\ \citenamefont {Grioni}}]{Moser_2014}%
  \BibitemOpen
  \bibfield  {author} {\bibinfo {author} {\bibfnamefont {S.}~\bibnamefont {Moser}}, \bibinfo {author} {\bibfnamefont {L.}~\bibnamefont {Moreschini}}, \bibinfo {author} {\bibfnamefont {A.}~\bibnamefont {Ebrahimi}}, \bibinfo {author} {\bibfnamefont {B.~D.}\ \bibnamefont {Piazza}}, \bibinfo {author} {\bibfnamefont {M.}~\bibnamefont {Isobe}}, \bibinfo {author} {\bibfnamefont {H.}~\bibnamefont {Okabe}}, \bibinfo {author} {\bibfnamefont {J.}~\bibnamefont {Akimitsu}}, \bibinfo {author} {\bibfnamefont {V.~V.}\ \bibnamefont {Mazurenko}}, \bibinfo {author} {\bibfnamefont {K.~S.}\ \bibnamefont {Kim}}, \bibinfo {author} {\bibfnamefont {A.}~\bibnamefont {Bostwick}}, \bibinfo {author} {\bibfnamefont {E.}~\bibnamefont {Rotenberg}}, \bibinfo {author} {\bibfnamefont {J.}~\bibnamefont {Chang}}, \bibinfo {author} {\bibfnamefont {H.~M.}\ \bibnamefont {R{\o}nnow}},\ and\ \bibinfo {author} {\bibfnamefont {M.}~\bibnamefont {Grioni}},\ }\bibfield  {title} {\bibinfo {title} {{The electronic structure of the high-symmetry perovskite
  iridate $\text{Ba}_2\text{IrO}_4$}},\ }\href {https://iopscience.iop.org/article/10.1088/1367-2630/16/1/013008} {\bibfield  {journal} {\bibinfo  {journal} {New Journal of Physics}\ }\textbf {\bibinfo {volume} {16}},\ \bibinfo {pages} {013008} (\bibinfo {year} {2014})}\BibitemShut {NoStop}%
\bibitem [{\citenamefont {Hou}\ \emph {et~al.}(2016)\citenamefont {Hou}, \citenamefont {Xiang},\ and\ \citenamefont {Gong}}]{Hou_2016}%
  \BibitemOpen
  \bibfield  {author} {\bibinfo {author} {\bibfnamefont {Y.~S.}\ \bibnamefont {Hou}}, \bibinfo {author} {\bibfnamefont {H.~J.}\ \bibnamefont {Xiang}},\ and\ \bibinfo {author} {\bibfnamefont {X.~G.}\ \bibnamefont {Gong}},\ }\bibfield  {title} {\bibinfo {title} {{Unveiling the origin of the basal-plane antiferromagnetism in the spin–orbit Mott insulator $\text{Ba}_2 \text{IrO}_4$: a density functional and model Hamiltonian study}},\ }\href {https://doi.org/10.1088/1367-2630/18/4/043007} {\bibfield  {journal} {\bibinfo  {journal} {New Journal of Physics}\ }\textbf {\bibinfo {volume} {18}},\ \bibinfo {pages} {043007} (\bibinfo {year} {2016})}\BibitemShut {NoStop}%
\bibitem [{\citenamefont {Katukuri}\ \emph {et~al.}(2014)\citenamefont {Katukuri}, \citenamefont {Yushankhai}, \citenamefont {Siurakshina}, \citenamefont {{van den Brink}}, \citenamefont {Hozoi},\ and\ \citenamefont {Rousochatzakis}}]{Katukuri_2014}%
  \BibitemOpen
  \bibfield  {author} {\bibinfo {author} {\bibfnamefont {V.~M.}\ \bibnamefont {Katukuri}}, \bibinfo {author} {\bibfnamefont {V.}~\bibnamefont {Yushankhai}}, \bibinfo {author} {\bibfnamefont {L.}~\bibnamefont {Siurakshina}}, \bibinfo {author} {\bibfnamefont {J.}~\bibnamefont {{van den Brink}}}, \bibinfo {author} {\bibfnamefont {L.}~\bibnamefont {Hozoi}},\ and\ \bibinfo {author} {\bibfnamefont {I.}~\bibnamefont {Rousochatzakis}},\ }\bibfield  {title} {\bibinfo {title} {{Mechanism of Basal-Plane Antiferromagnetism in the Spin-Orbit Driven Iridate $\text{Ba}_2 \text{IrO}_4$}},\ }\href {https://doi.org/10.1103/PhysRevX.4.021051} {\bibfield  {journal} {\bibinfo  {journal} {Phys Rev X}\ }\textbf {\bibinfo {volume} {4}},\ \bibinfo {pages} {021051} (\bibinfo {year} {2014})}\BibitemShut {NoStop}%
\bibitem [{\citenamefont {Wang}\ \emph {et~al.}(2015)\citenamefont {Wang}, \citenamefont {Wohlfeld}, \citenamefont {Moritz}, \citenamefont {Jia}, \citenamefont {van Veenendaal}, \citenamefont {Wu}, \citenamefont {Chen},\ and\ \citenamefont {Devereaux}}]{Wang_2015}%
  \BibitemOpen
  \bibfield  {author} {\bibinfo {author} {\bibfnamefont {Y.}~\bibnamefont {Wang}}, \bibinfo {author} {\bibfnamefont {K.}~\bibnamefont {Wohlfeld}}, \bibinfo {author} {\bibfnamefont {B.}~\bibnamefont {Moritz}}, \bibinfo {author} {\bibfnamefont {C.~J.}\ \bibnamefont {Jia}}, \bibinfo {author} {\bibfnamefont {M.}~\bibnamefont {van Veenendaal}}, \bibinfo {author} {\bibfnamefont {K.}~\bibnamefont {Wu}}, \bibinfo {author} {\bibfnamefont {C.-C.}\ \bibnamefont {Chen}},\ and\ \bibinfo {author} {\bibfnamefont {T.~P.}\ \bibnamefont {Devereaux}},\ }\bibfield  {title} {\bibinfo {title} {{Origin of strong dispersion in Hubbard insulators}},\ }\href {https://doi.org/10.1103/PhysRevB.92.075119} {\bibfield  {journal} {\bibinfo  {journal} {Phys. Rev. B}\ }\textbf {\bibinfo {volume} {92}},\ \bibinfo {pages} {075119} (\bibinfo {year} {2015})}\BibitemShut {NoStop}%
\bibitem [{\citenamefont {Bacq-Labreuil}\ \emph {et~al.}(2025)\citenamefont {Bacq-Labreuil}, \citenamefont {Fawaz}, \citenamefont {Okazaki}, \citenamefont {Obata}, \citenamefont {Cercellier}, \citenamefont {Le~F\`evre}, \citenamefont {Bertran}, \citenamefont {Santos-Cottin}, \citenamefont {Yamamoto}, \citenamefont {Yamada}, \citenamefont {Azuma}, \citenamefont {Horiba}, \citenamefont {Kumigashira}, \citenamefont {d'Astuto}, \citenamefont {Biermann},\ and\ \citenamefont {Lenz}}]{Bacq_2025}%
  \BibitemOpen
  \bibfield  {author} {\bibinfo {author} {\bibfnamefont {B.}~\bibnamefont {Bacq-Labreuil}}, \bibinfo {author} {\bibfnamefont {C.}~\bibnamefont {Fawaz}}, \bibinfo {author} {\bibfnamefont {Y.}~\bibnamefont {Okazaki}}, \bibinfo {author} {\bibfnamefont {Y.}~\bibnamefont {Obata}}, \bibinfo {author} {\bibfnamefont {H.}~\bibnamefont {Cercellier}}, \bibinfo {author} {\bibfnamefont {P.}~\bibnamefont {Le~F\`evre}}, \bibinfo {author} {\bibfnamefont {F.~m.~c.}\ \bibnamefont {Bertran}}, \bibinfo {author} {\bibfnamefont {D.}~\bibnamefont {Santos-Cottin}}, \bibinfo {author} {\bibfnamefont {H.}~\bibnamefont {Yamamoto}}, \bibinfo {author} {\bibfnamefont {I.}~\bibnamefont {Yamada}}, \bibinfo {author} {\bibfnamefont {M.}~\bibnamefont {Azuma}}, \bibinfo {author} {\bibfnamefont {K.}~\bibnamefont {Horiba}}, \bibinfo {author} {\bibfnamefont {H.}~\bibnamefont {Kumigashira}}, \bibinfo {author} {\bibfnamefont {M.}~\bibnamefont {d'Astuto}}, \bibinfo {author} {\bibfnamefont {S.}~\bibnamefont {Biermann}},\ and\ \bibinfo {author}
  {\bibfnamefont {B.}~\bibnamefont {Lenz}},\ }\bibfield  {title} {\bibinfo {title} {{Universal Waterfall Feature in Cuprate Superconductors: Evidence of a Momentum-Driven Crossover}},\ }\href {https://doi.org/10.1103/PhysRevLett.134.016502} {\bibfield  {journal} {\bibinfo  {journal} {Phys. Rev. Lett.}\ }\textbf {\bibinfo {volume} {134}},\ \bibinfo {pages} {016502} (\bibinfo {year} {2025})}\BibitemShut {NoStop}%
\bibitem [{\citenamefont {de~la Torre}\ \emph {et~al.}(2015)\citenamefont {de~la Torre}, \citenamefont {McKeown~Walker}, \citenamefont {Bruno}, \citenamefont {Ricc\'o}, \citenamefont {Wang}, \citenamefont {Gutierrez~Lezama}, \citenamefont {Scheerer}, \citenamefont {Giriat}, \citenamefont {Jaccard}, \citenamefont {Berthod}, \citenamefont {Kim}, \citenamefont {Hoesch}, \citenamefont {Hunter}, \citenamefont {Perry}, \citenamefont {Tamai},\ and\ \citenamefont {Baumberger}}]{DelaTorre_2015}%
  \BibitemOpen
  \bibfield  {author} {\bibinfo {author} {\bibfnamefont {A.}~\bibnamefont {de~la Torre}}, \bibinfo {author} {\bibfnamefont {S.}~\bibnamefont {McKeown~Walker}}, \bibinfo {author} {\bibfnamefont {F.~Y.}\ \bibnamefont {Bruno}}, \bibinfo {author} {\bibfnamefont {S.}~\bibnamefont {Ricc\'o}}, \bibinfo {author} {\bibfnamefont {Z.}~\bibnamefont {Wang}}, \bibinfo {author} {\bibfnamefont {I.}~\bibnamefont {Gutierrez~Lezama}}, \bibinfo {author} {\bibfnamefont {G.}~\bibnamefont {Scheerer}}, \bibinfo {author} {\bibfnamefont {G.}~\bibnamefont {Giriat}}, \bibinfo {author} {\bibfnamefont {D.}~\bibnamefont {Jaccard}}, \bibinfo {author} {\bibfnamefont {C.}~\bibnamefont {Berthod}}, \bibinfo {author} {\bibfnamefont {T.~K.}\ \bibnamefont {Kim}}, \bibinfo {author} {\bibfnamefont {M.}~\bibnamefont {Hoesch}}, \bibinfo {author} {\bibfnamefont {E.~C.}\ \bibnamefont {Hunter}}, \bibinfo {author} {\bibfnamefont {R.~S.}\ \bibnamefont {Perry}}, \bibinfo {author} {\bibfnamefont {A.}~\bibnamefont {Tamai}},\ and\ \bibinfo {author}
  {\bibfnamefont {F.}~\bibnamefont {Baumberger}},\ }\bibfield  {title} {\bibinfo {title} {{Collapse of the Mott Gap and Emergence of a Nodal Liquid in Lightly Doped $\text{Ba}_2 \text{IrO}_4$}},\ }\href {https://doi.org/10.1103/PhysRevLett.115.176402} {\bibfield  {journal} {\bibinfo  {journal} {Phys. Rev. Lett.}\ }\textbf {\bibinfo {volume} {115}},\ \bibinfo {pages} {176402} (\bibinfo {year} {2015})}\BibitemShut {NoStop}%
\bibitem [{\citenamefont {Kotliar}\ \emph {et~al.}(2006)\citenamefont {Kotliar}, \citenamefont {Savrasov}, \citenamefont {Haule}, \citenamefont {Oudovenko}, \citenamefont {Parcollet},\ and\ \citenamefont {Marianetti}}]{Kotliar_2006}%
  \BibitemOpen
  \bibfield  {author} {\bibinfo {author} {\bibfnamefont {G.}~\bibnamefont {Kotliar}}, \bibinfo {author} {\bibfnamefont {S.~Y.}\ \bibnamefont {Savrasov}}, \bibinfo {author} {\bibfnamefont {K.}~\bibnamefont {Haule}}, \bibinfo {author} {\bibfnamefont {V.~S.}\ \bibnamefont {Oudovenko}}, \bibinfo {author} {\bibfnamefont {O.}~\bibnamefont {Parcollet}},\ and\ \bibinfo {author} {\bibfnamefont {C.~A.}\ \bibnamefont {Marianetti}},\ }\bibfield  {title} {\bibinfo {title} {{Electronic structure calculations with dynamical mean-field theory}},\ }\href {https://doi.org/10.1103/RevModPhys.78.865} {\bibfield  {journal} {\bibinfo  {journal} {Rev. Mod. Phys.}\ }\textbf {\bibinfo {volume} {78}},\ \bibinfo {pages} {865} (\bibinfo {year} {2006})}\BibitemShut {NoStop}%
\bibitem [{\citenamefont {Cassol}\ \emph {et~al.}(2024)\citenamefont {Cassol}, \citenamefont {Gaspard}, \citenamefont {Casula}, \citenamefont {Martins},\ and\ \citenamefont {Lenz}}]{Cassol_2024}%
  \BibitemOpen
  \bibfield  {author} {\bibinfo {author} {\bibfnamefont {F.}~\bibnamefont {Cassol}}, \bibinfo {author} {\bibfnamefont {L.}~\bibnamefont {Gaspard}}, \bibinfo {author} {\bibfnamefont {M.}~\bibnamefont {Casula}}, \bibinfo {author} {\bibfnamefont {C.}~\bibnamefont {Martins}},\ and\ \bibinfo {author} {\bibfnamefont {B.}~\bibnamefont {Lenz}},\ }\bibfield  {title} {\bibinfo {title} {{Rich phase diagram of the prototypical iridate $\text{Ba}_2 \text{IrO}_4$: Effective low-energy models and metal-insulator transition}},\ }\href {https://doi.org/10.1103/PhysRevB.109.155120} {\bibfield  {journal} {\bibinfo  {journal} {Phys. Rev. B}\ }\textbf {\bibinfo {volume} {109}},\ \bibinfo {pages} {155120} (\bibinfo {year} {2024})}\BibitemShut {NoStop}%
\bibitem [{SM()}]{SM}%
  \BibitemOpen
  \bibinfo {title} {{See Supplemental Material for more details on the construction of the effective low-energy models, the role of the Ir-$e_g$ bands, details of the calculation of the optical conductivity and absorption as well as numerical details of the DMFT and SCBA simulations, which includes Refs. \cite{QuantumEspresso1,QuantumEspresso2,PBE1996,ONCVPSP2013,PseudoDojo2018,MLWF,wannier90,TRIQSDFTTOOLS2016,MQEM2018,RESPACK,Ye_2013, Qiangqiang2023,Klett2020,Schaefer2021,Werner2010,Pauli2025,propper_2016}.}}\BibitemShut {Stop}%
\bibitem [{\citenamefont {Zhang}\ and\ \citenamefont {Pavarini}(2021)}]{Pavarini_2021}%
  \BibitemOpen
\bibfield  {title} {  }\bibfield  {author} {\bibinfo {author} {\bibfnamefont {G.}~\bibnamefont {Zhang}}\ and\ \bibinfo {author} {\bibfnamefont {E.}~\bibnamefont {Pavarini}},\ }\bibfield  {title} {\bibinfo {title} {{Magnetic superexchange couplings in ${\mathrm{Sr}}_{2}{\mathrm{IrO}}_{4}$}},\ }\href {https://doi.org/10.1103/PhysRevB.104.125116} {\bibfield  {journal} {\bibinfo  {journal} {Phys. Rev. B}\ }\textbf {\bibinfo {volume} {104}},\ \bibinfo {pages} {125116} (\bibinfo {year} {2021})}\BibitemShut {NoStop}%
\bibitem [{\citenamefont {Georges}\ \emph {et~al.}(2013)\citenamefont {Georges}, \citenamefont {Medici},\ and\ \citenamefont {Mravlje}}]{Georges_2013}%
  \BibitemOpen
  \bibfield  {author} {\bibinfo {author} {\bibfnamefont {A.}~\bibnamefont {Georges}}, \bibinfo {author} {\bibfnamefont {L.~d.}\ \bibnamefont {Medici}},\ and\ \bibinfo {author} {\bibfnamefont {J.}~\bibnamefont {Mravlje}},\ }\bibfield  {title} {\bibinfo {title} {{Strong Correlations from Hund’s Coupling}},\ }\href {https://doi.org/10.1146/annurev-conmatphys-020911-125045} {\bibfield  {journal} {\bibinfo  {journal} {Annual Review of Condensed Matter Physics}\ }\textbf {\bibinfo {volume} {4}},\ \bibinfo {pages} {137} (\bibinfo {year} {2013})}\BibitemShut {NoStop}%
\bibitem [{\citenamefont {Aryasetiawan}\ \emph {et~al.}(2004)\citenamefont {Aryasetiawan}, \citenamefont {Imada}, \citenamefont {Georges}, \citenamefont {Kotliar}, \citenamefont {Biermann},\ and\ \citenamefont {Lichtenstein}}]{cRPA}%
  \BibitemOpen
  \bibfield  {author} {\bibinfo {author} {\bibfnamefont {F.}~\bibnamefont {Aryasetiawan}}, \bibinfo {author} {\bibfnamefont {M.}~\bibnamefont {Imada}}, \bibinfo {author} {\bibfnamefont {A.}~\bibnamefont {Georges}}, \bibinfo {author} {\bibfnamefont {G.}~\bibnamefont {Kotliar}}, \bibinfo {author} {\bibfnamefont {S.}~\bibnamefont {Biermann}},\ and\ \bibinfo {author} {\bibfnamefont {A.~I.}\ \bibnamefont {Lichtenstein}},\ }\bibfield  {title} {\bibinfo {title} {{Frequency-dependent local interactions and low-energy effective models from electronic structure calculations}},\ }\href {https://link.aps.org/doi/10.1103/PhysRevB.70.195104} {\bibfield  {journal} {\bibinfo  {journal} {Phys. Rev. B}\ }\textbf {\bibinfo {volume} {70}},\ \bibinfo {pages} {195104} (\bibinfo {year} {2004})}\BibitemShut {NoStop}%
\bibitem [{\citenamefont {Shinaoka}\ \emph {et~al.}(2015)\citenamefont {Shinaoka}, \citenamefont {Troyer},\ and\ \citenamefont {Werner}}]{cRPA2}%
  \BibitemOpen
  \bibfield  {author} {\bibinfo {author} {\bibfnamefont {H.}~\bibnamefont {Shinaoka}}, \bibinfo {author} {\bibfnamefont {M.}~\bibnamefont {Troyer}},\ and\ \bibinfo {author} {\bibfnamefont {P.}~\bibnamefont {Werner}},\ }\bibfield  {title} {\bibinfo {title} {{Accuracy of downfolding based on the constrained random-phase approximation}},\ }\href {https://doi.org/10.1103/PhysRevB.91.245156} {\bibfield  {journal} {\bibinfo  {journal} {Phys. Rev. B}\ }\textbf {\bibinfo {volume} {91}},\ \bibinfo {pages} {245156} (\bibinfo {year} {2015})}\BibitemShut {NoStop}%
\bibitem [{\citenamefont {Martins}\ \emph {et~al.}(2011)\citenamefont {Martins}, \citenamefont {Aichhorn}, \citenamefont {Vaugier},\ and\ \citenamefont {Biermann}}]{Martins_2011}%
  \BibitemOpen
  \bibfield  {author} {\bibinfo {author} {\bibfnamefont {C.}~\bibnamefont {Martins}}, \bibinfo {author} {\bibfnamefont {M.}~\bibnamefont {Aichhorn}}, \bibinfo {author} {\bibfnamefont {L.}~\bibnamefont {Vaugier}},\ and\ \bibinfo {author} {\bibfnamefont {S.}~\bibnamefont {Biermann}},\ }\bibfield  {title} {\bibinfo {title} {{Reduced Effective Spin-Orbital Degeneracy and Spin-Orbital Ordering in Paramagnetic Transition-Metal Oxides: ${\mathrm{Sr}}_{2}{\mathrm{IrO}}_{4}$ versus $\text{Sr}_2 \text{IrO}_4$}},\ }\href {https://doi.org/10.1103/PhysRevLett.107.266404} {\bibfield  {journal} {\bibinfo  {journal} {Phys. Rev. Lett.}\ }\textbf {\bibinfo {volume} {107}},\ \bibinfo {pages} {266404} (\bibinfo {year} {2011})}\BibitemShut {NoStop}%
\bibitem [{\citenamefont {Arita}\ \emph {et~al.}(2012)\citenamefont {Arita}, \citenamefont {Kune\ifmmode~\check{s}\else \v{s}\fi{}}, \citenamefont {Kozhevnikov}, \citenamefont {Eguiluz},\ and\ \citenamefont {Imada}}]{Arita_2012}%
  \BibitemOpen
  \bibfield  {author} {\bibinfo {author} {\bibfnamefont {R.}~\bibnamefont {Arita}}, \bibinfo {author} {\bibfnamefont {J.}~\bibnamefont {Kune\ifmmode~\check{s}\else \v{s}\fi{}}}, \bibinfo {author} {\bibfnamefont {A.~V.}\ \bibnamefont {Kozhevnikov}}, \bibinfo {author} {\bibfnamefont {A.~G.}\ \bibnamefont {Eguiluz}},\ and\ \bibinfo {author} {\bibfnamefont {M.}~\bibnamefont {Imada}},\ }\bibfield  {title} {\bibinfo {title} {{Ab initio Studies on the Interplay between Spin-Orbit Interaction and Coulomb Correlation in $\text{Sr}_2 \text{IrO}$ and $\text{Ba}_2 \text{IrO}_4$}},\ }\href {https://doi.org/10.1103/PhysRevLett.108.086403} {\bibfield  {journal} {\bibinfo  {journal} {Phys. Rev. Lett.}\ }\textbf {\bibinfo {volume} {108}},\ \bibinfo {pages} {086403} (\bibinfo {year} {2012})}\BibitemShut {NoStop}%
\bibitem [{\citenamefont {Lenz}\ \emph {et~al.}(2019)\citenamefont {Lenz}, \citenamefont {Martins},\ and\ \citenamefont {Biermann}}]{Lenz_2019}%
  \BibitemOpen
  \bibfield  {author} {\bibinfo {author} {\bibfnamefont {B.}~\bibnamefont {Lenz}}, \bibinfo {author} {\bibfnamefont {C.}~\bibnamefont {Martins}},\ and\ \bibinfo {author} {\bibfnamefont {S.}~\bibnamefont {Biermann}},\ }\bibfield  {title} {\bibinfo {title} {{Spectral functions of $\text{Sr}_2 \text{IrO}_4$: theory versus experiment}},\ }\href {https://doi.org/10.1088/1361-648X/ab146a} {\bibfield  {journal} {\bibinfo  {journal} {Journal of Physics: Condensed Matter}\ }\textbf {\bibinfo {volume} {31}},\ \bibinfo {pages} {293001} (\bibinfo {year} {2019})}\BibitemShut {NoStop}%
\bibitem [{\citenamefont {Zhang}\ and\ \citenamefont {Pavarini}(2023)}]{Pavarini_2023}%
  \BibitemOpen
  \bibfield  {author} {\bibinfo {author} {\bibfnamefont {G.}~\bibnamefont {Zhang}}\ and\ \bibinfo {author} {\bibfnamefont {E.}~\bibnamefont {Pavarini}},\ }\bibfield  {title} {\bibinfo {title} {{Multiorbital Nature of Doped $\text{Sr}_2 \text{IrO}_4$}},\ }\href {https://doi.org/10.1103/PhysRevLett.131.036504} {\bibfield  {journal} {\bibinfo  {journal} {Phys. Rev. Lett.}\ }\textbf {\bibinfo {volume} {131}},\ \bibinfo {pages} {036504} (\bibinfo {year} {2023})}\BibitemShut {NoStop}%
\bibitem [{\citenamefont {Choi}\ \emph {et~al.}(2024)\citenamefont {Choi}, \citenamefont {Yue}, \citenamefont {Azoury}, \citenamefont {Porter}, \citenamefont {Chen}, \citenamefont {Petocchi}, \citenamefont {Baldini}, \citenamefont {Lv}, \citenamefont {Mogi}, \citenamefont {Su}, \citenamefont {Wilson}, \citenamefont {Eckstein}, \citenamefont {Werner},\ and\ \citenamefont {Gedik}}]{Choi_2024}%
  \BibitemOpen
  \bibfield  {author} {\bibinfo {author} {\bibfnamefont {D.}~\bibnamefont {Choi}}, \bibinfo {author} {\bibfnamefont {C.}~\bibnamefont {Yue}}, \bibinfo {author} {\bibfnamefont {D.}~\bibnamefont {Azoury}}, \bibinfo {author} {\bibfnamefont {Z.}~\bibnamefont {Porter}}, \bibinfo {author} {\bibfnamefont {J.}~\bibnamefont {Chen}}, \bibinfo {author} {\bibfnamefont {F.}~\bibnamefont {Petocchi}}, \bibinfo {author} {\bibfnamefont {E.}~\bibnamefont {Baldini}}, \bibinfo {author} {\bibfnamefont {B.}~\bibnamefont {Lv}}, \bibinfo {author} {\bibfnamefont {M.}~\bibnamefont {Mogi}}, \bibinfo {author} {\bibfnamefont {Y.}~\bibnamefont {Su}}, \bibinfo {author} {\bibfnamefont {S.~D.}\ \bibnamefont {Wilson}}, \bibinfo {author} {\bibfnamefont {M.}~\bibnamefont {Eckstein}}, \bibinfo {author} {\bibfnamefont {P.}~\bibnamefont {Werner}},\ and\ \bibinfo {author} {\bibfnamefont {N.}~\bibnamefont {Gedik}},\ }\bibfield  {title} {\bibinfo {title} {{Light-induced insulator-metal transition in Sr$_2$IrO$_4$ reveals the nature of the insulating
  ground state}},\ }\href {https://doi.org/10.1073/pnas.2323013121} {\bibfield  {journal} {\bibinfo  {journal} {Proceedings of the National Academy of Sciences}\ }\textbf {\bibinfo {volume} {121}},\ \bibinfo {pages} {e2323013121} (\bibinfo {year} {2024})},\ \Eprint {https://arxiv.org/abs/https://www.pnas.org/doi/pdf/10.1073/pnas.2323013121} {https://www.pnas.org/doi/pdf/10.1073/pnas.2323013121} \BibitemShut {NoStop}%
\bibitem [{\citenamefont {Gull}\ \emph {et~al.}(2011)\citenamefont {Gull}, \citenamefont {Millis}, \citenamefont {Lichtenstein}, \citenamefont {Rubtsov}, \citenamefont {Troyer},\ and\ \citenamefont {Werner}}]{CTQMC}%
  \BibitemOpen
  \bibfield  {author} {\bibinfo {author} {\bibfnamefont {E.}~\bibnamefont {Gull}}, \bibinfo {author} {\bibfnamefont {A.~J.}\ \bibnamefont {Millis}}, \bibinfo {author} {\bibfnamefont {A.~I.}\ \bibnamefont {Lichtenstein}}, \bibinfo {author} {\bibfnamefont {A.~N.}\ \bibnamefont {Rubtsov}}, \bibinfo {author} {\bibfnamefont {M.}~\bibnamefont {Troyer}},\ and\ \bibinfo {author} {\bibfnamefont {P.}~\bibnamefont {Werner}},\ }\bibfield  {title} {\bibinfo {title} {{Continuous-time Monte Carlo methods for quantum impurity models}},\ }\href {https://doi.org/10.1103/RevModPhys.83.349} {\bibfield  {journal} {\bibinfo  {journal} {Rev. Mod. Phys.}\ }\textbf {\bibinfo {volume} {83}},\ \bibinfo {pages} {349} (\bibinfo {year} {2011})}\BibitemShut {NoStop}%
\bibitem [{\citenamefont {Seth}\ \emph {et~al.}(2016)\citenamefont {Seth}, \citenamefont {Krivenko}, \citenamefont {Ferrero},\ and\ \citenamefont {Parcollet}}]{TRIQSCTHYB2016}%
  \BibitemOpen
  \bibfield  {author} {\bibinfo {author} {\bibfnamefont {P.}~\bibnamefont {Seth}}, \bibinfo {author} {\bibfnamefont {I.}~\bibnamefont {Krivenko}}, \bibinfo {author} {\bibfnamefont {M.}~\bibnamefont {Ferrero}},\ and\ \bibinfo {author} {\bibfnamefont {O.}~\bibnamefont {Parcollet}},\ }\bibfield  {title} {\bibinfo {title} {{{{TRIQS}}/{{CTHYB}}: {{A}} Continuous-Time Quantum {{Monte Carlo}} Hybridisation Expansion Solver for Quantum Impurity Problems}},\ }\href {https://doi.org/10.1016/j.cpc.2015.10.023} {\bibfield  {journal} {\bibinfo  {journal} {Computer Physics Communications}\ }\textbf {\bibinfo {volume} {200}},\ \bibinfo {pages} {274} (\bibinfo {year} {2016})}\BibitemShut {NoStop}%
\bibitem [{\citenamefont {Parcollet}\ \emph {et~al.}(2015)\citenamefont {Parcollet}, \citenamefont {Ferrero}, \citenamefont {Ayral}, \citenamefont {Hafermann}, \citenamefont {Krivenko}, \citenamefont {Messio},\ and\ \citenamefont {Seth}}]{TRIQS2015}%
  \BibitemOpen
  \bibfield  {author} {\bibinfo {author} {\bibfnamefont {O.}~\bibnamefont {Parcollet}}, \bibinfo {author} {\bibfnamefont {M.}~\bibnamefont {Ferrero}}, \bibinfo {author} {\bibfnamefont {T.}~\bibnamefont {Ayral}}, \bibinfo {author} {\bibfnamefont {H.}~\bibnamefont {Hafermann}}, \bibinfo {author} {\bibfnamefont {I.}~\bibnamefont {Krivenko}}, \bibinfo {author} {\bibfnamefont {L.}~\bibnamefont {Messio}},\ and\ \bibinfo {author} {\bibfnamefont {P.}~\bibnamefont {Seth}},\ }\bibfield  {title} {\bibinfo {title} {{{{TRIQS}}: {{A}} Toolbox for Research on Interacting Quantum Systems}},\ }\href {https://doi.org/https://doi.org/10.1016/j.cpc.2015.04.023} {\bibfield  {journal} {\bibinfo  {journal} {Computer Physics Communications}\ }\textbf {\bibinfo {volume} {196}},\ \bibinfo {pages} {398} (\bibinfo {year} {2015})}\BibitemShut {NoStop}%
\bibitem [{\citenamefont {Sangiovanni}\ \emph {et~al.}(2006)\citenamefont {Sangiovanni}, \citenamefont {Toschi}, \citenamefont {Koch}, \citenamefont {Held}, \citenamefont {Capone}, \citenamefont {Castellani}, \citenamefont {Gunnarsson}, \citenamefont {Mo}, \citenamefont {Allen}, \citenamefont {Kim}, \citenamefont {Sekiyama}, \citenamefont {Yamasaki}, \citenamefont {Suga},\ and\ \citenamefont {Metcalf}}]{Sangiovanni_2006}%
  \BibitemOpen
  \bibfield  {author} {\bibinfo {author} {\bibfnamefont {G.}~\bibnamefont {Sangiovanni}}, \bibinfo {author} {\bibfnamefont {A.}~\bibnamefont {Toschi}}, \bibinfo {author} {\bibfnamefont {E.}~\bibnamefont {Koch}}, \bibinfo {author} {\bibfnamefont {K.}~\bibnamefont {Held}}, \bibinfo {author} {\bibfnamefont {M.}~\bibnamefont {Capone}}, \bibinfo {author} {\bibfnamefont {C.}~\bibnamefont {Castellani}}, \bibinfo {author} {\bibfnamefont {O.}~\bibnamefont {Gunnarsson}}, \bibinfo {author} {\bibfnamefont {S.-K.}\ \bibnamefont {Mo}}, \bibinfo {author} {\bibfnamefont {J.~W.}\ \bibnamefont {Allen}}, \bibinfo {author} {\bibfnamefont {H.-D.}\ \bibnamefont {Kim}}, \bibinfo {author} {\bibfnamefont {A.}~\bibnamefont {Sekiyama}}, \bibinfo {author} {\bibfnamefont {A.}~\bibnamefont {Yamasaki}}, \bibinfo {author} {\bibfnamefont {S.}~\bibnamefont {Suga}},\ and\ \bibinfo {author} {\bibfnamefont {P.}~\bibnamefont {Metcalf}},\ }\bibfield  {title} {\bibinfo {title} {{Static versus dynamical mean-field theory of Mott antiferromagnets}},\
  }\href {https://doi.org/10.1103/PhysRevB.73.205121} {\bibfield  {journal} {\bibinfo  {journal} {Phys. Rev. B}\ }\textbf {\bibinfo {volume} {73}},\ \bibinfo {pages} {205121} (\bibinfo {year} {2006})}\BibitemShut {NoStop}%
\bibitem [{\citenamefont {Khurana}(1990)}]{Khurana_1990}%
  \BibitemOpen
  \bibfield  {author} {\bibinfo {author} {\bibfnamefont {A.}~\bibnamefont {Khurana}},\ }\bibfield  {title} {\bibinfo {title} {{Electrical conductivity in the infinite-dimensional Hubbard model}},\ }\href {https://doi.org/10.1103/PhysRevLett.64.1990} {\bibfield  {journal} {\bibinfo  {journal} {Phys. Rev. Lett.}\ }\textbf {\bibinfo {volume} {64}},\ \bibinfo {pages} {1990} (\bibinfo {year} {1990})}\BibitemShut {NoStop}%
\bibitem [{\citenamefont {Tomczak}\ and\ \citenamefont {Biermann}(2009)}]{Tomczak_2009}%
  \BibitemOpen
  \bibfield  {author} {\bibinfo {author} {\bibfnamefont {J.~M.}\ \bibnamefont {Tomczak}}\ and\ \bibinfo {author} {\bibfnamefont {S.}~\bibnamefont {Biermann}},\ }\bibfield  {title} {\bibinfo {title} {{Optical properties of correlated materials: Generalized Peierls approach and its application to $\text{V}\text{O}_2$}},\ }\href {https://doi.org/10.1103/PhysRevB.80.085117} {\bibfield  {journal} {\bibinfo  {journal} {Phys. Rev. B}\ }\textbf {\bibinfo {volume} {80}},\ \bibinfo {pages} {085117} (\bibinfo {year} {2009})}\BibitemShut {NoStop}%
\bibitem [{\citenamefont {Martinez}\ and\ \citenamefont {Horsch}(1991)}]{Spin_polaron_t_J}%
  \BibitemOpen
  \bibfield  {author} {\bibinfo {author} {\bibfnamefont {G.}~\bibnamefont {Martinez}}\ and\ \bibinfo {author} {\bibfnamefont {P.}~\bibnamefont {Horsch}},\ }\bibfield  {title} {\bibinfo {title} {{Spin polarons in the t-J model}},\ }\href {https://doi.org/10.1103/PhysRevB.44.317} {\bibfield  {journal} {\bibinfo  {journal} {Phys. Rev. B}\ }\textbf {\bibinfo {volume} {44}},\ \bibinfo {pages} {317} (\bibinfo {year} {1991})}\BibitemShut {NoStop}%
\bibitem [{\citenamefont {Bala}\ \emph {et~al.}(1995)\citenamefont {Bala}, \citenamefont {Ole\ifmmode~\acute{s}\else \'{s}\fi{}},\ and\ \citenamefont {Zaanen}}]{Spin_polaron_t-t1_J}%
  \BibitemOpen
  \bibfield  {author} {\bibinfo {author} {\bibfnamefont {J.}~\bibnamefont {Bala}}, \bibinfo {author} {\bibfnamefont {A.~M.}\ \bibnamefont {Ole\ifmmode~\acute{s}\else \'{s}\fi{}}},\ and\ \bibinfo {author} {\bibfnamefont {J.}~\bibnamefont {Zaanen}},\ }\bibfield  {title} {\bibinfo {title} {{Spin polarons in the t-t\ensuremath{'}-J model}},\ }\href {https://doi.org/10.1103/PhysRevB.52.4597} {\bibfield  {journal} {\bibinfo  {journal} {Phys. Rev. B}\ }\textbf {\bibinfo {volume} {52}},\ \bibinfo {pages} {4597} (\bibinfo {year} {1995})}\BibitemShut {NoStop}%
\bibitem [{\citenamefont {Clancy}\ \emph {et~al.}(2023)\citenamefont {Clancy}, \citenamefont {Gretarsson}, \citenamefont {Lupascu}, \citenamefont {Sears}, \citenamefont {Nie}, \citenamefont {Upton}, \citenamefont {Kim}, \citenamefont {Islam}, \citenamefont {Uchida}, \citenamefont {Schlom}, \citenamefont {Shen},\ and\ \citenamefont {Kim}}]{Clancy_2023}%
  \BibitemOpen
  \bibfield  {author} {\bibinfo {author} {\bibfnamefont {J.~P.}\ \bibnamefont {Clancy}}, \bibinfo {author} {\bibfnamefont {H.}~\bibnamefont {Gretarsson}}, \bibinfo {author} {\bibfnamefont {A.}~\bibnamefont {Lupascu}}, \bibinfo {author} {\bibfnamefont {J.~A.}\ \bibnamefont {Sears}}, \bibinfo {author} {\bibfnamefont {Z.}~\bibnamefont {Nie}}, \bibinfo {author} {\bibfnamefont {M.~H.}\ \bibnamefont {Upton}}, \bibinfo {author} {\bibfnamefont {J.}~\bibnamefont {Kim}}, \bibinfo {author} {\bibfnamefont {Z.}~\bibnamefont {Islam}}, \bibinfo {author} {\bibfnamefont {M.}~\bibnamefont {Uchida}}, \bibinfo {author} {\bibfnamefont {D.~G.}\ \bibnamefont {Schlom}}, \bibinfo {author} {\bibfnamefont {K.~M.}\ \bibnamefont {Shen}},\ and\ \bibinfo {author} {\bibfnamefont {Y.-J.}\ \bibnamefont {Kim}},\ }\bibfield  {title} {\bibinfo {title} {{Magnetic excitations in the square-lattice iridate $\text{Ba}_2 \text{IrO}_4$}},\ }\href {https://doi.org/10.1103/PhysRevB.107.054423} {\bibfield  {journal} {\bibinfo  {journal} {Phys. Rev. B}\
  }\textbf {\bibinfo {volume} {107}},\ \bibinfo {pages} {054423} (\bibinfo {year} {2023})}\BibitemShut {NoStop}%
\bibitem [{\citenamefont {J.~Nichols}\ \emph {et~al.}(2014)\citenamefont {J.~Nichols}, \citenamefont {Korneta}, \citenamefont {Terzic}, \citenamefont {Cao}, \citenamefont {Brill},\ and\ \citenamefont {Seo}}]{Nichols_2014}%
  \BibitemOpen
  \bibfield  {author} {\bibinfo {author} {\bibfnamefont {J.}~\bibnamefont {J.~Nichols}}, \bibinfo {author} {\bibfnamefont {O.~B.}\ \bibnamefont {Korneta}}, \bibinfo {author} {\bibfnamefont {J.}~\bibnamefont {Terzic}}, \bibinfo {author} {\bibfnamefont {G.}~\bibnamefont {Cao}}, \bibinfo {author} {\bibfnamefont {J.~W.}\ \bibnamefont {Brill}},\ and\ \bibinfo {author} {\bibfnamefont {S.~S.~A.}\ \bibnamefont {Seo}},\ }\bibfield  {title} {\bibinfo {title} {{Epitaxial $\text{Ba}_2 \text{IrO}_4$ thin-films grown on SrTiO 3 substrates by pulsed laser deposition}},\ }\href {https://pubs.aip.org/aip/apl/article-abstract/104/12/121913/24554/Epitaxial-Ba2IrO4-thin-films-grown-on-SrTiO3?redirectedFrom=fulltext} {\bibfield  {journal} {\bibinfo  {journal} {Appl. Phys. Lett.}\ }\textbf {\bibinfo {volume} {104}},\ \bibinfo {pages} {121913} (\bibinfo {year} {2014})}\BibitemShut {NoStop}%
\bibitem [{\citenamefont {Dasari}\ \emph {et~al.}(2026)\citenamefont {Dasari}, \citenamefont {Strand}, \citenamefont {Eckstein}, \citenamefont {Lichtenstein},\ and\ \citenamefont {Stepanov}}]{Dasari_2026}%
  \BibitemOpen
  \bibfield  {author} {\bibinfo {author} {\bibfnamefont {N.}~\bibnamefont {Dasari}}, \bibinfo {author} {\bibfnamefont {H.~U.~R.}\ \bibnamefont {Strand}}, \bibinfo {author} {\bibfnamefont {M.}~\bibnamefont {Eckstein}}, \bibinfo {author} {\bibfnamefont {A.~I.}\ \bibnamefont {Lichtenstein}},\ and\ \bibinfo {author} {\bibfnamefont {E.~A.}\ \bibnamefont {Stepanov}},\ }\bibfield  {title} {\bibinfo {title} {Nonlocal correlation effects in dc and optical conductivity of the hubbard model},\ }\href {https://doi.org/10.1103/gzwp-zd6t} {\bibfield  {journal} {\bibinfo  {journal} {Phys. Rev. Lett.}\ }\textbf {\bibinfo {volume} {136}},\ \bibinfo {pages} {106905} (\bibinfo {year} {2026})}\BibitemShut {NoStop}%
\bibitem [{\citenamefont {Uchida}\ \emph {et~al.}(2014)\citenamefont {Uchida}, \citenamefont {Nie}, \citenamefont {King}, \citenamefont {Kim}, \citenamefont {Fennie}, \citenamefont {Schlom},\ and\ \citenamefont {Shen}}]{Uchida_2014}%
  \BibitemOpen
  \bibfield  {author} {\bibinfo {author} {\bibfnamefont {M.}~\bibnamefont {Uchida}}, \bibinfo {author} {\bibfnamefont {Y.~F.}\ \bibnamefont {Nie}}, \bibinfo {author} {\bibfnamefont {P.~D.~C.}\ \bibnamefont {King}}, \bibinfo {author} {\bibfnamefont {C.~H.}\ \bibnamefont {Kim}}, \bibinfo {author} {\bibfnamefont {C.~J.}\ \bibnamefont {Fennie}}, \bibinfo {author} {\bibfnamefont {D.~G.}\ \bibnamefont {Schlom}},\ and\ \bibinfo {author} {\bibfnamefont {K.~M.}\ \bibnamefont {Shen}},\ }\bibfield  {title} {\bibinfo {title} {{Correlated vs. conventional insulating behavior in the ${\mathrm{J}}_{\mathrm{eff}}=\frac{1}{2}$ vs. $\frac{3}{2}$ bands in the layered iridate $\text{Ba}_2 \text{IrO}_4$}},\ }\href {https://link.aps.org/doi/10.1103/PhysRevB.90.075142} {\bibfield  {journal} {\bibinfo  {journal} {Phys. Rev. B}\ }\textbf {\bibinfo {volume} {90}},\ \bibinfo {pages} {075142} (\bibinfo {year} {2014})}\BibitemShut {NoStop}%
\bibitem [{\citenamefont {Nie}\ \emph {et~al.}(2015)\citenamefont {Nie}, \citenamefont {King}, \citenamefont {Kim}, \citenamefont {Uchida}, \citenamefont {Wei}, \citenamefont {Faeth}, \citenamefont {Ruf}, \citenamefont {Ruff}, \citenamefont {Xie}, \citenamefont {Pan}, \citenamefont {Fennie}, \citenamefont {Schlom},\ and\ \citenamefont {Shen}}]{Nie_2015}%
  \BibitemOpen
  \bibfield  {author} {\bibinfo {author} {\bibfnamefont {Y.~F.}\ \bibnamefont {Nie}}, \bibinfo {author} {\bibfnamefont {P.~D.~C.}\ \bibnamefont {King}}, \bibinfo {author} {\bibfnamefont {C.~H.}\ \bibnamefont {Kim}}, \bibinfo {author} {\bibfnamefont {M.}~\bibnamefont {Uchida}}, \bibinfo {author} {\bibfnamefont {H.~I.}\ \bibnamefont {Wei}}, \bibinfo {author} {\bibfnamefont {B.~D.}\ \bibnamefont {Faeth}}, \bibinfo {author} {\bibfnamefont {J.~P.}\ \bibnamefont {Ruf}}, \bibinfo {author} {\bibfnamefont {J.~P.~C.}\ \bibnamefont {Ruff}}, \bibinfo {author} {\bibfnamefont {L.}~\bibnamefont {Xie}}, \bibinfo {author} {\bibfnamefont {X.}~\bibnamefont {Pan}}, \bibinfo {author} {\bibfnamefont {C.~J.}\ \bibnamefont {Fennie}}, \bibinfo {author} {\bibfnamefont {D.~G.}\ \bibnamefont {Schlom}},\ and\ \bibinfo {author} {\bibfnamefont {K.~M.}\ \bibnamefont {Shen}},\ }\bibfield  {title} {\bibinfo {title} {{Interplay of Spin-Orbit Interactions, Dimensionality, and Octahedral Rotations in Semimetallic
  ${\mathbf{\text{SrIrO}}}_{3}$}},\ }\href {https://doi.org/10.1103/PhysRevLett.114.016401} {\bibfield  {journal} {\bibinfo  {journal} {Phys. Rev. Lett.}\ }\textbf {\bibinfo {volume} {114}},\ \bibinfo {pages} {016401} (\bibinfo {year} {2015})}\BibitemShut {NoStop}%
\bibitem [{\citenamefont {Liu}\ \emph {et~al.}(2015)\citenamefont {Liu}, \citenamefont {Yu}, \citenamefont {Jia}, \citenamefont {Zhao}, \citenamefont {Weng}, \citenamefont {Peng}, \citenamefont {Chen}, \citenamefont {Xie}, \citenamefont {Mou}, \citenamefont {He}, \citenamefont {Liu}, \citenamefont {Feng}, \citenamefont {Yi}, \citenamefont {Zhao}, \citenamefont {Liu}, \citenamefont {He}, \citenamefont {Dong}, \citenamefont {Zhang}, \citenamefont {Xu}, \citenamefont {Chen}, \citenamefont {Cao}, \citenamefont {Dai}, \citenamefont {Fang},\ and\ \citenamefont {Zhou}}]{Liu_2015}%
  \BibitemOpen
  \bibfield  {author} {\bibinfo {author} {\bibfnamefont {Y.}~\bibnamefont {Liu}}, \bibinfo {author} {\bibfnamefont {L.}~\bibnamefont {Yu}}, \bibinfo {author} {\bibfnamefont {X.}~\bibnamefont {Jia}}, \bibinfo {author} {\bibfnamefont {J.}~\bibnamefont {Zhao}}, \bibinfo {author} {\bibfnamefont {H.}~\bibnamefont {Weng}}, \bibinfo {author} {\bibfnamefont {Y.}~\bibnamefont {Peng}}, \bibinfo {author} {\bibfnamefont {C.}~\bibnamefont {Chen}}, \bibinfo {author} {\bibfnamefont {Z.}~\bibnamefont {Xie}}, \bibinfo {author} {\bibfnamefont {D.}~\bibnamefont {Mou}}, \bibinfo {author} {\bibfnamefont {J.}~\bibnamefont {He}}, \bibinfo {author} {\bibfnamefont {X.}~\bibnamefont {Liu}}, \bibinfo {author} {\bibfnamefont {Y.}~\bibnamefont {Feng}}, \bibinfo {author} {\bibfnamefont {H.}~\bibnamefont {Yi}}, \bibinfo {author} {\bibfnamefont {L.}~\bibnamefont {Zhao}}, \bibinfo {author} {\bibfnamefont {G.}~\bibnamefont {Liu}}, \bibinfo {author} {\bibfnamefont {S.}~\bibnamefont {He}}, \bibinfo {author} {\bibfnamefont {X.}~\bibnamefont
  {Dong}}, \bibinfo {author} {\bibfnamefont {J.}~\bibnamefont {Zhang}}, \bibinfo {author} {\bibfnamefont {Z.}~\bibnamefont {Xu}}, \bibinfo {author} {\bibfnamefont {C.}~\bibnamefont {Chen}}, \bibinfo {author} {\bibfnamefont {G.}~\bibnamefont {Cao}}, \bibinfo {author} {\bibfnamefont {X.}~\bibnamefont {Dai}}, \bibinfo {author} {\bibfnamefont {Z.}~\bibnamefont {Fang}},\ and\ \bibinfo {author} {\bibfnamefont {X.~J.}\ \bibnamefont {Zhou}},\ }\bibfield  {title} {\bibinfo {title} {{Anomalous High-Energy Waterfall-Like Electronic Structure in 5d Transition Metal Oxide $\text{Sr}_2\text{IrO}_4$ with a Strong Spin-Orbit Coupling}},\ }\href {https://doi.org/10.1038/srep13036} {\bibfield  {journal} {\bibinfo  {journal} {Scientific Reports}\ }\textbf {\bibinfo {volume} {5}},\ \bibinfo {pages} {13036} (\bibinfo {year} {2015})}\BibitemShut {NoStop}%
\bibitem [{\citenamefont {Cao}\ \emph {et~al.}(2016)\citenamefont {Cao}, \citenamefont {Wang}, \citenamefont {Waugh}, \citenamefont {Reber}, \citenamefont {Li}, \citenamefont {Zhou}, \citenamefont {Parham}, \citenamefont {Park}, \citenamefont {Plumb}, \citenamefont {Rotenberg}, \citenamefont {Bostwick}, \citenamefont {Denlinger}, \citenamefont {Qi}, \citenamefont {Hermele}, \citenamefont {Cao},\ and\ \citenamefont {Dessau}}]{Cao_2016}%
  \BibitemOpen
  \bibfield  {author} {\bibinfo {author} {\bibfnamefont {Y.}~\bibnamefont {Cao}}, \bibinfo {author} {\bibfnamefont {Q.}~\bibnamefont {Wang}}, \bibinfo {author} {\bibfnamefont {J.~A.}\ \bibnamefont {Waugh}}, \bibinfo {author} {\bibfnamefont {T.~J.}\ \bibnamefont {Reber}}, \bibinfo {author} {\bibfnamefont {H.}~\bibnamefont {Li}}, \bibinfo {author} {\bibfnamefont {X.}~\bibnamefont {Zhou}}, \bibinfo {author} {\bibfnamefont {S.}~\bibnamefont {Parham}}, \bibinfo {author} {\bibfnamefont {S.-R.}\ \bibnamefont {Park}}, \bibinfo {author} {\bibfnamefont {N.~C.}\ \bibnamefont {Plumb}}, \bibinfo {author} {\bibfnamefont {E.}~\bibnamefont {Rotenberg}}, \bibinfo {author} {\bibfnamefont {A.}~\bibnamefont {Bostwick}}, \bibinfo {author} {\bibfnamefont {J.~D.}\ \bibnamefont {Denlinger}}, \bibinfo {author} {\bibfnamefont {T.}~\bibnamefont {Qi}}, \bibinfo {author} {\bibfnamefont {M.~A.}\ \bibnamefont {Hermele}}, \bibinfo {author} {\bibfnamefont {G.}~\bibnamefont {Cao}},\ and\ \bibinfo {author} {\bibfnamefont {D.~S.}\ \bibnamefont
  {Dessau}},\ }\bibfield  {title} {\bibinfo {title} {{Hallmarks of the Mott-metal crossover in the hole-doped pseudospin-$1/2$ Mott insulator Sr$_2$IrO$_4$}},\ }\href {https://doi.org/10.1038/ncomms11367} {\bibfield  {journal} {\bibinfo  {journal} {Nature Communications}\ }\textbf {\bibinfo {volume} {7}},\ \bibinfo {pages} {11367} (\bibinfo {year} {2016})}\BibitemShut {NoStop}%
\bibitem [{\citenamefont {Li}\ \emph {et~al.}(2013)\citenamefont {Li}, \citenamefont {Cao}, \citenamefont {Okamoto}, \citenamefont {Yi}, \citenamefont {Lin}, \citenamefont {Sales}, \citenamefont {Yan}, \citenamefont {Arita}, \citenamefont {Kuneš}, \citenamefont {Kozhevnikov}, \citenamefont {Eguiluz}, \citenamefont {Imada}, \citenamefont {Gai}, \citenamefont {Pan},\ and\ \citenamefont {Mandrus}}]{Li_2013}%
  \BibitemOpen
  \bibfield  {author} {\bibinfo {author} {\bibfnamefont {Q.}~\bibnamefont {Li}}, \bibinfo {author} {\bibfnamefont {G.}~\bibnamefont {Cao}}, \bibinfo {author} {\bibfnamefont {S.}~\bibnamefont {Okamoto}}, \bibinfo {author} {\bibfnamefont {J.}~\bibnamefont {Yi}}, \bibinfo {author} {\bibfnamefont {W.}~\bibnamefont {Lin}}, \bibinfo {author} {\bibfnamefont {B.~C.}\ \bibnamefont {Sales}}, \bibinfo {author} {\bibfnamefont {J.}~\bibnamefont {Yan}}, \bibinfo {author} {\bibfnamefont {R.}~\bibnamefont {Arita}}, \bibinfo {author} {\bibfnamefont {J.}~\bibnamefont {Kuneš}}, \bibinfo {author} {\bibfnamefont {A.~V.}\ \bibnamefont {Kozhevnikov}}, \bibinfo {author} {\bibfnamefont {A.~G.}\ \bibnamefont {Eguiluz}}, \bibinfo {author} {\bibfnamefont {M.}~\bibnamefont {Imada}}, \bibinfo {author} {\bibfnamefont {Z.}~\bibnamefont {Gai}}, \bibinfo {author} {\bibfnamefont {M.}~\bibnamefont {Pan}},\ and\ \bibinfo {author} {\bibfnamefont {D.~G.}\ \bibnamefont {Mandrus}},\ }\bibfield  {title} {\bibinfo {title} {{Atomically resolved
  spectroscopic study of Sr$_2$IrO$_4$: Experiment and theory}},\ }\href {https://doi.org/10.1038/srep03073} {\bibfield  {journal} {\bibinfo  {journal} {Scientific Reports}\ }\textbf {\bibinfo {volume} {3}},\ \bibinfo {pages} {3073} (\bibinfo {year} {2013})}\BibitemShut {NoStop}%
\bibitem [{\citenamefont {Yan}\ \emph {et~al.}(2015)\citenamefont {Yan}, \citenamefont {Ren}, \citenamefont {Xu}, \citenamefont {Xie}, \citenamefont {Tao}, \citenamefont {Choi}, \citenamefont {Lee}, \citenamefont {Choi}, \citenamefont {Zhang},\ and\ \citenamefont {Feng}}]{Yan_2015}%
  \BibitemOpen
  \bibfield  {author} {\bibinfo {author} {\bibfnamefont {Y.~J.}\ \bibnamefont {Yan}}, \bibinfo {author} {\bibfnamefont {M.~Q.}\ \bibnamefont {Ren}}, \bibinfo {author} {\bibfnamefont {H.~C.}\ \bibnamefont {Xu}}, \bibinfo {author} {\bibfnamefont {B.~P.}\ \bibnamefont {Xie}}, \bibinfo {author} {\bibfnamefont {R.}~\bibnamefont {Tao}}, \bibinfo {author} {\bibfnamefont {H.~Y.}\ \bibnamefont {Choi}}, \bibinfo {author} {\bibfnamefont {N.}~\bibnamefont {Lee}}, \bibinfo {author} {\bibfnamefont {Y.~J.}\ \bibnamefont {Choi}}, \bibinfo {author} {\bibfnamefont {T.}~\bibnamefont {Zhang}},\ and\ \bibinfo {author} {\bibfnamefont {D.~L.}\ \bibnamefont {Feng}},\ }\bibfield  {title} {\bibinfo {title} {{Electron-Doped ${\mathrm{Sr}}_{2}{\mathrm{IrO}}_{4}$: An Analogue of Hole-Doped Cuprate Superconductors Demonstrated by Scanning Tunneling Microscopy}},\ }\href {https://doi.org/10.1103/PhysRevX.5.041018} {\bibfield  {journal} {\bibinfo  {journal} {Phys. Rev. X}\ }\textbf {\bibinfo {volume} {5}},\ \bibinfo {pages} {041018}
  (\bibinfo {year} {2015})}\BibitemShut {NoStop}%
\bibitem [{\citenamefont {Martins}\ \emph {et~al.}(2018)\citenamefont {Martins}, \citenamefont {Lenz}, \citenamefont {Perfetti}, \citenamefont {Brouet}, \citenamefont {Bertran},\ and\ \citenamefont {Biermann}}]{Martins_2018}%
  \BibitemOpen
  \bibfield  {author} {\bibinfo {author} {\bibfnamefont {C.}~\bibnamefont {Martins}}, \bibinfo {author} {\bibfnamefont {B.}~\bibnamefont {Lenz}}, \bibinfo {author} {\bibfnamefont {L.}~\bibnamefont {Perfetti}}, \bibinfo {author} {\bibfnamefont {V.}~\bibnamefont {Brouet}}, \bibinfo {author} {\bibfnamefont {F.~m.~c.}\ \bibnamefont {Bertran}},\ and\ \bibinfo {author} {\bibfnamefont {S.}~\bibnamefont {Biermann}},\ }\bibfield  {title} {\bibinfo {title} {{Nonlocal Coulomb correlations in pure and electron-doped $\text{Sr}_2 \text{IrO}_4$: Spectral functions, Fermi surface, and pseudo-gap-like spectral weight distributions from oriented cluster dynamical mean-field theory}},\ }\href {https://doi.org/10.1103/PhysRevMaterials.2.032001} {\bibfield  {journal} {\bibinfo  {journal} {Phys. Rev. Mater.}\ }\textbf {\bibinfo {volume} {2}},\ \bibinfo {pages} {032001} (\bibinfo {year} {2018})}\BibitemShut {NoStop}%
\bibitem [{\citenamefont {Pärschke}\ \emph {et~al.}(2017)\citenamefont {Pärschke}, \citenamefont {Wohlfeld}, \citenamefont {Foyevtsova},\ and\ \citenamefont {van~den Brink}}]{parschke_2017}%
  \BibitemOpen
  \bibfield  {author} {\bibinfo {author} {\bibfnamefont {E.~M.}\ \bibnamefont {Pärschke}}, \bibinfo {author} {\bibfnamefont {K.}~\bibnamefont {Wohlfeld}}, \bibinfo {author} {\bibfnamefont {K.}~\bibnamefont {Foyevtsova}},\ and\ \bibinfo {author} {\bibfnamefont {J.}~\bibnamefont {van~den Brink}},\ }\bibfield  {title} {\bibinfo {title} {{Correlation induced electron-hole asymmetry in quasi- two-dimensional iridates}},\ }\href {https://doi.org/10.1038/s41467-017-00818-8} {\bibfield  {journal} {\bibinfo  {journal} {Nature Communications}\ }\textbf {\bibinfo {volume} {8}},\ \bibinfo {pages} {686} (\bibinfo {year} {2017})},\ \bibinfo {note} {publisher: Nature Publishing Group}\BibitemShut {NoStop}%
\bibitem [{\citenamefont {Pärschke}\ and\ \citenamefont {Ray}(2018)}]{parschke_2018}%
  \BibitemOpen
  \bibfield  {author} {\bibinfo {author} {\bibfnamefont {E.~M.}\ \bibnamefont {Pärschke}}\ and\ \bibinfo {author} {\bibfnamefont {R.}~\bibnamefont {Ray}},\ }\bibfield  {title} {\bibinfo {title} {{Influence of the multiplet structure on the photoemission spectra of spin-orbit driven {Mott} insulators: {Application} to $\text{Sr}_2 \text{IrO}_4$}},\ }\href {https://doi.org/10.1103/PhysRevB.98.064422} {\bibfield  {journal} {\bibinfo  {journal} {Physical Review B}\ }\textbf {\bibinfo {volume} {98}},\ \bibinfo {pages} {064422} (\bibinfo {year} {2018})},\ \bibinfo {note} {publisher: American Physical Society}\BibitemShut {NoStop}%
\bibitem [{\citenamefont {Pärschke}\ \emph {et~al.}(2022)\citenamefont {Pärschke}, \citenamefont {Chen}, \citenamefont {Ray},\ and\ \citenamefont {Chen}}]{parschke_2022}%
  \BibitemOpen
  \bibfield  {author} {\bibinfo {author} {\bibfnamefont {E.~M.}\ \bibnamefont {Pärschke}}, \bibinfo {author} {\bibfnamefont {W.-C.}\ \bibnamefont {Chen}}, \bibinfo {author} {\bibfnamefont {R.}~\bibnamefont {Ray}},\ and\ \bibinfo {author} {\bibfnamefont {C.-C.}\ \bibnamefont {Chen}},\ }\bibfield  {title} {\bibinfo {title} {{Evolution of electronic and magnetic properties of $\text{Sr}_2 \text{IrO}_4$ under strain}},\ }\href {https://doi.org/10.1038/s41535-022-00496-w} {\bibfield  {journal} {\bibinfo  {journal} {npj Quantum Materials}\ }\textbf {\bibinfo {volume} {7}},\ \bibinfo {pages} {1} (\bibinfo {year} {2022})},\ \bibinfo {note} {publisher: Nature Publishing Group}\BibitemShut {NoStop}%
\bibitem [{\citenamefont {Taranto}\ \emph {et~al.}(2012)\citenamefont {Taranto}, \citenamefont {Sangiovanni}, \citenamefont {Held}, \citenamefont {Capone}, \citenamefont {Georges},\ and\ \citenamefont {Toschi}}]{Taranto_2012}%
  \BibitemOpen
  \bibfield  {author} {\bibinfo {author} {\bibfnamefont {C.}~\bibnamefont {Taranto}}, \bibinfo {author} {\bibfnamefont {G.}~\bibnamefont {Sangiovanni}}, \bibinfo {author} {\bibfnamefont {K.}~\bibnamefont {Held}}, \bibinfo {author} {\bibfnamefont {M.}~\bibnamefont {Capone}}, \bibinfo {author} {\bibfnamefont {A.}~\bibnamefont {Georges}},\ and\ \bibinfo {author} {\bibfnamefont {A.}~\bibnamefont {Toschi}},\ }\bibfield  {title} {\bibinfo {title} {Signature of antiferromagnetic long-range order in the optical spectrum of strongly correlated electron systems},\ }\href {https://doi.org/10.1103/PhysRevB.85.085124} {\bibfield  {journal} {\bibinfo  {journal} {Phys. Rev. B}\ }\textbf {\bibinfo {volume} {85}},\ \bibinfo {pages} {085124} (\bibinfo {year} {2012})}\BibitemShut {NoStop}%
\bibitem [{BZp()}]{BZpath}%
  \BibitemOpen
  \href@noop {} {}\bibinfo {note} {Around each of the points in the $k_x-k_y$ plane, we define infinitesimally narrow cylindrical regions with a radius of $ 0.025 \times 2\pi/a$ [1/\AA] and a height of $ 3.30 \times 2\pi/a$ [1/\AA], where $a$ is the shorter Ir–Ir distance. The height is chosen to fully cover the extent of the conventional Brillouin zone along the $k_z$ direction. The selected $\k$-points for this analysis are P1 $=$ X$^*$, P2 $=$ ($[0.4; 0; 0] \times 2\pi /a$), and P3 $=$ ($[0.29; 0; 0] \times 2\pi /a$).}\BibitemShut {Stop}%
\bibitem [{\citenamefont {Zhang}\ \emph {et~al.}(2013)\citenamefont {Zhang}, \citenamefont {Haule},\ and\ \citenamefont {Vanderbilt}}]{Zhang_2013}%
  \BibitemOpen
  \bibfield  {author} {\bibinfo {author} {\bibfnamefont {H.}~\bibnamefont {Zhang}}, \bibinfo {author} {\bibfnamefont {K.}~\bibnamefont {Haule}},\ and\ \bibinfo {author} {\bibfnamefont {D.}~\bibnamefont {Vanderbilt}},\ }\bibfield  {title} {\bibinfo {title} {{Effective $J\mathbf{=}1/2$ Insulating State in Ruddlesden-Popper Iridates: An $\mathrm{LDA}\mathbf{+}\mathrm{DMFT}$ Study}},\ }\href {https://doi.org/10.1103/PhysRevLett.111.246402} {\bibfield  {journal} {\bibinfo  {journal} {Phys. Rev. Lett.}\ }\textbf {\bibinfo {volume} {111}},\ \bibinfo {pages} {246402} (\bibinfo {year} {2013})}\BibitemShut {NoStop}%
\bibitem [{\citenamefont {Louat}\ \emph {et~al.}(2019)\citenamefont {Louat}, \citenamefont {Lenz}, \citenamefont {Biermann}, \citenamefont {Martins}, \citenamefont {Bertran}, \citenamefont {Le~F\`evre}, \citenamefont {Rault}, \citenamefont {Bert},\ and\ \citenamefont {Brouet}}]{Louat_2019}%
  \BibitemOpen
  \bibfield  {author} {\bibinfo {author} {\bibfnamefont {A.}~\bibnamefont {Louat}}, \bibinfo {author} {\bibfnamefont {B.}~\bibnamefont {Lenz}}, \bibinfo {author} {\bibfnamefont {S.}~\bibnamefont {Biermann}}, \bibinfo {author} {\bibfnamefont {C.}~\bibnamefont {Martins}}, \bibinfo {author} {\bibfnamefont {F.~m.~c.}\ \bibnamefont {Bertran}}, \bibinfo {author} {\bibfnamefont {P.}~\bibnamefont {Le~F\`evre}}, \bibinfo {author} {\bibfnamefont {J.~E.}\ \bibnamefont {Rault}}, \bibinfo {author} {\bibfnamefont {F.}~\bibnamefont {Bert}},\ and\ \bibinfo {author} {\bibfnamefont {V.}~\bibnamefont {Brouet}},\ }\bibfield  {title} {\bibinfo {title} {{ARPES study of orbital character, symmetry breaking, and pseudogaps in doped and pure $\text{Sr}_2 \text{IrO}_4$}},\ }\href {https://doi.org/10.1103/PhysRevB.100.205135} {\bibfield  {journal} {\bibinfo  {journal} {Phys. Rev. B}\ }\textbf {\bibinfo {volume} {100}},\ \bibinfo {pages} {205135} (\bibinfo {year} {2019})}\BibitemShut {NoStop}%
\bibitem [{\citenamefont {Brouet}\ \emph {et~al.}(2021)\citenamefont {Brouet}, \citenamefont {Foulquier}, \citenamefont {Louat}, \citenamefont {Bertran}, \citenamefont {Le~F\`evre}, \citenamefont {Rault},\ and\ \citenamefont {Colson}}]{Brouet_2021}%
  \BibitemOpen
  \bibfield  {author} {\bibinfo {author} {\bibfnamefont {V.}~\bibnamefont {Brouet}}, \bibinfo {author} {\bibfnamefont {P.}~\bibnamefont {Foulquier}}, \bibinfo {author} {\bibfnamefont {A.}~\bibnamefont {Louat}}, \bibinfo {author} {\bibfnamefont {F.~m.~c.}\ \bibnamefont {Bertran}}, \bibinfo {author} {\bibfnamefont {P.}~\bibnamefont {Le~F\`evre}}, \bibinfo {author} {\bibfnamefont {J.~E.}\ \bibnamefont {Rault}},\ and\ \bibinfo {author} {\bibfnamefont {D.}~\bibnamefont {Colson}},\ }\bibfield  {title} {\bibinfo {title} {{Origin of the different electronic structure of Rh- and Ru-doped $\text{Sr}_2 \text{IrO}_4$}},\ }\href {https://link.aps.org/doi/10.1103/PhysRevB.104.L121104} {\bibfield  {journal} {\bibinfo  {journal} {Phys. Rev. B}\ }\textbf {\bibinfo {volume} {104}},\ \bibinfo {pages} {L121104} (\bibinfo {year} {2021})}\BibitemShut {NoStop}%
\bibitem [{\citenamefont {Jarrell}\ \emph {et~al.}(1995)\citenamefont {Jarrell}, \citenamefont {Freericks},\ and\ \citenamefont {Pruschke}}]{Jarrel1995}%
  \BibitemOpen
  \bibfield  {author} {\bibinfo {author} {\bibfnamefont {M.}~\bibnamefont {Jarrell}}, \bibinfo {author} {\bibfnamefont {J.~K.}\ \bibnamefont {Freericks}},\ and\ \bibinfo {author} {\bibfnamefont {T.}~\bibnamefont {Pruschke}},\ }\bibfield  {title} {\bibinfo {title} {{Optical conductivity of the infinite-dimensional Hubbard model}},\ }\href {https://doi.org/10.1103/PhysRevB.51.11704} {\bibfield  {journal} {\bibinfo  {journal} {Phys. Rev. B}\ }\textbf {\bibinfo {volume} {51}},\ \bibinfo {pages} {11704} (\bibinfo {year} {1995})}\BibitemShut {NoStop}%
\bibitem [{\citenamefont {Rozenberg}\ \emph {et~al.}(1995)\citenamefont {Rozenberg}, \citenamefont {Kotliar}, \citenamefont {Kajueter}, \citenamefont {Thomas}, \citenamefont {Rapkine}, \citenamefont {Honig},\ and\ \citenamefont {Metcalf}}]{Rozenberg95}%
  \BibitemOpen
  \bibfield  {author} {\bibinfo {author} {\bibfnamefont {M.~J.}\ \bibnamefont {Rozenberg}}, \bibinfo {author} {\bibfnamefont {G.}~\bibnamefont {Kotliar}}, \bibinfo {author} {\bibfnamefont {H.}~\bibnamefont {Kajueter}}, \bibinfo {author} {\bibfnamefont {G.~A.}\ \bibnamefont {Thomas}}, \bibinfo {author} {\bibfnamefont {D.~H.}\ \bibnamefont {Rapkine}}, \bibinfo {author} {\bibfnamefont {J.~M.}\ \bibnamefont {Honig}},\ and\ \bibinfo {author} {\bibfnamefont {P.}~\bibnamefont {Metcalf}},\ }\bibfield  {title} {\bibinfo {title} {{Optical Conductivity in Mott-Hubbard Systems}},\ }\href {https://doi.org/10.1103/PhysRevLett.75.105} {\bibfield  {journal} {\bibinfo  {journal} {Phys. Rev. Lett.}\ }\textbf {\bibinfo {volume} {75}},\ \bibinfo {pages} {105} (\bibinfo {year} {1995})}\BibitemShut {NoStop}%
\bibitem [{\citenamefont {Toschi}\ \emph {et~al.}(2005)\citenamefont {Toschi}, \citenamefont {Capone}, \citenamefont {Ortolani}, \citenamefont {Calvani}, \citenamefont {Lupi},\ and\ \citenamefont {Castellani}}]{Toschi2005}%
  \BibitemOpen
  \bibfield  {author} {\bibinfo {author} {\bibfnamefont {A.}~\bibnamefont {Toschi}}, \bibinfo {author} {\bibfnamefont {M.}~\bibnamefont {Capone}}, \bibinfo {author} {\bibfnamefont {M.}~\bibnamefont {Ortolani}}, \bibinfo {author} {\bibfnamefont {P.}~\bibnamefont {Calvani}}, \bibinfo {author} {\bibfnamefont {S.}~\bibnamefont {Lupi}},\ and\ \bibinfo {author} {\bibfnamefont {C.}~\bibnamefont {Castellani}},\ }\bibfield  {title} {\bibinfo {title} {{Temperature Dependence of the Optical Spectral Weight in the Cuprates: Role of Electron Correlations}},\ }\href {https://doi.org/10.1103/PhysRevLett.95.097002} {\bibfield  {journal} {\bibinfo  {journal} {Phys. Rev. Lett.}\ }\textbf {\bibinfo {volume} {95}},\ \bibinfo {pages} {097002} (\bibinfo {year} {2005})}\BibitemShut {NoStop}%
\bibitem [{\citenamefont {Toschi}\ and\ \citenamefont {Capone}(2008)}]{Toschi2008}%
  \BibitemOpen
  \bibfield  {author} {\bibinfo {author} {\bibfnamefont {A.}~\bibnamefont {Toschi}}\ and\ \bibinfo {author} {\bibfnamefont {M.}~\bibnamefont {Capone}},\ }\bibfield  {title} {\bibinfo {title} {{Optical sum rule anomalies in the cuprates: Interplay between strong correlation and electronic band structure}},\ }\href {https://doi.org/10.1103/PhysRevB.77.014518} {\bibfield  {journal} {\bibinfo  {journal} {Phys. Rev. B}\ }\textbf {\bibinfo {volume} {77}},\ \bibinfo {pages} {014518} (\bibinfo {year} {2008})}\BibitemShut {NoStop}%
\bibitem [{\citenamefont {Nicoletti}\ \emph {et~al.}(2010)\citenamefont {Nicoletti}, \citenamefont {Limaj}, \citenamefont {Calvani}, \citenamefont {Rohringer}, \citenamefont {Toschi}, \citenamefont {Sangiovanni}, \citenamefont {Capone}, \citenamefont {Held}, \citenamefont {Ono}, \citenamefont {Ando},\ and\ \citenamefont {Lupi}}]{Nicoletti2010}%
  \BibitemOpen
  \bibfield  {author} {\bibinfo {author} {\bibfnamefont {D.}~\bibnamefont {Nicoletti}}, \bibinfo {author} {\bibfnamefont {O.}~\bibnamefont {Limaj}}, \bibinfo {author} {\bibfnamefont {P.}~\bibnamefont {Calvani}}, \bibinfo {author} {\bibfnamefont {G.}~\bibnamefont {Rohringer}}, \bibinfo {author} {\bibfnamefont {A.}~\bibnamefont {Toschi}}, \bibinfo {author} {\bibfnamefont {G.}~\bibnamefont {Sangiovanni}}, \bibinfo {author} {\bibfnamefont {M.}~\bibnamefont {Capone}}, \bibinfo {author} {\bibfnamefont {K.}~\bibnamefont {Held}}, \bibinfo {author} {\bibfnamefont {S.}~\bibnamefont {Ono}}, \bibinfo {author} {\bibfnamefont {Y.}~\bibnamefont {Ando}},\ and\ \bibinfo {author} {\bibfnamefont {S.}~\bibnamefont {Lupi}},\ }\bibfield  {title} {\bibinfo {title} {{High-Temperature Optical Spectral Weight and Fermi-liquid Renormalization in Bi-Based Cuprate Superconductors}},\ }\href {https://doi.org/10.1103/PhysRevLett.105.077002} {\bibfield  {journal} {\bibinfo  {journal} {Phys. Rev. Lett.}\ }\textbf {\bibinfo {volume} {105}},\
  \bibinfo {pages} {077002} (\bibinfo {year} {2010})}\BibitemShut {NoStop}%
\bibitem [{\citenamefont {Kim}\ \emph {et~al.}(2014{\natexlab{b}})\citenamefont {Kim}, \citenamefont {Krupin}, \citenamefont {Denlinger}, \citenamefont {Bostwick}, \citenamefont {Rotenberg}, \citenamefont {Zhao}, \citenamefont {Mitchell}, \citenamefont {Allen},\ and\ \citenamefont {Kim}}]{Kim2014F}%
  \BibitemOpen
  \bibfield  {author} {\bibinfo {author} {\bibfnamefont {Y.~K.}\ \bibnamefont {Kim}}, \bibinfo {author} {\bibfnamefont {O.}~\bibnamefont {Krupin}}, \bibinfo {author} {\bibfnamefont {J.~D.}\ \bibnamefont {Denlinger}}, \bibinfo {author} {\bibfnamefont {A.}~\bibnamefont {Bostwick}}, \bibinfo {author} {\bibfnamefont {E.}~\bibnamefont {Rotenberg}}, \bibinfo {author} {\bibfnamefont {Q.}~\bibnamefont {Zhao}}, \bibinfo {author} {\bibfnamefont {J.~F.}\ \bibnamefont {Mitchell}}, \bibinfo {author} {\bibfnamefont {J.~W.}\ \bibnamefont {Allen}},\ and\ \bibinfo {author} {\bibfnamefont {B.~J.}\ \bibnamefont {Kim}},\ }\bibfield  {title} {\bibinfo {title} {Fermi arcs in a doped pseudospin-1/2 heisenberg antiferromagnet},\ }\href {https://doi.org/10.1126/science.1251151} {\bibfield  {journal} {\bibinfo  {journal} {Science}\ }\textbf {\bibinfo {volume} {345}},\ \bibinfo {pages} {187} (\bibinfo {year} {2014}{\natexlab{b}})}\BibitemShut {NoStop}%
\bibitem [{\citenamefont {Peng}\ \emph {et~al.}(2022)\citenamefont {Peng}, \citenamefont {Lane}, \citenamefont {Hu}, \citenamefont {Guo}, \citenamefont {Chen}, \citenamefont {Sun}, \citenamefont {Hashimoto}, \citenamefont {Lu}, \citenamefont {Shen}, \citenamefont {Wu}, \citenamefont {Chen}, \citenamefont {Markiewicz}, \citenamefont {Wang}, \citenamefont {Bansil}, \citenamefont {Wilson},\ and\ \citenamefont {He}}]{Peng_2022}%
  \BibitemOpen
  \bibfield  {author} {\bibinfo {author} {\bibfnamefont {S.}~\bibnamefont {Peng}}, \bibinfo {author} {\bibfnamefont {C.}~\bibnamefont {Lane}}, \bibinfo {author} {\bibfnamefont {Y.}~\bibnamefont {Hu}}, \bibinfo {author} {\bibfnamefont {M.}~\bibnamefont {Guo}}, \bibinfo {author} {\bibfnamefont {X.}~\bibnamefont {Chen}}, \bibinfo {author} {\bibfnamefont {Z.}~\bibnamefont {Sun}}, \bibinfo {author} {\bibfnamefont {M.}~\bibnamefont {Hashimoto}}, \bibinfo {author} {\bibfnamefont {D.}~\bibnamefont {Lu}}, \bibinfo {author} {\bibfnamefont {Z.-X.}\ \bibnamefont {Shen}}, \bibinfo {author} {\bibfnamefont {T.}~\bibnamefont {Wu}}, \bibinfo {author} {\bibfnamefont {X.}~\bibnamefont {Chen}}, \bibinfo {author} {\bibfnamefont {R.~S.}\ \bibnamefont {Markiewicz}}, \bibinfo {author} {\bibfnamefont {Y.}~\bibnamefont {Wang}}, \bibinfo {author} {\bibfnamefont {A.}~\bibnamefont {Bansil}}, \bibinfo {author} {\bibfnamefont {S.~D.}\ \bibnamefont {Wilson}},\ and\ \bibinfo {author} {\bibfnamefont {J.}~\bibnamefont {He}},\ }\bibfield
  {title} {\bibinfo {title} {{Electronic nature of the pseudogap in electron-doped Sr$_2$IrO$_4$}},\ }\href {https://doi.org/10.1038/s41535-022-00467-1} {\bibfield  {journal} {\bibinfo  {journal} {npj Quantum Materials}\ }\textbf {\bibinfo {volume} {7}},\ \bibinfo {pages} {58} (\bibinfo {year} {2022})}\BibitemShut {NoStop}%
\bibitem [{\citenamefont {Gr\"ober}\ \emph {et~al.}(2000)\citenamefont {Gr\"ober}, \citenamefont {Eder},\ and\ \citenamefont {Hanke}}]{Grober_2000}%
  \BibitemOpen
  \bibfield  {author} {\bibinfo {author} {\bibfnamefont {C.}~\bibnamefont {Gr\"ober}}, \bibinfo {author} {\bibfnamefont {R.}~\bibnamefont {Eder}},\ and\ \bibinfo {author} {\bibfnamefont {W.}~\bibnamefont {Hanke}},\ }\bibfield  {title} {\bibinfo {title} {Anomalous low-doping phase of the hubbard model},\ }\href {https://doi.org/10.1103/PhysRevB.62.4336} {\bibfield  {journal} {\bibinfo  {journal} {Phys. Rev. B}\ }\textbf {\bibinfo {volume} {62}},\ \bibinfo {pages} {4336} (\bibinfo {year} {2000})}\BibitemShut {NoStop}%
\bibitem [{\citenamefont {Manousakis}(2007)}]{Manousakis_2007}%
  \BibitemOpen
  \bibfield  {author} {\bibinfo {author} {\bibfnamefont {E.}~\bibnamefont {Manousakis}},\ }\bibfield  {title} {\bibinfo {title} {String excitations of a hole in a quantum antiferromagnet and photoelectron spectroscopy},\ }\href {https://doi.org/10.1103/PhysRevB.75.035106} {\bibfield  {journal} {\bibinfo  {journal} {Phys. Rev. B}\ }\textbf {\bibinfo {volume} {75}},\ \bibinfo {pages} {035106} (\bibinfo {year} {2007})}\BibitemShut {NoStop}%
\bibitem [{\citenamefont {Cassol}\ \emph {et~al.}(2025)\citenamefont {Cassol}, \citenamefont {Gaspard}, \citenamefont {Martins}, \citenamefont {Casula},\ and\ \citenamefont {Lenz}}]{cassol_2025_Zenodo}%
  \BibitemOpen
  \bibfield  {author} {\bibinfo {author} {\bibfnamefont {F.}~\bibnamefont {Cassol}}, \bibinfo {author} {\bibfnamefont {L.}~\bibnamefont {Gaspard}}, \bibinfo {author} {\bibfnamefont {C.}~\bibnamefont {Martins}}, \bibinfo {author} {\bibfnamefont {M.}~\bibnamefont {Casula}},\ and\ \bibinfo {author} {\bibfnamefont {B.}~\bibnamefont {Lenz}},\ }\href {https://doi.org/10.5281/zenodo.17186418} {\bibinfo {title} {Spin-polaron fingerprints in the optical conductivity of iridates}} (\bibinfo {year} {2025})\BibitemShut {NoStop}%
\bibitem [{\citenamefont {Giannozzi}\ \emph {et~al.}(2009)\citenamefont {Giannozzi}, \citenamefont {Baroni}, \citenamefont {Bonini}, \citenamefont {Calandra}, \citenamefont {Car}, \citenamefont {Cavazzoni}, \citenamefont {Ceresoli}, \citenamefont {Chiarotti}, \citenamefont {Cococcioni}, \citenamefont {Dabo}, \citenamefont {Dal~Corso}, \citenamefont {{de Gironcoli}}, \citenamefont {Fabris}, \citenamefont {Fratesi}, \citenamefont {Gebauer}, \citenamefont {Gerstmann}, \citenamefont {Gougoussis}, \citenamefont {Kokalj}, \citenamefont {Lazzeri}, \citenamefont {{Martin-Samos}}, \citenamefont {Marzari}, \citenamefont {Mauri}, \citenamefont {Mazzarello}, \citenamefont {Paolini}, \citenamefont {Pasquarello}, \citenamefont {Paulatto}, \citenamefont {Sbraccia}, \citenamefont {Scandolo}, \citenamefont {Sclauzero}, \citenamefont {Seitsonen}, \citenamefont {Smogunov}, \citenamefont {Umari},\ and\ \citenamefont {Wentzcovitch}}]{QuantumEspresso1}%
  \BibitemOpen
  \bibfield  {author} {\bibinfo {author} {\bibfnamefont {P.}~\bibnamefont {Giannozzi}}, \bibinfo {author} {\bibfnamefont {S.}~\bibnamefont {Baroni}}, \bibinfo {author} {\bibfnamefont {N.}~\bibnamefont {Bonini}}, \bibinfo {author} {\bibfnamefont {M.}~\bibnamefont {Calandra}}, \bibinfo {author} {\bibfnamefont {R.}~\bibnamefont {Car}}, \bibinfo {author} {\bibfnamefont {C.}~\bibnamefont {Cavazzoni}}, \bibinfo {author} {\bibfnamefont {D.}~\bibnamefont {Ceresoli}}, \bibinfo {author} {\bibfnamefont {G.~L.}\ \bibnamefont {Chiarotti}}, \bibinfo {author} {\bibfnamefont {M.}~\bibnamefont {Cococcioni}}, \bibinfo {author} {\bibfnamefont {I.}~\bibnamefont {Dabo}}, \bibinfo {author} {\bibfnamefont {A.}~\bibnamefont {Dal~Corso}}, \bibinfo {author} {\bibfnamefont {S.}~\bibnamefont {{de Gironcoli}}}, \bibinfo {author} {\bibfnamefont {S.}~\bibnamefont {Fabris}}, \bibinfo {author} {\bibfnamefont {G.}~\bibnamefont {Fratesi}}, \bibinfo {author} {\bibfnamefont {R.}~\bibnamefont {Gebauer}}, \bibinfo {author} {\bibfnamefont
  {U.}~\bibnamefont {Gerstmann}}, \bibinfo {author} {\bibfnamefont {C.}~\bibnamefont {Gougoussis}}, \bibinfo {author} {\bibfnamefont {A.}~\bibnamefont {Kokalj}}, \bibinfo {author} {\bibfnamefont {M.}~\bibnamefont {Lazzeri}}, \bibinfo {author} {\bibfnamefont {L.}~\bibnamefont {{Martin-Samos}}}, \bibinfo {author} {\bibfnamefont {N.}~\bibnamefont {Marzari}}, \bibinfo {author} {\bibfnamefont {F.}~\bibnamefont {Mauri}}, \bibinfo {author} {\bibfnamefont {R.}~\bibnamefont {Mazzarello}}, \bibinfo {author} {\bibfnamefont {S.}~\bibnamefont {Paolini}}, \bibinfo {author} {\bibfnamefont {A.}~\bibnamefont {Pasquarello}}, \bibinfo {author} {\bibfnamefont {L.}~\bibnamefont {Paulatto}}, \bibinfo {author} {\bibfnamefont {C.}~\bibnamefont {Sbraccia}}, \bibinfo {author} {\bibfnamefont {S.}~\bibnamefont {Scandolo}}, \bibinfo {author} {\bibfnamefont {G.}~\bibnamefont {Sclauzero}}, \bibinfo {author} {\bibfnamefont {A.~P.}\ \bibnamefont {Seitsonen}}, \bibinfo {author} {\bibfnamefont {A.}~\bibnamefont {Smogunov}}, \bibinfo {author}
  {\bibfnamefont {P.}~\bibnamefont {Umari}},\ and\ \bibinfo {author} {\bibfnamefont {R.~M.}\ \bibnamefont {Wentzcovitch}},\ }\bibfield  {title} {\bibinfo {title} {{{QUANTUM ESPRESSO}}: A modular and open-source software project for quantum simulations of materials},\ }\href {https://doi.org/https://doi.org/10.1088/0953-8984/21/39/395502} {\bibfield  {journal} {\bibinfo  {journal} {Journal of Physics. Condensed Matter: An Institute of Physics Journal}\ }\textbf {\bibinfo {volume} {21}},\ \bibinfo {pages} {395502} (\bibinfo {year} {2009})}\BibitemShut {NoStop}%
\bibitem [{\citenamefont {Giannozzi}\ \emph {et~al.}(2017)\citenamefont {Giannozzi}, \citenamefont {Andreussi}, \citenamefont {Brumme}, \citenamefont {Bunau}, \citenamefont {Buongiorno~Nardelli}, \citenamefont {Calandra}, \citenamefont {Car}, \citenamefont {Cavazzoni}, \citenamefont {Ceresoli}, \citenamefont {Cococcioni}, \citenamefont {Colonna}, \citenamefont {Carnimeo}, \citenamefont {Dal~Corso}, \citenamefont {{de Gironcoli}}, \citenamefont {Delugas}, \citenamefont {DiStasio}, \citenamefont {Ferretti}, \citenamefont {Floris}, \citenamefont {Fratesi}, \citenamefont {Fugallo}, \citenamefont {Gebauer}, \citenamefont {Gerstmann}, \citenamefont {Giustino}, \citenamefont {Gorni}, \citenamefont {Jia}, \citenamefont {Kawamura}, \citenamefont {Ko}, \citenamefont {Kokalj}, \citenamefont {Kucukbenli}, \citenamefont {Lazzeri}, \citenamefont {Marsili}, \citenamefont {Marzari}, \citenamefont {Mauri}, \citenamefont {Nguyen}, \citenamefont {Nguyen}, \citenamefont {{Otero-de-la-Roza}}, \citenamefont {Paulatto},
  \citenamefont {Ponce}, \citenamefont {Rocca}, \citenamefont {Sabatini}, \citenamefont {Santra}, \citenamefont {Schlipf}, \citenamefont {Seitsonen}, \citenamefont {Smogunov}, \citenamefont {Timrov}, \citenamefont {Thonhauser}, \citenamefont {Umari}, \citenamefont {Vast}, \citenamefont {Wu},\ and\ \citenamefont {Baroni}}]{QuantumEspresso2}%
  \BibitemOpen
  \bibfield  {author} {\bibinfo {author} {\bibfnamefont {P.}~\bibnamefont {Giannozzi}}, \bibinfo {author} {\bibfnamefont {O.}~\bibnamefont {Andreussi}}, \bibinfo {author} {\bibfnamefont {T.}~\bibnamefont {Brumme}}, \bibinfo {author} {\bibfnamefont {O.}~\bibnamefont {Bunau}}, \bibinfo {author} {\bibfnamefont {M.}~\bibnamefont {Buongiorno~Nardelli}}, \bibinfo {author} {\bibfnamefont {M.}~\bibnamefont {Calandra}}, \bibinfo {author} {\bibfnamefont {R.}~\bibnamefont {Car}}, \bibinfo {author} {\bibfnamefont {C.}~\bibnamefont {Cavazzoni}}, \bibinfo {author} {\bibfnamefont {D.}~\bibnamefont {Ceresoli}}, \bibinfo {author} {\bibfnamefont {M.}~\bibnamefont {Cococcioni}}, \bibinfo {author} {\bibfnamefont {N.}~\bibnamefont {Colonna}}, \bibinfo {author} {\bibfnamefont {I.}~\bibnamefont {Carnimeo}}, \bibinfo {author} {\bibfnamefont {A.}~\bibnamefont {Dal~Corso}}, \bibinfo {author} {\bibfnamefont {S.}~\bibnamefont {{de Gironcoli}}}, \bibinfo {author} {\bibfnamefont {P.}~\bibnamefont {Delugas}}, \bibinfo {author}
  {\bibfnamefont {R.~A.}\ \bibnamefont {DiStasio}}, \bibinfo {author} {\bibfnamefont {A.}~\bibnamefont {Ferretti}}, \bibinfo {author} {\bibfnamefont {A.}~\bibnamefont {Floris}}, \bibinfo {author} {\bibfnamefont {G.}~\bibnamefont {Fratesi}}, \bibinfo {author} {\bibfnamefont {G.}~\bibnamefont {Fugallo}}, \bibinfo {author} {\bibfnamefont {R.}~\bibnamefont {Gebauer}}, \bibinfo {author} {\bibfnamefont {U.}~\bibnamefont {Gerstmann}}, \bibinfo {author} {\bibfnamefont {F.}~\bibnamefont {Giustino}}, \bibinfo {author} {\bibfnamefont {T.}~\bibnamefont {Gorni}}, \bibinfo {author} {\bibfnamefont {J.}~\bibnamefont {Jia}}, \bibinfo {author} {\bibfnamefont {M.}~\bibnamefont {Kawamura}}, \bibinfo {author} {\bibfnamefont {H.-Y.}\ \bibnamefont {Ko}}, \bibinfo {author} {\bibfnamefont {A.}~\bibnamefont {Kokalj}}, \bibinfo {author} {\bibfnamefont {E.}~\bibnamefont {Kucukbenli}}, \bibinfo {author} {\bibfnamefont {M.}~\bibnamefont {Lazzeri}}, \bibinfo {author} {\bibfnamefont {M.}~\bibnamefont {Marsili}}, \bibinfo {author}
  {\bibfnamefont {N.}~\bibnamefont {Marzari}}, \bibinfo {author} {\bibfnamefont {F.}~\bibnamefont {Mauri}}, \bibinfo {author} {\bibfnamefont {N.~L.}\ \bibnamefont {Nguyen}}, \bibinfo {author} {\bibfnamefont {H.-V.}\ \bibnamefont {Nguyen}}, \bibinfo {author} {\bibfnamefont {A.}~\bibnamefont {{Otero-de-la-Roza}}}, \bibinfo {author} {\bibfnamefont {L.}~\bibnamefont {Paulatto}}, \bibinfo {author} {\bibfnamefont {S.}~\bibnamefont {Ponce}}, \bibinfo {author} {\bibfnamefont {D.}~\bibnamefont {Rocca}}, \bibinfo {author} {\bibfnamefont {R.}~\bibnamefont {Sabatini}}, \bibinfo {author} {\bibfnamefont {B.}~\bibnamefont {Santra}}, \bibinfo {author} {\bibfnamefont {M.}~\bibnamefont {Schlipf}}, \bibinfo {author} {\bibfnamefont {A.~P.}\ \bibnamefont {Seitsonen}}, \bibinfo {author} {\bibfnamefont {A.}~\bibnamefont {Smogunov}}, \bibinfo {author} {\bibfnamefont {I.}~\bibnamefont {Timrov}}, \bibinfo {author} {\bibfnamefont {T.}~\bibnamefont {Thonhauser}}, \bibinfo {author} {\bibfnamefont {P.}~\bibnamefont {Umari}}, \bibinfo
  {author} {\bibfnamefont {N.}~\bibnamefont {Vast}}, \bibinfo {author} {\bibfnamefont {X.}~\bibnamefont {Wu}},\ and\ \bibinfo {author} {\bibfnamefont {S.}~\bibnamefont {Baroni}},\ }\bibfield  {title} {\bibinfo {title} {Advanced capabilities for materials modelling with {{Quantum ESPRESSO}}},\ }\href {https://doi.org/https://doi.org/10.1088/1361-648X/aa8f79} {\bibfield  {journal} {\bibinfo  {journal} {Journal of Physics. Condensed Matter: An Institute of Physics Journal}\ }\textbf {\bibinfo {volume} {29}},\ \bibinfo {pages} {465901} (\bibinfo {year} {2017})}\BibitemShut {NoStop}%
\bibitem [{\citenamefont {Perdew}\ \emph {et~al.}(1996)\citenamefont {Perdew}, \citenamefont {Burke},\ and\ \citenamefont {Ernzerhof}}]{PBE1996}%
  \BibitemOpen
  \bibfield  {author} {\bibinfo {author} {\bibfnamefont {J.~P.}\ \bibnamefont {Perdew}}, \bibinfo {author} {\bibfnamefont {K.}~\bibnamefont {Burke}},\ and\ \bibinfo {author} {\bibfnamefont {M.}~\bibnamefont {Ernzerhof}},\ }\bibfield  {title} {\bibinfo {title} {{Generalized {{Gradient Approximation Made Simple}}}},\ }\href {https://doi.org/https://doi.org/10.1103/PhysRevLett.77.3865} {\bibfield  {journal} {\bibinfo  {journal} {Phys. Rev. Lett.}\ }\textbf {\bibinfo {volume} {77}},\ \bibinfo {pages} {3865} (\bibinfo {year} {1996})}\BibitemShut {NoStop}%
\bibitem [{\citenamefont {Hamann}(2013)}]{ONCVPSP2013}%
  \BibitemOpen
  \bibfield  {author} {\bibinfo {author} {\bibfnamefont {D.~R.}\ \bibnamefont {Hamann}},\ }\bibfield  {title} {\bibinfo {title} {{Optimized Norm-Conserving {{Vanderbilt}} Pseudopotentials}},\ }\href {https://doi.org/https://doi.org/10.1103/PhysRevB.88.085117} {\bibfield  {journal} {\bibinfo  {journal} {Physical Review B}\ }\textbf {\bibinfo {volume} {88}},\ \bibinfo {pages} {085117} (\bibinfo {year} {2013})}\BibitemShut {NoStop}%
\bibitem [{\citenamefont {{van Setten}}\ \emph {et~al.}(2018)\citenamefont {{van Setten}}, \citenamefont {Giantomassi}, \citenamefont {Bousquet}, \citenamefont {Verstraete}, \citenamefont {Hamann}, \citenamefont {Gonze},\ and\ \citenamefont {Rignanese}}]{PseudoDojo2018}%
  \BibitemOpen
  \bibfield  {author} {\bibinfo {author} {\bibfnamefont {M.~J.}\ \bibnamefont {{van Setten}}}, \bibinfo {author} {\bibfnamefont {M.}~\bibnamefont {Giantomassi}}, \bibinfo {author} {\bibfnamefont {E.}~\bibnamefont {Bousquet}}, \bibinfo {author} {\bibfnamefont {M.~J.}\ \bibnamefont {Verstraete}}, \bibinfo {author} {\bibfnamefont {D.~R.}\ \bibnamefont {Hamann}}, \bibinfo {author} {\bibfnamefont {X.}~\bibnamefont {Gonze}},\ and\ \bibinfo {author} {\bibfnamefont {G.~M.}\ \bibnamefont {Rignanese}},\ }\bibfield  {title} {\bibinfo {title} {{The {{PseudoDojo}}: {{Training}} and Grading a 85 Element Optimized Norm-Conserving Pseudopotential Table}},\ }\href {https://doi.org/https://doi.org/10.1016/j.cpc.2018.01.012} {\bibfield  {journal} {\bibinfo  {journal} {Computer Physics Communications}\ }\textbf {\bibinfo {volume} {226}},\ \bibinfo {pages} {39} (\bibinfo {year} {2018})}\BibitemShut {NoStop}%
\bibitem [{\citenamefont {Marzari}\ \emph {et~al.}(2012)\citenamefont {Marzari}, \citenamefont {Mostofi}, \citenamefont {Yates}, \citenamefont {Souza},\ and\ \citenamefont {Vanderbilt}}]{MLWF}%
  \BibitemOpen
  \bibfield  {author} {\bibinfo {author} {\bibfnamefont {N.}~\bibnamefont {Marzari}}, \bibinfo {author} {\bibfnamefont {A.~A.}\ \bibnamefont {Mostofi}}, \bibinfo {author} {\bibfnamefont {J.~R.}\ \bibnamefont {Yates}}, \bibinfo {author} {\bibfnamefont {I.}~\bibnamefont {Souza}},\ and\ \bibinfo {author} {\bibfnamefont {D.}~\bibnamefont {Vanderbilt}},\ }\bibfield  {title} {\bibinfo {title} {{Maximally localized Wannier functions: Theory and applications}},\ }\href {https://doi.org/https://doi.org/10.1103/RevModPhys.84.1419} {\bibfield  {journal} {\bibinfo  {journal} {Rev. Mod. Phys.}\ }\textbf {\bibinfo {volume} {84}},\ \bibinfo {pages} {1419} (\bibinfo {year} {2012})}\BibitemShut {NoStop}%
\bibitem [{\citenamefont {Pizzi}\ \emph {et~al.}(2020)\citenamefont {Pizzi}, \citenamefont {Vitale}, \citenamefont {Arita}, \citenamefont {Bl{\"u}gel}, \citenamefont {Freimuth}, \citenamefont {G{\'e}ranton}, \citenamefont {Gibertini}, \citenamefont {Gresch}, \citenamefont {Johnson}, \citenamefont {Koretsune} \emph {et~al.}}]{wannier90}%
  \BibitemOpen
  \bibfield  {author} {\bibinfo {author} {\bibfnamefont {G.}~\bibnamefont {Pizzi}}, \bibinfo {author} {\bibfnamefont {V.}~\bibnamefont {Vitale}}, \bibinfo {author} {\bibfnamefont {R.}~\bibnamefont {Arita}}, \bibinfo {author} {\bibfnamefont {S.}~\bibnamefont {Bl{\"u}gel}}, \bibinfo {author} {\bibfnamefont {F.}~\bibnamefont {Freimuth}}, \bibinfo {author} {\bibfnamefont {G.}~\bibnamefont {G{\'e}ranton}}, \bibinfo {author} {\bibfnamefont {M.}~\bibnamefont {Gibertini}}, \bibinfo {author} {\bibfnamefont {D.}~\bibnamefont {Gresch}}, \bibinfo {author} {\bibfnamefont {C.}~\bibnamefont {Johnson}}, \bibinfo {author} {\bibfnamefont {T.}~\bibnamefont {Koretsune}}, \emph {et~al.},\ }\bibfield  {title} {\bibinfo {title} {{Wannier90 as a community code: new features and applications}},\ }\href@noop {} {\bibfield  {journal} {\bibinfo  {journal} {Journal of Physics: Condensed Matter}\ }\textbf {\bibinfo {volume} {32}},\ \bibinfo {pages} {165902} (\bibinfo {year} {2020})}\BibitemShut {NoStop}%
\bibitem [{\citenamefont {Aichhorn}\ \emph {et~al.}(2016)\citenamefont {Aichhorn}, \citenamefont {Pourovskii}, \citenamefont {Seth}, \citenamefont {Vildosola}, \citenamefont {Zingl}, \citenamefont {Peil}, \citenamefont {Deng}, \citenamefont {Mravlje}, \citenamefont {Kraberger}, \citenamefont {Martins}, \citenamefont {Ferrero},\ and\ \citenamefont {Parcollet}}]{TRIQSDFTTOOLS2016}%
  \BibitemOpen
  \bibfield  {author} {\bibinfo {author} {\bibfnamefont {M.}~\bibnamefont {Aichhorn}}, \bibinfo {author} {\bibfnamefont {L.}~\bibnamefont {Pourovskii}}, \bibinfo {author} {\bibfnamefont {P.}~\bibnamefont {Seth}}, \bibinfo {author} {\bibfnamefont {V.}~\bibnamefont {Vildosola}}, \bibinfo {author} {\bibfnamefont {M.}~\bibnamefont {Zingl}}, \bibinfo {author} {\bibfnamefont {O.~E.}\ \bibnamefont {Peil}}, \bibinfo {author} {\bibfnamefont {X.}~\bibnamefont {Deng}}, \bibinfo {author} {\bibfnamefont {J.}~\bibnamefont {Mravlje}}, \bibinfo {author} {\bibfnamefont {G.~J.}\ \bibnamefont {Kraberger}}, \bibinfo {author} {\bibfnamefont {C.}~\bibnamefont {Martins}}, \bibinfo {author} {\bibfnamefont {M.}~\bibnamefont {Ferrero}},\ and\ \bibinfo {author} {\bibfnamefont {O.}~\bibnamefont {Parcollet}},\ }\bibfield  {title} {\bibinfo {title} {{{{TRIQS}}/{{DFTTools}}: {{A TRIQS}} Application for Ab Initio Calculations of Correlated Materials}},\ }\href {https://doi.org/https://doi.org/10.1016/j.cpc.2016.03.014} {\bibfield  {journal}
  {\bibinfo  {journal} {Computer Physics Communications}\ }\textbf {\bibinfo {volume} {204}},\ \bibinfo {pages} {200} (\bibinfo {year} {2016})}\BibitemShut {NoStop}%
\bibitem [{\citenamefont {Sim}\ and\ \citenamefont {Han}(2018)}]{MQEM2018}%
  \BibitemOpen
  \bibfield  {author} {\bibinfo {author} {\bibfnamefont {J.-H.}\ \bibnamefont {Sim}}\ and\ \bibinfo {author} {\bibfnamefont {M.~J.}\ \bibnamefont {Han}},\ }\bibfield  {title} {\bibinfo {title} {{Maximum Quantum Entropy Method}},\ }\href {https://doi.org/10.1103/PhysRevB.98.205102} {\bibfield  {journal} {\bibinfo  {journal} {Phys. Rev. B}\ }\textbf {\bibinfo {volume} {98}},\ \bibinfo {pages} {205102} (\bibinfo {year} {2018})}\BibitemShut {NoStop}%
\bibitem [{\citenamefont {Nakamura}\ \emph {et~al.}(2021)\citenamefont {Nakamura}, \citenamefont {Yoshimoto}, \citenamefont {Nomura}, \citenamefont {Tadano}, \citenamefont {Kawamura}, \citenamefont {Kosugi}, \citenamefont {Yoshimi}, \citenamefont {Misawa},\ and\ \citenamefont {Motoyama}}]{RESPACK}%
  \BibitemOpen
  \bibfield  {author} {\bibinfo {author} {\bibfnamefont {K.}~\bibnamefont {Nakamura}}, \bibinfo {author} {\bibfnamefont {Y.}~\bibnamefont {Yoshimoto}}, \bibinfo {author} {\bibfnamefont {Y.}~\bibnamefont {Nomura}}, \bibinfo {author} {\bibfnamefont {T.}~\bibnamefont {Tadano}}, \bibinfo {author} {\bibfnamefont {M.}~\bibnamefont {Kawamura}}, \bibinfo {author} {\bibfnamefont {T.}~\bibnamefont {Kosugi}}, \bibinfo {author} {\bibfnamefont {K.}~\bibnamefont {Yoshimi}}, \bibinfo {author} {\bibfnamefont {T.}~\bibnamefont {Misawa}},\ and\ \bibinfo {author} {\bibfnamefont {Y.}~\bibnamefont {Motoyama}},\ }\bibfield  {title} {\bibinfo {title} {Respack: An ab initio tool for derivation of effective low-energy model of material},\ }\href@noop {} {\bibfield  {journal} {\bibinfo  {journal} {Computer Physics Communications}\ }\textbf {\bibinfo {volume} {261}},\ \bibinfo {pages} {107781} (\bibinfo {year} {2021})}\BibitemShut {NoStop}%
\bibitem [{\citenamefont {Ye}\ \emph {et~al.}(2013)\citenamefont {Ye}, \citenamefont {Chi}, \citenamefont {Chakoumakos}, \citenamefont {Fernandez-Baca}, \citenamefont {Qi},\ and\ \citenamefont {Cao}}]{Ye_2013}%
  \BibitemOpen
  \bibfield  {author} {\bibinfo {author} {\bibfnamefont {F.}~\bibnamefont {Ye}}, \bibinfo {author} {\bibfnamefont {S.}~\bibnamefont {Chi}}, \bibinfo {author} {\bibfnamefont {B.~C.}\ \bibnamefont {Chakoumakos}}, \bibinfo {author} {\bibfnamefont {J.~A.}\ \bibnamefont {Fernandez-Baca}}, \bibinfo {author} {\bibfnamefont {T.}~\bibnamefont {Qi}},\ and\ \bibinfo {author} {\bibfnamefont {G.}~\bibnamefont {Cao}},\ }\bibfield  {title} {\bibinfo {title} {{Magnetic and crystal structures of $\mathrm{Sr}_{2}\mathrm{IrO}_{4}$: A neutron diffraction study}},\ }\href {https://doi.org/10.1103/PhysRevB.87.140406} {\bibfield  {journal} {\bibinfo  {journal} {Phys. Rev. B}\ }\textbf {\bibinfo {volume} {87}},\ \bibinfo {pages} {140406} (\bibinfo {year} {2013})}\BibitemShut {NoStop}%
\bibitem [{\citenamefont {Gu}\ \emph {et~al.}(2023)\citenamefont {Gu}, \citenamefont {Pandey},\ and\ \citenamefont {Tiwari}}]{Qiangqiang2023}%
  \BibitemOpen
  \bibfield  {author} {\bibinfo {author} {\bibfnamefont {Q.}~\bibnamefont {Gu}}, \bibinfo {author} {\bibfnamefont {S.~K.}\ \bibnamefont {Pandey}},\ and\ \bibinfo {author} {\bibfnamefont {R.}~\bibnamefont {Tiwari}},\ }\bibfield  {title} {\bibinfo {title} {{A Computational Method to Estimate Spin\textendash Orbital Interaction Strength in Solid State Systems}},\ }\href {https://www.sciencedirect.com/science/article/abs/pii/S0927025623000848} {\bibfield  {journal} {\bibinfo  {journal} {Computational Materials Science}\ }\textbf {\bibinfo {volume} {221}},\ \bibinfo {pages} {112090} (\bibinfo {year} {2023})}\BibitemShut {NoStop}%
\bibitem [{\citenamefont {Klett}\ \emph {et~al.}(2020)\citenamefont {Klett}, \citenamefont {Wentzell}, \citenamefont {Sch\"afer}, \citenamefont {Simkovic}, \citenamefont {Parcollet}, \citenamefont {Andergassen},\ and\ \citenamefont {Hansmann}}]{Klett2020}%
  \BibitemOpen
  \bibfield  {author} {\bibinfo {author} {\bibfnamefont {M.}~\bibnamefont {Klett}}, \bibinfo {author} {\bibfnamefont {N.}~\bibnamefont {Wentzell}}, \bibinfo {author} {\bibfnamefont {T.}~\bibnamefont {Sch\"afer}}, \bibinfo {author} {\bibfnamefont {F.}~\bibnamefont {Simkovic}}, \bibinfo {author} {\bibfnamefont {O.}~\bibnamefont {Parcollet}}, \bibinfo {author} {\bibfnamefont {S.}~\bibnamefont {Andergassen}},\ and\ \bibinfo {author} {\bibfnamefont {P.}~\bibnamefont {Hansmann}},\ }\bibfield  {title} {\bibinfo {title} {Real-space cluster dynamical mean-field theory: Center-focused extrapolation on the one- and two particle-levels},\ }\href {https://doi.org/10.1103/PhysRevResearch.2.033476} {\bibfield  {journal} {\bibinfo  {journal} {Phys. Rev. Res.}\ }\textbf {\bibinfo {volume} {2}},\ \bibinfo {pages} {033476} (\bibinfo {year} {2020})}\BibitemShut {NoStop}%
\bibitem [{\citenamefont {Sch\"afer}\ \emph {et~al.}(2021)\citenamefont {Sch\"afer}, \citenamefont {Wentzell}, \citenamefont {\ifmmode~\check{S}\else \v{S}\fi{}imkovic}, \citenamefont {He}, \citenamefont {Hille}, \citenamefont {Klett}, \citenamefont {Eckhardt}, \citenamefont {Arzhang}, \citenamefont {Harkov}, \citenamefont {Le~R\'egent}, \citenamefont {Kirsch}, \citenamefont {Wang}, \citenamefont {Kim}, \citenamefont {Kozik}, \citenamefont {Stepanov}, \citenamefont {Kauch}, \citenamefont {Andergassen}, \citenamefont {Hansmann}, \citenamefont {Rohe}, \citenamefont {Vilk}, \citenamefont {LeBlanc}, \citenamefont {Zhang}, \citenamefont {Tremblay}, \citenamefont {Ferrero}, \citenamefont {Parcollet},\ and\ \citenamefont {Georges}}]{Schaefer2021}%
  \BibitemOpen
  \bibfield  {author} {\bibinfo {author} {\bibfnamefont {T.}~\bibnamefont {Sch\"afer}}, \bibinfo {author} {\bibfnamefont {N.}~\bibnamefont {Wentzell}}, \bibinfo {author} {\bibfnamefont {F.}~\bibnamefont {\ifmmode~\check{S}\else \v{S}\fi{}imkovic}}, \bibinfo {author} {\bibfnamefont {Y.-Y.}\ \bibnamefont {He}}, \bibinfo {author} {\bibfnamefont {C.}~\bibnamefont {Hille}}, \bibinfo {author} {\bibfnamefont {M.}~\bibnamefont {Klett}}, \bibinfo {author} {\bibfnamefont {C.~J.}\ \bibnamefont {Eckhardt}}, \bibinfo {author} {\bibfnamefont {B.}~\bibnamefont {Arzhang}}, \bibinfo {author} {\bibfnamefont {V.}~\bibnamefont {Harkov}}, \bibinfo {author} {\bibfnamefont {F.~m. c.-M.}\ \bibnamefont {Le~R\'egent}}, \bibinfo {author} {\bibfnamefont {A.}~\bibnamefont {Kirsch}}, \bibinfo {author} {\bibfnamefont {Y.}~\bibnamefont {Wang}}, \bibinfo {author} {\bibfnamefont {A.~J.}\ \bibnamefont {Kim}}, \bibinfo {author} {\bibfnamefont {E.}~\bibnamefont {Kozik}}, \bibinfo {author} {\bibfnamefont {E.~A.}\ \bibnamefont {Stepanov}},
  \bibinfo {author} {\bibfnamefont {A.}~\bibnamefont {Kauch}}, \bibinfo {author} {\bibfnamefont {S.}~\bibnamefont {Andergassen}}, \bibinfo {author} {\bibfnamefont {P.}~\bibnamefont {Hansmann}}, \bibinfo {author} {\bibfnamefont {D.}~\bibnamefont {Rohe}}, \bibinfo {author} {\bibfnamefont {Y.~M.}\ \bibnamefont {Vilk}}, \bibinfo {author} {\bibfnamefont {J.~P.~F.}\ \bibnamefont {LeBlanc}}, \bibinfo {author} {\bibfnamefont {S.}~\bibnamefont {Zhang}}, \bibinfo {author} {\bibfnamefont {A.-M.~S.}\ \bibnamefont {Tremblay}}, \bibinfo {author} {\bibfnamefont {M.}~\bibnamefont {Ferrero}}, \bibinfo {author} {\bibfnamefont {O.}~\bibnamefont {Parcollet}},\ and\ \bibinfo {author} {\bibfnamefont {A.}~\bibnamefont {Georges}},\ }\bibfield  {title} {\bibinfo {title} {Tracking the footprints of spin fluctuations: A multimethod, multimessenger study of the two-dimensional hubbard model},\ }\href {https://doi.org/10.1103/PhysRevX.11.011058} {\bibfield  {journal} {\bibinfo  {journal} {Phys. Rev. X}\ }\textbf {\bibinfo {volume}
  {11}},\ \bibinfo {pages} {011058} (\bibinfo {year} {2021})}\BibitemShut {NoStop}%
\bibitem [{\citenamefont {Werner}\ and\ \citenamefont {Millis}(2010)}]{Werner2010}%
  \BibitemOpen
  \bibfield  {author} {\bibinfo {author} {\bibfnamefont {P.}~\bibnamefont {Werner}}\ and\ \bibinfo {author} {\bibfnamefont {A.~J.}\ \bibnamefont {Millis}},\ }\bibfield  {title} {\bibinfo {title} {Dynamical screening in correlated electron materials},\ }\href {https://doi.org/10.1103/PhysRevLett.104.146401} {\bibfield  {journal} {\bibinfo  {journal} {Phys. Rev. Lett.}\ }\textbf {\bibinfo {volume} {104}},\ \bibinfo {pages} {146401} (\bibinfo {year} {2010})}\BibitemShut {NoStop}%
\bibitem [{\citenamefont {Pauli}\ \emph {et~al.}(2025)\citenamefont {Pauli}, \citenamefont {Mishra}, \citenamefont {Rösner},\ and\ \citenamefont {van Loon}}]{Pauli2025}%
  \BibitemOpen
  \bibfield  {author} {\bibinfo {author} {\bibfnamefont {A.}~\bibnamefont {Pauli}}, \bibinfo {author} {\bibfnamefont {A.}~\bibnamefont {Mishra}}, \bibinfo {author} {\bibfnamefont {M.}~\bibnamefont {Rösner}},\ and\ \bibinfo {author} {\bibfnamefont {E.~G. C.~P.}\ \bibnamefont {van Loon}},\ }\href {https://arxiv.org/abs/2507.05974} {\bibinfo {title} {Static treatment of dynamic interactions in correlated electron systems}} (\bibinfo {year} {2025}),\ \Eprint {https://arxiv.org/abs/2507.05974} {arXiv:2507.05974 [cond-mat.str-el]} \BibitemShut {NoStop}%
\bibitem [{\citenamefont {Pr\"opper}\ \emph {et~al.}(2016)\citenamefont {Pr\"opper}, \citenamefont {Yaresko}, \citenamefont {H\"oppner}, \citenamefont {Matiks}, \citenamefont {Mathis}, \citenamefont {Takayama}, \citenamefont {Matsumoto}, \citenamefont {Takagi}, \citenamefont {Keimer},\ and\ \citenamefont {Boris}}]{propper_2016}%
  \BibitemOpen
  \bibfield  {author} {\bibinfo {author} {\bibfnamefont {D.}~\bibnamefont {Pr\"opper}}, \bibinfo {author} {\bibfnamefont {A.~N.}\ \bibnamefont {Yaresko}}, \bibinfo {author} {\bibfnamefont {M.}~\bibnamefont {H\"oppner}}, \bibinfo {author} {\bibfnamefont {Y.}~\bibnamefont {Matiks}}, \bibinfo {author} {\bibfnamefont {Y.-L.}\ \bibnamefont {Mathis}}, \bibinfo {author} {\bibfnamefont {T.}~\bibnamefont {Takayama}}, \bibinfo {author} {\bibfnamefont {A.}~\bibnamefont {Matsumoto}}, \bibinfo {author} {\bibfnamefont {H.}~\bibnamefont {Takagi}}, \bibinfo {author} {\bibfnamefont {B.}~\bibnamefont {Keimer}},\ and\ \bibinfo {author} {\bibfnamefont {A.~V.}\ \bibnamefont {Boris}},\ }\bibfield  {title} {\bibinfo {title} {Optical anisotropy of the ${J}_{\mathrm{eff}}=1/2$ mott insulator \text{Sr}$_2$\text{IrO}$_4$},\ }\href {https://doi.org/10.1103/PhysRevB.94.035158} {\bibfield  {journal} {\bibinfo  {journal} {Phys. Rev. B}\ }\textbf {\bibinfo {volume} {94}},\ \bibinfo {pages} {035158} (\bibinfo {year} {2016})}\BibitemShut
  {NoStop}%
\end{thebibliography}
\end{document}